\documentclass[showpacs,graphicx,floatfix,eqsecnum]{revtex4}
\usepackage{amssymb}
\usepackage{amsmath}
\usepackage{epsf}
\begin{document}
\title{Singular perturbation theory for interacting fermions in two
dimensions}
\author{Andrey V. Chubukov}
\vspace{0.5cm} \affiliation{Department of Physics, University of
Maryland, College Park, MD
20742-4111\\
and  Department of Physics, University of Wisconsin-Madison, 1150
Univ. Ave., Madison, WI 53706-1390}
\author{Dmitrii L. Maslov}
\affiliation{\cite{perm}Department of Physics, University of
Florida, P. O. Box 118440,
Gainesville, FL 32611-8440\\
and Abdus Salam International Centre for Theoretical Physics, 11
Strada Costiera, Trieste 34014, Italy}
\author{Suhas Gangadharaiah}
\affiliation{Department of Physics, University of Florida, P. O.
Box 118440, Gainesville, FL 32611-8440}
\author{Leonid I. Glazman}
\affiliation{Theoretical Physics Institute, University of
Minnesota, Minneapolis, MN 55455}
\begin{abstract}
We consider a system of interacting fermions in two dimensions
beyond the second-order perturbation theory in the interaction. It
is shown that the mass-shell singularities in the self-energy,
arising already at the second order of the perturbation theory,
manifest a non-perturbative effect: an interaction with the
zero-sound mode. Resumming the perturbation theory for a weak,
short-range interaction and accounting for a finite curvature of
the fermion spectrum, we eliminate the singularities and obtain
the results for the quasi-particle self-energy and the spectral
function to all orders in the interaction with the zero-sound
mode. A threshold for emission of zero-sound waves leads a
non-monotonic variation of the self-energy with energy (or
momentum) near the mass shell. Consequently, the spectral function
has a kink-like feature. We also study in detail a non-analytic
temperature dependence of the specific heat, $C(T)\propto T^2$. It
turns out that although the interaction with the collective mode
results in an enhancement of the fermion self-energy, this
interaction does not affect the non-analytic term in $C(T)$ due to
a subtle cancellation between the contributions from the real and
imaginary parts of the self-energy. For a short-range and weak
interaction, this implies that the second-order perturbation
theory suffices to determine the non-analytic part of $C(T)$.
 We also obtain a general form
of the non-analytic term in $C(T)$, valid for the case of a
generic Fermi liquid, \emph{i.e.}, beyond the perturbation theory.

\end{abstract}
\maketitle
\section{Introduction}
\label{sec:intro} The validity of the Landau Fermi liquid (FL)
theory continues to be a subject of intense discussions over the
last five decades. In essence, the FL theory states that the
behavior of interacting fermions at the lowest energies is similar
to that of non-interacting fermions~\cite{agd,statphys}. In
particular, specific heat, $C\left( T\right) ,$ scales linearly
with temperature $T$ at $T\rightarrow 0$, whereas the spin
susceptibility, $\chi \left( T\right) , $ approaches a finite
value at $T\rightarrow 0.$ This theory has been enormously
successful in describing He$^{3}$ and a large number of metals.
Yet, it fails to explain the properties of high-temperature
superconductors~\cite{tom} and quite a few heavy-fermion
materials~\cite{piers}. Given that the FL theory is essentially a
phenomenological one, built on a number of appealing albeit
unproven assumptions, the interest in its relation to microscopic
models, which allow for controllable perturbative treatment, has
always been intense since the times of the FL's inception.
Perturbative calculations of the 1950s within a model of a
low-density 3D Fermi gas with a repulsive, short-range interaction
(a ``non-ideal Fermi gas model''~\cite{agd,statphys}) reproduced
the Fermi-liquid results for thermodynamic quantities up to
leading terms in parameter $ T/E_{F}$, where $E_{F}$ is the Fermi
energy (we set $\hbar =k_{\rm{B}}=1$ throughout the paper). These
calculations were later extended to arbitrary dimensionality $D$
with the result that the FL theory is valid for $D>1,$ provided
that the interaction falls off with the distance rapidly enough,
but becomes invalid in one dimension. The energy range for the FL
theory, however, may shrink as the system approaches a quantum
phase transition of some kind~\cite{abanov}.

At the same time, perturbative calculations show that the
similarity between a Fermi liquid and ideal Fermi gas does not go
beyond the leading order in $ T/E_{F}$. In a Fermi gas,
sub-leading terms in $C\left( T\right) /T$ and $\chi \left(
T\right) $ form regular series in powers of $T^{2}$. For an
interacting system, the second-order perturbation theory shows
that already next-to-leading terms in $ T/E_{F}$
 are non-analytic in $D \leq 3$. In $D=3$, the sub-leading term in $%
C\left( T\right) /T$ is $T^{2}\ln T$ \cite
{eliashberg,doniach,brinkman,amit,comment_1}. \ In addition, a
non-uniform spin susceptibility, $\chi \left( Q\right)$, scales as
$Q^{2}\ln Q,$ where $Q$ is the boson momentum \cite{belitz}. The
non-analytic behavior becomes more pronounced as the system
dimensionality is reduced. In 2D, the sub-leading terms in
$C\left( T\right) /T$
 and in $\chi \left( Q,T\right)$ scale as  $T$ \cite{bedell,chm,dassarma}
and
 $\max \{Q,T\} $ \cite{belitz,marenko,millis,chm,pepin}, respectively.
  Non-analytic corrections to the FL
behavior have been observed in a number of experiments
\cite{experiment}. Quite recently, the interest to these
corrections has been revitalized due to their importance for the
effective theories of quantum critical phenomena in itinerant
ferromagnets \cite{belitz,belitz_qc,pepin} of the
Hertz-Millis-Moriya type \cite{qc}.

The non-analytic behavior of the thermodynamic quantities is
related to the long-range \emph{effective} interaction between
fermions which falls off as a power law of the inter-fermion
separation, even if the nominal interaction is short-range. An
example of such an interaction is scattering from the Friedel
oscillation imposed by a static local perturbation, \emph{e.g.},
an impurity. This interaction falls off as $r^{-D} $ with the
distance from the impurity. Scattering from the $r^{-D}-$
potential of the Friedel oscillation leads to a non-analytic
energy dependence of the scattering amplitude near the Fermi level
and, as a result, to non-analytic corrections to the tunneling
density of states \cite{rudin} and conductivity \cite{zna}. In a
disorder-free-system, fermions interact via \emph{dynamic }charge-
and spin-density fluctuations. The non-analyticities in the
dynamic density-density correlation function (polarization bubble)
also give rise to a long-range {\it retarded} interaction
 falling off as $
r^{-(D-1)}$ \cite{bedell,chm}. These non-analyticities are due to
the  processes with both small and large ($2k_{F}$) momentum
transfers, where $k_{F}$ is the Fermi momentum.

The second order perturbative analysis in 2D not only reveals a
non-analytic behavior of $C(T)$ and $\chi (Q,T)$, but also brings
about an unexpected result. Namely, the imaginary part of the
fermion self-energy diverges logarithmically on the mass shell
$\omega =\epsilon _{k}$~
\cite{castellani,fukuyama,metzner}, \cite{chm}, if one linearizes
the single-particle spectrum, $\epsilon _{k},$ near the Fermi
level. This log-singularity is the 2D analog of a
stronger--power-law--singularity in 1D (\lq\lq infrared
catastrophe\rq\rq ) \cite{1D}. Although the mass-shell divergence
does not affect the specific heat and the spin susceptibility to
second order in the interaction, it does signal a potential
breakdown of the perturbation theory in 2D. ~

The purpose of this paper is to analyze the non-analytic
corrections to the FL\ behavior beyond the second-order
perturbation theory. Specifically, we focus on two issues. The
first one is what happens to the mass-shell singularity beyond
second order. Power counting shows that the mass-shell
singularities proliferate with the order of the perturbation
theory. At first glance, this confirms Anderson's conjecture that
the FL is destroyed in 2D \cite{anderson}. However, this issue can
be addressed properly only after a re-summation of the
perturbation theory, which is what we will do here. The second
issue is whether collective modes, which emerge once the
perturbation theory is summed up to all orders, give rise to extra
non-analytic corrections to thermodynamic quantities. The role of
collective modes -- zero sound for neutral fermions and plasmon
for electrons -- is especially intriguing for $D=2$. In this case,
power counting combined with the assumption that a collective mode
is a free excitation (similar to a phonon) shows that the
collective mode contribution to $C\left( T\right) /T$ scales as
$T,$ \emph{i.e.,} it has the same form as the perturbative
correction \cite {bedell}. This argument needs to be treated with
caution, however. Indeed, since a collective mode arises at the
infinite order in the interaction between
fermions, it is unclear whether it can be treated as a free boson mode.%
\emph{\ }We will see that the issues of the collective-mode
contribution to $C(T)$ and mass-shell singularities in the
self-energy are related; namely, a contribution to the specific
heat from the collective mode can be viewed as coming from the
self-energy obtained by summing up mass-shell singularities to all
orders in the perturbation theory.

This paper is organized as follows. In Sec.~\ref{sec:process} we
introduce relevant scattering processes. In Sec.~\ref{sec:sigma},
we discuss the self-energy of 2D fermions with both contact and
finite-range interactions. In Sec.~\ref{sec:imsigma} and
~\ref{sec:imsigma_1}  we analyze the mass-shell singularities in
the imaginary part of the self-energy arising in the
order-by-order perturbation theory. Re-summation of the
perturbation series for the vertex part is performed in
Sec.~\ref{sec:resumation}. The imaginary and real parts of the
self-energy upon re-summation are discussed in
Secs.~\ref{sec:imsigma_all} and ~\ref {sec:resigma}, respectively.
In Sec.~\ref{sec:spectral}, we demonstrate that the spectral
function exhibits a non-monotonic variation near the mass shell
due to the interaction of fermions with the zero-sound mode. In
Sec.~%
\ref{sec:sh}, we discuss the  non-analytic contribution to the
specific heat in two ways. First, in Sec.~\ref{sec:cviasigma}, we
find $C\left( T\right) $ via the self-energy, utilizing the
results of
Sec.~\ref{sec:sigma}%
. Then, in Sec.~\ref{sec:sh_omega_m}, we evaluate the the
non-analytic part of the thermodynamic potential directly. In
Sec.~\ref{sec:generic}, we consider the specific heat in a generic
Fermi liquid. In Sec.~\ref{sec:coulomb}, we consider the case of a
Coulomb potential. Our conclusions are given in
Sec.~\ref{sec:conclusions}. Details of some of the calculations
are presented in Appendices A-F.

For the  convenience of a reader, we present below a summary of
the main results of this paper.

\subsection{Summary of the results}

\subsubsection{Self-energy}

In Secs. \ref{sec:process}-\ref{sec:sigma}, we consider mostly a
2D system of fermions with a weak, short-range
 repulsive interaction, specified by its
Fourier-transform $U(Q)$. The self-energy of such a system
consists of two parts: a analytic one and a non-analytic one. The
analytic part of the self-energy,
\begin{equation}
\Sigma _{\mathrm{an}}=a\omega +b\epsilon _{k}+c ~i\omega ^{2},
\label{sigma_an}\end{equation}
where $a,b$, and $c$ are real, is
determined by
 scattering events with large
 momentum transfers, of order $k_{F}$.
 In this
paper, we will be interested only in the non-analytic part of the
self-energy, which comes from two types of effectively 1D
scattering processes. In the first type (``forward scattering''),
all four momenta--two incoming and two outgoing-- align almost
along the same direction [cf. Fig.~\ref{fig:proc}(a)]. In the
second type (``backscattering''), both the initial and final
momenta of the fermion pair are close to zero, while the momentum
transfer can be near either zero or $2k_{F}$ [cf.
Figs.~\ref{fig:proc}(b) and (c)]. The angular spreading of the
trajectories shrinks in the low-energy limit in proportion to
 $\left| \omega \right| /E_{F}$ for both types of
scattering. To second order in the interaction and for a
linearized single-particle spectrum [$\epsilon
_{k}=v_{F}(k-k_{F})$], the forward (F) and backscattering (B)
contributions to the self-energy near the mass shell
are~\cite{castellani,fukuyama,metzner}, \cite{chm}
\begin{subequations}
\begin{eqnarray}
\text{Re}\Sigma _{\mathrm{F}}^{R}(\omega =\epsilon _{k}) &=&0,~~~\text{Im}%
\Sigma _{\mathrm{F}}^{R}(\omega ,k)=\frac{u^{2}}{8\pi }~\frac{\omega ^{2}}{%
E_{F}}~\ln {\frac{E_{F}}{|\Delta |}};  \label{may_31x} \\
\text{Re}\Sigma _{\mathrm{B}}^{R}(\omega =\epsilon _{k}) &=&-\frac{%
u_{0}^{2}+u_{2k_{F}}^{2}-u_{0}u_{2k_{F}}}{8}~\frac{\omega |\omega |}{E_{F}}%
,~~~\text{Im}\Sigma _{\mathrm{B}}^{R}(\omega ,k)=\frac{%
u_{0}^{2}+u_{2k_{F}}^{2}-u_{0}u_{2k_{F}}}{4\pi }~\frac{\omega ^{2}}{E_{F}}%
~\ln {\frac{E_{F}}{|\omega |}}.  \label{may_31y}
\end{eqnarray}
Here $E_{F}=k_{F}v_{F}/2$ is the Fermi energy,
\end{subequations}
\begin{equation}
u_{0}\equiv \frac{mU(0)}{2\pi },~~~u_{2k_{F}}\equiv
\frac{mU(2k_{F})}{2\pi } \label{i1}
\end{equation}
are the dimensionless coupling constants which are assumed to be
small, and
\begin{equation}
\Delta \equiv \omega -\epsilon _{k},  \label{delta}
\end{equation}
is the ``distance''\/ to the mass shell.  On the Fermi surface
($\epsilon _{k}=0$), $\text{Im} \Sigma^{R} = \text{Im} \Sigma
_{\mathrm{F}}^{R} + \text{Im} \Sigma _{\mathrm{B}}^{R}$ reduces to
a familiar form $\text{Im} \Sigma^{R} (\omega, k_F) \propto \omega
^{2}\ln |\omega |$ \cite{2D}.

The special role of backscattering processes for the non-analytic
corrections to thermodynamic variables of a FL has been considered
earlier by two of us~\cite{chm}. In this paper, we present a
complete description of forward-scattering processes. The
peculiarities of these processes show up already at the second
order: we see from Eq.~(\ref{may_31x}) that
 on the mass shell, where $\Delta =0$,
$\mathrm{Im}\Sigma _{\mathrm{F}}^{R}$ diverges logarithmically. The
 divergence  is regularized~\cite{castellani,fukuyama,metzner}, \cite{chm} by restoring a finite
curvature of the single-particle spectrum, $m_{c}^{-1}\equiv
\partial ^{2}\epsilon _{k}/\partial k_{\perp }^{2},$ where
$\mathbf{{k}_{\perp }}$ is the
component of $\mathbf{k}$ transverse to the local
Fermi velocity $\mathbf{v}%
_{F}(\mathbf{k})$. Finite curvature brings in a new scale
\begin{equation}
\Delta _{c}\equiv \omega ^{2}/W,  \label{delta_c}
\end{equation}
where $W\equiv m_{c}v_{F}^{2}/2,$ and the logarithmic singularity
in Eq.~(\ref {may_31x}) is rounded off at $\Delta \simeq \Delta
_{c}$.

At a first glance, this regularization stabilizes the perturbation
theory. However, starting from the third order in the interaction,
the divergences due to forward scattering become of a power-law
form; for a linearized
 spectrum, we find
\begin{equation}
\text{\textrm{Im}}\Sigma _{\mathrm{F}}^{R}\left( \omega ,k\right)
\propto (u_{0}^{2}|\omega /\Delta |)^{n/2-1},  \label{pert>2}
\end{equation}
where $n$ is the order of the perturbation theory. Finite
curvature rounds off the power-law singularity on a scale $\Delta
\simeq \Delta_c$ at every given order, but the resulting series
for the self-energy holds in parameter $\omega_c/|\omega|$, where
 $\omega _{c}\equiv u_{0}^{2} W/2$, and
 does not converge for $|\omega | < \omega _{c}$.
\emph{Therefore, the perturbation theory in 2D must be re-summed
even for an infinitesimally weak interaction and realistic fermion
spectrum}.
 We show that the most divergent
contributions to the forward-scattering part of the self-energy
can be re-summed exactly to all orders in $u_{0}$, without
exploiting RPA-type approximations.

Upon re-summation, the origin of the mass-shell singularities in
the perturbation theory becomes clear: they correspond to the
interaction between fermions and the zero-sound (ZS) collective
mode. At every finite order of the perturbation theory, the
collective mode coincides with the upper edge of the particle-hole
continuum, and this degeneracy generates divergences in $\Sigma
_{\mathrm{F}}^{R}$. Once the perturbations are summed up to all
orders, the ZS mode splits off from the continuum, and the
power-law divergences disappear. The remaining logarithmic
singularity is eliminated by the finite curvature.

The  total self-energy after re-summation is
described by a scaling function of two variables $\Delta /\Delta _{c}$ and $%
\Delta /\Delta ^{\ast },$ where $\Delta _{c}$ is defined in Eq.~(\ref{delta_c}),
and
\begin{equation}
\Delta ^{\ast }\equiv u_{0}^{2}\omega /2.  \label{delta_star}
\end{equation}
is the scale at which perturbation series Eq.~(\ref{pert>2}) diverges
 for a linearized spectrum.
A general form of the scaling function is rather complicated, and
will be discussed in the main text of the paper. In the limit
$\Delta \rightarrow 0$,
 when both scaling variables are small, and also for low frequencies,
$\Sigma ^{R}$ reduces to
\begin{subequations}
\begin{eqnarray}
\mathrm{Re}\Sigma ^{R}(\omega &=&\epsilon
_{k})=\frac{u_{2k_{F}}\left( u_{0}-u_{2k_{F}}\right)
}{8}\frac{\omega \left| \omega \right| }{E_{F}};
\label{may31_a} \\
\mathrm{Im}\Sigma ^{R}(\omega  &=& \epsilon_k) = \frac{u_{0}^{2}}{4\pi }\frac{\omega ^{2}%
}{E_{F}}\ln \frac{W}{|\omega |},  \label{may31_1b}
\end{eqnarray}
\end{subequations}
Comparing Eq.~(\ref{may31_a}), (\ref{may31_1b})  with the second-order self-energy,
[Eq.~(\ref{may_31x}), (\ref{may_31y})], we see that the re-summation (i) eliminates
the divergence in $\mathrm{Im}\Sigma ^{R}(\omega =\epsilon_k)$ and
(ii)  drastically changes the
 result for $\mathrm{Re}\Sigma ^{R}(\omega =\epsilon_k)$.
 In particular, for a constant interaction ($u_{0}=u_{2k_{F}}$),
re-summation of higher-order terms in the self-energy
 cancels out the second-order term, so that full
$\mathrm{Re}\Sigma ^{R}(\omega =\epsilon _{k})$ vanishes.

Perhaps the most essential
 result of our analysis is that the interaction
 with the zero-sound not only leads to a scaling behavior of the
self-energy, but also results in a singularity of the
 self-energy: the derivative $d
\mathrm{Im}\Sigma ^{R}/d\Delta$ diverges as $1/\sqrt{\Delta
-\Delta^\ast}$ at $\Delta = \Delta^\ast$. This singularity is
present in the non-perturbative regime, \emph{i.e.}, for
$|\omega|<\omega_c$. Physically, it corresponds to a change
 in kinematics of ZS waves emission.
On the mass-shell ($\Delta =0$), emission of ZS waves by fermions
is impossible as the zero-sound velocity is larger than the Fermi
one. For $0<\Delta < \Delta^\ast$, emission is possible but it is
subject to a Cherenkov-type restriction: a fermion with frequency
$\omega >0$ can only emit a ZS wave in the frequency interval
$\Omega< \omega \Delta/\Delta^\ast$. For $\Delta> \Delta^\ast$,
emission of ZS waves
 in the whole interval $0<\Omega<\omega$ becomes possible.
 The self-energy is singular right
 at the onset of the Cherenkov-type restriction for emission of
 ZS waves.

The singularity in the self-energy
 translates into a kink in the spectral function,
$A(\omega ,k)=-\pi ^{-1}\text{Im}G^{R}(\omega ,k)$ at $\Delta
=\Delta ^{\ast }$ (cf. Fig.~\ref{fig:spectral}). This effect is
actually present for both short-range and Coulomb interaction. For
the latter, the collective mode is a plasmon, and the
 kink is positioned near the Fermi surface, where $\epsilon_k=0$ (or $\Delta =\omega$).
 The
 prediction for a kink $A(\omega ,k)$
 can be verified in
angle-resolved photoemission on layered compounds \cite{photo} or
in momentum-conserving tunneling between two parallel layers of 2D
electron gas \cite{eisenstein}.

\subsubsection{specific heat}

In the second part of the paper (Sec.~\ref{sec:sh}), we analyze
the non-analytic behavior of the specific heat, $C(T)$,  for three
types of interaction: i) short-range, weak repulsion; ii) Coulomb
interaction; and iii) generic Fermi-liquid interaction. For all
three cases, we find that a non-analytic term in $C(T)$ behaves as
$T^2$. In the perturbation theory, an origin of this term can be
simply related to a non-analytic, $\omega |\omega |$ form of the real
part of the self-energy. To second order in the interaction, a
non-analytic part of $C(T)$
\begin{equation}
\delta C(T)\equiv C(T)-\gamma T,
\end{equation}
where $\gamma $ is the (interaction-dependent) Sommerfeld factor,
was shown earlier~\cite{chm} to be
\begin{equation}
\delta C\left( T\right) /T=-\left(
u_{0}^{2}+u_{2k_{F}}^{2}-u_{0}u_{2k_{F}}\right) ~\frac{9\zeta (3)}{\pi ^{2}}%
C_{\mathrm{FG}}/E_{F},  \label{genu_1}
\end{equation}
where
\begin{equation}
C_{\mathrm{FG}}=m\pi T/3  \label{FG}
\end{equation}
is the specific heat of a Fermi gas [Eq.~(\ref{FG})] (see also
Refs.~\cite {bedell,dassarma,aleiner}).

The issue considered in this paper is whether $\delta C(T)$ is
affected by the interactions of fermions with the zero-sound mode.
At a first glance, it
should be. Indeed, $C(T)$ is related to an exact retarded Green's function $%
G^{R}\left( \omega ,k\right) =\left[ \omega -\epsilon _{k}+\Sigma
^{R}(\omega ,k)\right] ^{-1}$ via \cite{agd}
\begin{equation}
C(T)/T=-\frac{2}{\pi }~\frac{\partial }{\partial T}\left[ \frac{1}{T}%
~\int \frac{d^{2}k}{(2\pi )^{2}}~\int_{-\infty
}^{\infty
}d\omega \omega \frac{\partial n_{0}}{\partial \omega }\arg G^{R}(\omega ,k)%
\right],   \label{feb5_2i}
\end{equation}
where $n_{0}$ is the Fermi distribution function. As the real part
of the self-energy is changed significantly by a non-perturbative
contribution from the ZS mode, the corresponding change in $G^{R}$
should \emph{a priori} affect $C(T). $ However, we see from
Eq.~(\ref{feb5_2i}) that--contrary to the common wisdom-- not only
the real but also the imaginary part of $\Sigma ^{R}$ affect
$C(T).$ Indeed, at low frequencies, which we only need at small
$T$,
  both perturbative and
non-perturbative parts of the self-energy are asymptotically smaller than $%
|\omega |;$ thus $G^{R}$ in Eq.~(\ref{feb5_2i}) can be expanded to
first order in $\Sigma ^{R}$ with the result
\begin{equation}
\delta C(T)/T=\frac{2}{\pi }~\frac{\partial }{\partial T}\left[ \frac{1}{T}%
~\int \frac{d^{2}k}{(2\pi )^{2}}~\int_{-\infty
}^{\infty
}d\omega \omega \frac{\partial n_{0}}{\partial \omega }\left\{ \text{Re}%
\Sigma ^{R}(\omega ,k)\text{Im}G_{0}^{R}(\omega
,k)+\text{Im}\Sigma ^{R}(\omega ,k)\text{Re}G_{0}^{R}(\omega
,k)\right\} \right] . \label{feb5_1}
\end{equation}
Substituting the second-order result for $\text{Re}\Sigma ^{R}$,
Eq.~(\ref {may_31y}), into Eq.~(\ref{feb5_1}), we indeed reproduce the
 $T$-dependence of $\delta C(T)/T$, as given by Eq.~(%
\ref{genu_1}) (more care is required to reproduce a numerical
prefactor as it turns out that one should use an expression for
Re$\Sigma ^{R}(\omega ,k)$ at finite temperatures-- cf.
Sec.~\ref{sec:cviasigma}). The perturbative part of Im$\Sigma
^{R}(\omega ,k)$ does not contribute to the specific heat as
it depends on $\omega $ but not on $k$;  as a result, the second term
in Eq.~(\ref{feb5_1}) vanishes by parity upon switching from
integration over $d^{2}k$ to that over $d\epsilon _{k}.$ However,
the non-perturbative  contribution to Im$\Sigma ^{R}(\omega ,k)$
due to the interaction with the zero-sound mode
 depends strongly on $k$. As a result, both \textrm{Re}$\Sigma^{R}(\omega
,k)$ and \textrm{Im}$\Sigma^{R}(\omega ,k)$ contribute to $%
C\left( T\right) .$ \ We show that non-perturbative terms in these
 two contributions
\textit{cancel} each other, \emph{i.e.}, there is no
non-perturbative contribution to the specific heat. A
non-analytic, $T^{2}$-term in $\delta C(T)$ then comes entirely
from the perturbative part of $\Sigma ^{R}$, and Eq.~(\ref{genu_1}) is the complete result for $\delta C(T)$ to second
order in the interaction.

Another way to understand an absence of the non-perturbative
contribution to the specific heat is to evaluate the thermodynamic
potential, $\Xi ,$ directly in Matsubara frequencies, and then use
the relation between $\Xi $ and $C\left( T\right) .$ This is done
in Sec.~\ref{sec:sh_omega_m}. In contrast to the real-frequency
description of the self-energy, there are no singularities at any
order of the perturbation theory for $\Xi $ in Matsubara
frequencies. This means that, as long as the interaction is weak,
one can truncate the perturbative series at an arbitrary order and
be sure that the higher-order terms would give only sub-leading
contributions. In this approach, \emph{there simply cannot be
non-perturbative contributions to the thermodynamic potential,}
and\emph{\ }hence to $C\left( T\right) $.
To second order, this procedure gives the same result as in Eq.~(\ref{genu_1}%
). For completeness, we also evaluate thermodynamic potential in
real frequencies and demonstrate explicitly how the
collective-mode contribution to the specific heat cancels out.

Finally, we extend our analysis to a Fermi liquid with not
necessarily weak interaction. We find that the $T^{2}$ term in the
specific heat for a generic Fermi liquid is expressed via the
charge (c) and spin (s) components, $f_{c,s}(\theta )$, of the
quasi-particle scattering amplitude between particles at the Fermi
surface, at angle $\theta =\pi $ between two incoming momenta as
\begin{equation}
\delta C(T)/T=-\frac{3\zeta (3)}{2\pi \left( v_{F}^{\ast }\right) ^{2}}~%
\left[ f_{c}^{2}(\pi )+3f_{s}^{2}(\pi )\right] T,
\label{jul3_4_1}
\end{equation}
where $v_{F}^{\ast }=k_{F}/m^{\ast }$ and $m^{\ast }$ is the
renormalized effective mass. We remind the reader that the
scattering amplitude (as a tensor in the spin space) is related to
a particular limiting form of the interaction vertex, $\Gamma
_{a}^{k}(\theta )$, as \cite{agd,statphys}
\begin{equation}
{\hat{f}}\left( \theta \right) =Z^{2}{\hat{\Gamma}}^{k}\left(
\theta \right) ,
\end{equation}
where
\begin{equation}
{\hat{\Gamma}}^{k}(\theta )=\lim_{|\Omega |/Q\rightarrow 0}{\hat{\Gamma}}%
(k_{F}{\hat{n}}_{1},0;k_{F}{\hat{n}}_{2},0|\mathbf{Q},\Omega ),
\label{gammak}
\end{equation}
where ${\hat{\Gamma}}(\mathbf{k}_{1},\omega _{1};\mathbf{k}_{2},\omega _{2}|%
\mathbf{Q},\Omega )$ is the vertex for a process $\left( \mathbf{k_{1}}%
,\omega _{1};\mathbf{k}_{2},\omega _{2}\rightarrow \mathbf{k}_{1}-\mathbf{Q}%
,\omega -\Omega ;\mathbf{k}_{2}+\mathbf{Q},\omega +\Omega \right) $, and $%
\theta $ is the angle between $\mathbf{k}_{1}$ and
$\mathbf{k}_{2}$. In the FL theory, the renormalizations of the
thermodynamic quantities are expressed via the angular moments of
the Landau interaction function, which is related to another
limiting form of the vertex,
\begin{equation}
{\hat{\Gamma}}^{\omega }(\theta )=\lim_{Q/|\Omega |\rightarrow 0}{\hat{\Gamma%
}}(k_{F}{\hat{n}}_{1},0;k_{F}{\hat{n}}_{2},0|\mathbf{Q},\Omega ).
\label{gammaomega}
\end{equation}
Simple algebraic relations between the \textit{partial components} of ${\hat{%
\Gamma}}^{k}(\theta )$ and ${\hat{\Gamma}}^{\omega }(\theta )$
enable one to express the analytic parts of thermodynamic
quantities either via the moments of the Landau interaction
function or that of the scattering amplitude. However, the
non-analytic, $T^{2}$- part of the specific
heat is related to ${\hat{\Gamma}}^{k}(\theta )$ at a particular angle ($%
\theta =\pi $), rather than to its angular average. As there is no
simple relation between ${\hat{\Gamma}}^{k}(\theta )$ and
${\hat{\Gamma}}^{\omega }(\theta )$ for any given angle, including
$\theta =\pi $, the non-analytic in the specific heat in general
\textit{cannot} be expressed in a compact form in terms of the
Landau interaction function without making additional model
approximations \cite{amit}. In this respect, our result for
$\delta C(T)$ differs from that of Ref.~\cite{aleiner}, where
$\delta C(T)$ was
expressed via the charge and spin components of a single Landau parameter ${%
\hat{F}}^{0}\propto \int d\theta {\hat{\Gamma}}^{\omega }(\theta
)$. We did, however, obtain $\delta C(T)$ in terms of
${\hat{\Gamma}}^{\omega }(\pi )$ in the limit when its charge
component is much larger than the spin one. The limit when the
charge component tends to infinity whereas the spin one tends to
zero describes the Coulomb interaction in the high-density limit.
In this case, we find that the $T^{2}$ term in $C(T)$ is universal
and independent of the electron charge [cf. Eq.~(\ref{genu_1111})],
in agreement with Ref.~\cite{aleiner}.

In the rest of the paper we present the details of our analysis.

 \section{Scattering processes}
\label{sec:process}
\begin{figure}[tbp]
\begin{center}
\epsfxsize=1.0 \columnwidth \epsffile{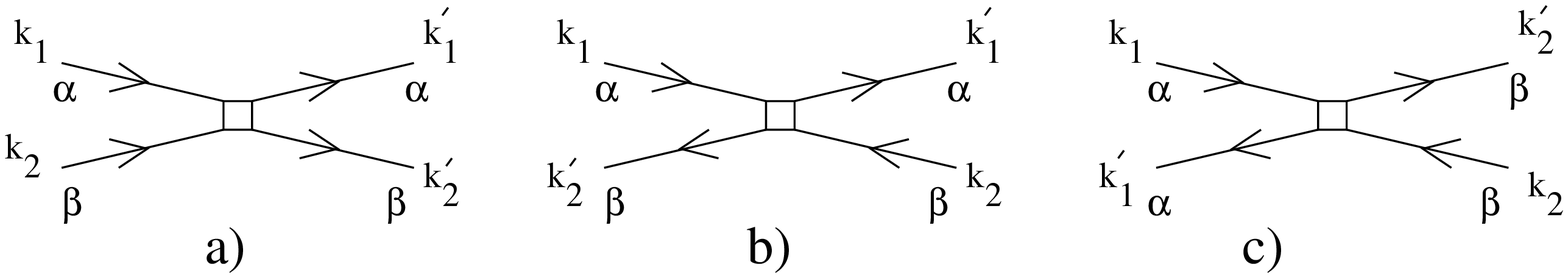}
\end{center}
\caption{Scattering processes responsible for divergent and/or
non-analytic corrections to the self-energy in 2D. a) ``Forward
scattering''--an analog of the ``$g_{4}$''-process in 1D. All four
fermion momenta are close to each other. b) Backscattering--an
analog of the ``$g_{2}$''-process in 1D. The net momentum before
and after collision is small. Initial momenta are close to the
final ones. Although the momentum transfer in such a process is
small, we still refer to this process as ``backscattering'' (see
the discussion in the main text). c) Another component of the
backscattering process: $2k_{F}-$ scattering.} \label{fig:proc}
\end{figure}

In a typical event of interaction between low-energy
quasi-particles with momenta $k_1\approx k_2\approx k_F$,
 the change in the
momentum of a given quasi-particle, $\delta k\equiv
|\mathbf{{k_1}-k_1^{\prime}|}$, is of order $k_F$, but not
necessarily close either to zero or to $2k_F$. These large-angle
scattering events are responsible for the analytic part of the
self-energy, Eq.~(\ref{sigma_an}).
 In addition, there are special scattering events in which either $%
\delta k\simeq |\Omega|/v_F \ll k_F$or $|\delta k-2k_F|\simeq
|\Omega|/v_F \ll k_F$, where $\Omega$ is the energy transfer.
Although the phase space associated with these events is small for
$D>1$, these processes give rise to non-analyticities in the
dynamic density-density correlation function  and eventually
determine non-analyticities in $\Sigma (\omega)$~\cite{chm}. The
role of these special processes increases as the dimensionality is
reduced. For $D=3$, the
non-analytic part of the self-energy $\Sigma_{\mathrm{na}%
}\propto \omega ^3\ln(-i\omega)$, resulting from the special
processes, is sub-leading to the analytic one,  resulting from the
generic processes. However, already for $D=2$, the non-analytic
part $\left(\Sigma \right)_{\mathrm{a}}\propto
i\omega^2\ln\left(-i\omega\right)$ dominates over the analytic
one.

In 2D, kinematics of processes with small momentum is essentially
{\it one-dimensional}, \emph{i.e.}, the initial and final momenta
of two interacting fermions are either almost parallel or
antiparallel to each other. (In 3D, both 1D and non-1D processes contribute to the non-analytic
behavior.) Accordingly, these processes can be
divided into two types. In the first type, the two colliding
particles move initially almost in the same direction ($\mathbf{k_{1}}%
\approx \mathbf{k_{2})}$ and retain their respective momenta after
the collision, so that all four momenta (two initial and two
final) are close to each other
\begin{equation}
\mathbf{k}^{\prime }\mathbf{_{1}}\approx \mathbf{k_{1}}\approx
\mathbf{k}_{2}\approx \mathbf{k_{2}}^{\prime }.
\end{equation}
 This type of process is
shown in Fig.~\ref{fig:proc}(a).  In $g$-ology \cite{solyom}, such
an event is called ``$g_{4}$-scattering''. The deviation from the
purely 1D kinematics is due to finite energy transfers: a typical
angle between momenta in Fig.~\ref{fig:proc}(a) is of order
$|\omega|/v_Fk_F\ll 1$. In what follows, we will refer to the
process in Fig.~\ref{fig:proc}(a) as simply  ``forward
scattering''\/. In the second type, the colliding particles move
initially in almost opposite directions ($\mathbf{k}_{1}\approx
-\mathbf{k}_{2})$ but, as for the forward scattering case, they
also retain their respective momenta after the collision. The
difference between such an event and the forward-scattering one is
that not only the transferred but also the \emph{total} initial
and final momenta are small.  This type of process is depicted in
Fig.~\ref{fig:proc}(b). In
$g-$%
ology notations, this is a ``$g_{2}$-process''.

Another process which contributes to the non-analytic part of
$\Sigma^R (\omega)$
 is   a ``$2k_{F}$- process'' (or
$g_1$-scattering, in $g$-ology notations), in which two fermions
moving initially in almost opposite directions, reverse their
respective momenta [see Fig.~\ref{fig:proc}(c)]. Since both
processes in Figs.~\ref{fig:proc}(b) and (c) contribute to the
same scattering
amplitude $%
f\left( \mathbf{k}_{1},\mathbf{k}_{2}\right) $ with the angle
between
\emph{%
initial} momenta $\mathbf{k}_{1}$ and $\mathbf{k}_{2}$ being close
to $\pi ,$ we will refer to both of them as  ``backscattering''. To
distinguish between the
two, we will refer to Fig.~\ref{fig:proc}(b)  as  ``$g_{2}-$%
backscattering'' and to Fig.~\ref{fig:proc}(c) as  ``$2k_{F}-$
backscattering''.

Scattering by $2k_F$ is one-dimensional in all dimensions. As we
just said, forward- and backscattering become one-dimensional in
$D=2$. We thus conclude \emph{that for $D=2$ the non-analytic part
of the self-energy comes from essentially 1D scattering processes,
embedded into the 2D phase space}.

 We pause
here for an important remark. Although there is a strong
similarity between special scattering processes, resulting in
non-analytic behavior in 2D, and their 1D analogs, there is also
an important difference. Namely, neither $g_2$ nor $g_4$ processes
lead to a non-analytic behavior of thermodynamic quantities in 1D.
This is already obvious from the fact that a 1D Hamiltonian with a
linearized spectrum and in the absence of $g_1$-scattering
(Tomonaga-Luttinger model) allows for an exact diagonalization in
terms of new excitations--free bosons. As a result, the specific
heat is strictly linear in $T$ and the spin susceptibility is
simply a constant within the Tomonaga-Luttinger model. However, if
$g_1$-scattering is present even as a marginally relevant
perturbation, exact diagonalization in terms of free bosons is no
longer possible, as the spin sector is now described by the
sine-Gordon rather than Gaussian theory. This results in strong
non-analyticities in both the specific heat $\delta C(T)$
($\propto T\ln T$) and spin susceptibility $\delta\chi_s$
($\propto |\ln H|$), where $H$ is the magnetic field
\cite{DL72,nersesyan}. It is possible to obtain these 1D
non-analyticities within the same fra<mework as their 2D analogs
are analyzed in this paper, but we defer this discussion to a
separate publication \cite{maslov}.

\section{Self-energy}
\label{sec:sigma} In this Section, we derive an expression for the
self-energy to all orders in the interaction. We analyze and
re-sum the mass-shell singularities in the forward-scattering part
of the self-energy, and also review the behavior of the
backscattering part. For the sake of completeness, however, we
start with the brief discussion of the second- order results for
the self-energy. The fermion self-energy is defined via the Dyson
equation
\begin{equation}
G^{-1}=G_{0}^{-1}+\Sigma,\label{aug13_1}
\end{equation}
where $G_0$ and $G$ are the bare  and exact Green's functions,
respectively. (Notice that we define $\Sigma $ with  an opposite
sign compared to Refs.~\cite{agd,statphys}.)

\subsection{second order}
\label{sec:imsigma}
\begin{figure}[tbp]
\begin{center}
\epsfxsize=0.8 \columnwidth \epsffile{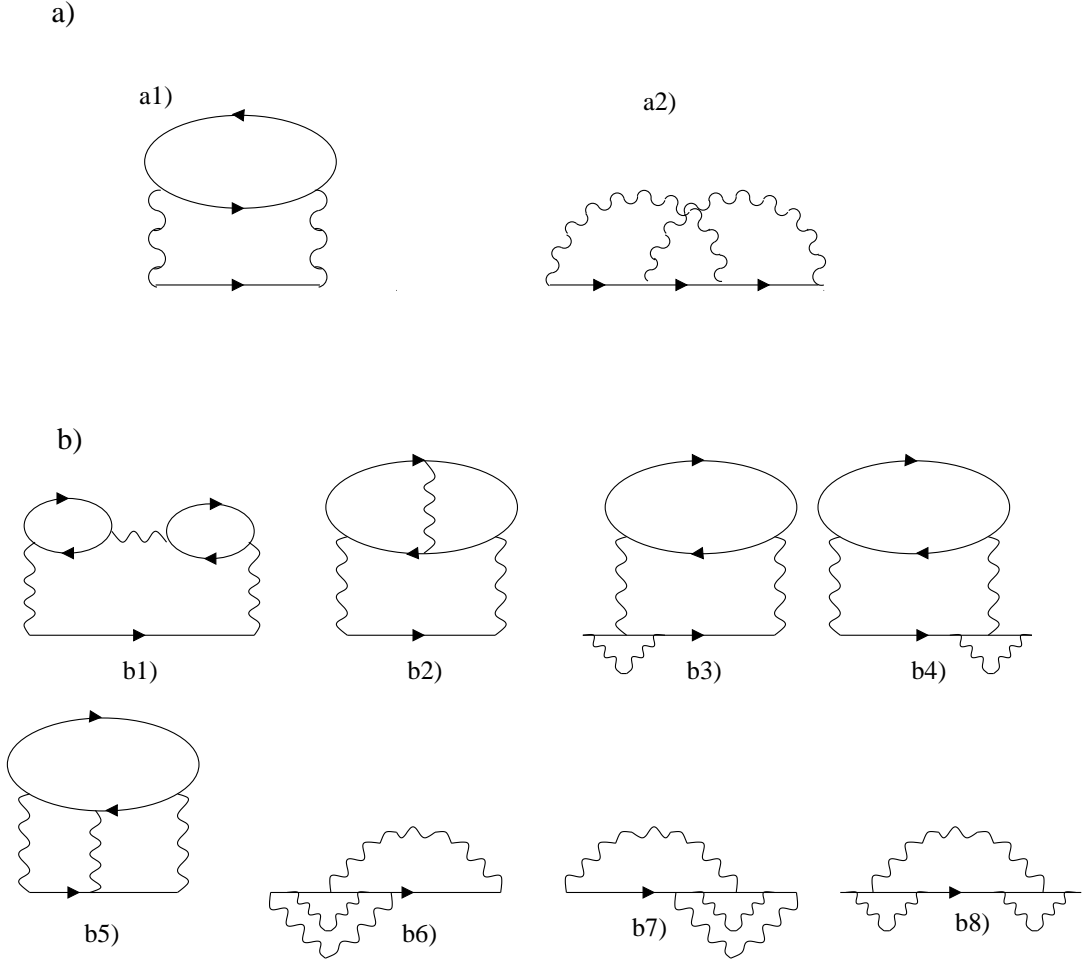}
\end{center}
\caption{ Non-trivial second (a) and third (b) order diagrams for
the self-energy.} \label{fig:selfenergy}
\end{figure}
To second order in the interaction, there are only two non-trivial
diagrams for $\Sigma $, which are shown in
Fig.~\ref{fig:selfenergy}%
(a). For a contact interaction, $V\left( r\right) =U\delta \left(
r\right) ,$ the contribution from diagram (a2) is $\left(
-1/2\right) $ of that from diagram (a1). The net contribution to
the imaginary part of the retarded self-energy at $T=0$ is given
by
\begin{equation}
\text{\textrm{Im}}\Sigma _{2}^{R}\left( \omega ,k\right)
=-U^{2}\int_{-\omega }^{0}\frac{d\Omega }{\pi }\int
\frac{d^{2}Q}{\left( 2\pi \right)
^{2}}\text{\textrm{Im}}G_{0}^{R}\left( \omega +\Omega
,\mathbf{%
k+Q}\right) ~\text{\textrm{Im}}\Pi ^{R}(\Omega ,Q),  \label{a1}
\end{equation}
where $G_{0}^{R}\left( \omega ,k\right) =\left( \omega -\epsilon
_{k}+i0^{+}\right) ^{-1}$ is the free retarded Green's function,
$\Pi ^{R}(\Omega ,Q)$ is the polarization bubble of free fermions,
and subindex $2 $ of the self-energy denotes the order of the
perturbation theory.  A general expression for $\Pi ^{R}(\Omega
,Q)$ is rather complicated \cite{stern}, but in what follows we
will need only its two asymptotic forms. The first of these forms
is valid for small $Q$
\begin{equation}
\Pi ^{R}(\Omega ,Q)=-\frac{m}{2\pi }\left( 1+\frac{i\Omega
}{\sqrt{\left( v_{F}Q\right) ^{2}-\left( \Omega +i0^{+}\right)
^{2}}}\right) , \label{feb6_1}
\end{equation}
and the other one is valid near $Q=2k_{F}$
\begin{equation}
\Pi ^{R}(\Omega ,Q)=-\frac{m}{2\pi }\left( 1-\left(
\frac{Q-2k_{F}}{2k_{F}}+%
\sqrt{\left( \frac{Q-2k_{F}}{2k_{F}}\right) ^{2}-\left(
\frac{\Omega
+i0^{+}%
}{2k_{F}v_{F}}\right) ^{2}}\right) ^{1/2}\right) .
\label{apr20_1}
\end{equation}
Here $m$ is the fermion's mass and $v_{F}$ is the Fermi velocity.
Non-analyticities in the two limiting forms of the bubble describe
Landau damping and Kohn anomaly, respectively. Landau damping of
an excitation with energy $\Omega $ and momentum $Q$ is possible
only within the particle-hole (PH) continuum, \emph{i.e.}, for
$\Omega <v_{F}Q$. For $\Omega \ll v_{F}Q$, the
non-analytic term in  Eq.~(\ref{feb6_1}) scales as $\Pi _{\text{sing}%
}^{R}(\Omega ,Q)\propto i\Omega /|Q|$. The Kohn anomaly near
$2k_{F}$ is static for $Q>2k_{F}$ [in this range, the non-analytic
part of $\Pi^R$ behaves as $\Pi _{\text{sing}}^{R}(0,Q)\propto
(Q-2k_{F})^{1/2}$%
] but is dynamic for $Q<2k_{F}$ [in this range, $\Pi
_{\text{sing}}^{R}(\Omega ,Q)\propto i\Omega /(2k_{F}-Q)^{1/2}$].

\subsubsection{backscattering}

To logarithmic accuracy, both $g_{2}-$ and $2k_{F}-$processes
[Fig.~\ref {fig:proc}(b) and (c), respectively] contribute
equally to the non-analytic part of the fermion self-energy. For a
contact interaction, the sum of the two contributions is
\cite{chm}
\begin{equation}
\text{\textrm{Im}}\Sigma _{2,\mathrm{B}}^{R}\left( \omega
,k\right) =\frac{%
u^{2}}{4\pi }~\frac{\omega ^{2}}{E_{F}}~\ln \frac{E_{F}}{|\omega
|}, \label{c1c}
\end{equation}
where $B$ stands for backscattering and $u$ is defined in
 Eq.~(\ref{i1}).
It is important in what follows that, to logarithmic accuracy, \textrm{Im}$%
\Sigma _{2,\mathrm{B}}^{R}$ depends only on $\omega $ but not on
$\epsilon _{k}.$ For the sake of completeness, we present the
derivation of  Eq.~(\ref{c1c}) in Appendix \ref{sec:appendix_se}. We
find that for $2k_{F}-$processes, a non-analytic part of the
self-energy originates from the \textit{dynamic} Kohn anomaly,
whereas the static Kohn anomaly contributes only to the regular
part.

Higher-order contributions to the backscattering part of the
self-energy form regular series in $u$ which result in the
renormalization of the prefactor. As we keep $u$ small, it
suffices to stop the perturbation theory at order $u^{2}.$
Consequently, we set
\begin{equation}
\Sigma^R _{\mathrm{B}}=\Sigma^R _{2,\mathrm{B}}
\end{equation}
in the rest of the paper.
\subsubsection{forward scattering}
The second-order forward-scattering contribution to the
self-energy is given by \cite{chm}
\begin{equation}
\text{\textrm{Im}}\Sigma _{2,\mathrm{F}}^{R}\left( \omega
,k\right) =\frac{%
u^{2}}{8\pi }\frac{\omega ^{2}}{E_{F}}\ln \frac{E_{F}}{|\Delta |},
\label{c1b}
\end{equation}
where $F$ stands for forward scattering, and we remind that
$\Delta \equiv \omega -\epsilon _{k}$ is the ``distance'' to the
mass shell. Away from the mass shell, this contribution behaves as
$\omega ^{2}\ln |\omega |$, \emph{i.e.}, it has the same
functional form as \textrm{Im}$\Sigma _{\mathrm{B}}^{R}$. In
contrast to the backscattering part, however, the
forward-scattering contribution diverges at the mass shell,
\emph{i.e., }for $\Delta \rightarrow 0$. The origin of this
divergence can be traced back to the form of the polarization
bubble at small momenta. From  Eq.~(\ref{feb6_1}), we find that
\begin{equation}
\text{\textrm{Im}}\Pi ^{R}(\Omega ,Q)=-\left( \frac{m}{2\pi
}\right) \frac{%
\Omega }{\sqrt{\left( v_{F}Q\right) ^{2}-\Omega ^{2}}}\theta
\left( v_{F}Q-\left| \Omega \right| \right) ,  \label{sqrt1}
\end{equation}
where $\theta \left( x\right) $ is the step
function.  $\text{\textrm{Im}}%
\Pi ^{R}(\Omega ,Q)$ has square-root singularities at $|\Omega
|=v_{F}Q$. On
the other hand, expanding $\epsilon _{\mathbf{k}+\mathbf{Q}}$ in $%
G_{0}^{R}(\omega +\Omega ,\mathbf{k}+\mathbf{Q})$ in  Eq.~(\ref{a1}) as $%
\epsilon _{\mathbf{k+Q}}=\epsilon _{k}+v_{F}Q\cos \theta $ and
integrating over $\theta $, we obtain another square-root
singularity
\begin{equation}
\int d\theta \text{\textrm{Im}}G_{0}^{R}=-~\frac{2\pi
}{\sqrt{\left( v_{F}Q\right) ^{2}-\left( \omega +\Omega -\epsilon
_{k}\right) ^{2}}}. \label{sqrt2}
\end{equation}
On the mass shell ($\omega =\epsilon _{k}$), the arguments of the
square roots in  Eq.~(\ref{sqrt1}) and  Eq.~(\ref{sqrt2}) coincide,
hence the integral over $d^{2}Q$ diverges logarithmically.
{\bf DM}
(\ref{c1b}) is valid only for a linearized fermion dispersion.
Two of us demonstrated in Ref.~\cite{chm} that finite curvature of
the dispersion
 eliminates the logarithmic singularity in
\textrm{Im}$\Sigma _{2,\mathrm{F}}^{R}$. To keep our presentation
uninterrupted, we continue to proceed with the analysis of the
singularities due to forward scattering, assuming that the
curvature is equal to zero, \emph{i.e}., the dispersion is
linear. We then discuss separately the modifications imposed by a
finite curvature of the dispersion (cf. Sec.~\ref{sec:curv}). We
emphasize again that  the logarithmic
 singularity in the self-energy arises from
essentially 1D scattering processes, embedded in a 2D phase space.
Therefore, this singularity can be viewed as a pre-cursor of a
stronger (power-law) singularity in 1D (``infrared catastrophe'')
\cite{1D}, \cite {chm}a.
\subsection{higher-order forward-scattering contributions}
\label{sec:imsigma_1}

 Higher orders of the perturbation theory
contain more bubbles with small momenta (``soft bubbles''). As a
result, the mass-shell singularities proliferate.
 The third-order diagrams, shown in Fig.~\ref {fig:selfenergy},
contain the square of the soft bubbles. These bubbles appear
either explicitly (as in diagram b1) or are generated upon
integrating over fermion energies/momenta in the rest of the
diagrams. The singular part of the self-energy at this order is
given by
\begin{equation}
\text{\textrm{Im}}\Sigma _{3,\mathrm{F}}^{R}\left( \omega
,k\right) =U^{3}\int_{-\omega }^{0}\frac{d\Omega }{2\pi }\int
\frac{d^{2}Q}{\left( 2\pi \right)
^{2}}\text{\textrm{Im}}G^{R}_0\left( \omega +\Omega
,\mathbf{k+Q%
}\right) \Pi _{2}^{R}(\Omega ,Q),  \notag  \label{a2}
\end{equation}
where
\begin{equation}
\Pi _{2}^{R}(\Omega ,Q)\equiv \text{\textrm{Im}}\Pi ^{2}\left(
\Omega +i0^{+},Q\right).  \label{a5}
\end{equation}
The most singular term in
$\Pi _{2}^{R}(\Omega ,Q)$ is given by
\begin{equation}
\left[ \Pi _{2}^{R}(\Omega ,Q)\right] _{\text{sing}}=-\left(
m/2\pi \right) ^{2}\pi \Omega \left| \Omega \right| \delta \left(
\Omega ^{2}-v_{F}^{2}Q^{2}\right) .  \label{a6}
\end{equation}
The product of the square-root and delta-function singularities
(from $\int G_{0}^{R}d\theta $ and $\Pi _{2}^{R}(\Omega ,Q)$,
respectively) gives rise to a one-sided, square-root singularity
on the mass shell:
\begin{equation}
\text{\textrm{Im}}\Sigma _{3,\mathrm{F}}^{R}\left( \omega
,k\right) =-\frac{%
\sqrt{2}u^{3}}{20}\frac{\omega ^{2}}{E_{F}}\sqrt{\frac{\omega
}{\Delta }}\theta\left(\frac{\omega}{\Delta}\right). \label{ims3}
\end{equation}
It can be readily verified  that at $n-$ th order
\begin{equation}
\text{\textrm{Im}}\Sigma _{n,\mathrm{F}}^{R}\propto U^{n}\omega
^{\frac{n}{2}%
+1}/\Delta ^{\frac{n}{2}-1},  \label{imsn}
\end{equation}
for $n>2.$ Collecting forward-scattering contributions to all
orders in $u$, we obtain
\begin{equation}
\text{\textrm{Im}}\Sigma _{\mathrm{F}}^{R}(\omega
)=\frac{u^{2}}{8\pi
}\frac{%
\omega ^{2}}{E_{F}}\left[ \ln \frac{E_{F}}{|\Delta
|}+\sum_{n=1}^{\infty }C_{n}\left( u^{2}\frac{\omega }{\Delta
}\right) ^{n/2}\right] , \label{b1}
\end{equation}
where $C_{n}$ are the numerical coefficients.
 We see that perturbative expansion in $u$ works only for
$u^{2}\omega /\Delta \ll 1$. Outside this range, series in $u$
does not converge, and one needs to re-sum the perturbation
theory.

\subsection{Re-summation of forward-scattering contributions}
\label{sec:resummaton} \label{sec:resumation}

To perform the re-summation of the perturbation theory, we need to select
diagrams with the maximum number of particle-hole bubbles at small
frequency/momentum. It is convenient to select first analogous diagrams
for
the four-fermion vertex, $\Gamma _{\alpha \beta ,\gamma \varepsilon
}(p_{1,}p;p_{3},p_{4}),$ and then relate $\Sigma $ to $\Gamma $ via the
Dyson equation. In this subsection, we will be using notations $p\equiv
\left( \omega _{n},\mathbf{k}\right) $ and $q\equiv \left( \Omega
_{m},Q\right)$, where $\omega_m=\pi(2m+1)T$ and $\Omega_m=2\pi mT$.

\subsubsection{four-fermion vertex, zero-sound mode}

\begin{figure}[tbp]
\begin{center}
\epsfxsize=0.7 \columnwidth
\epsffile{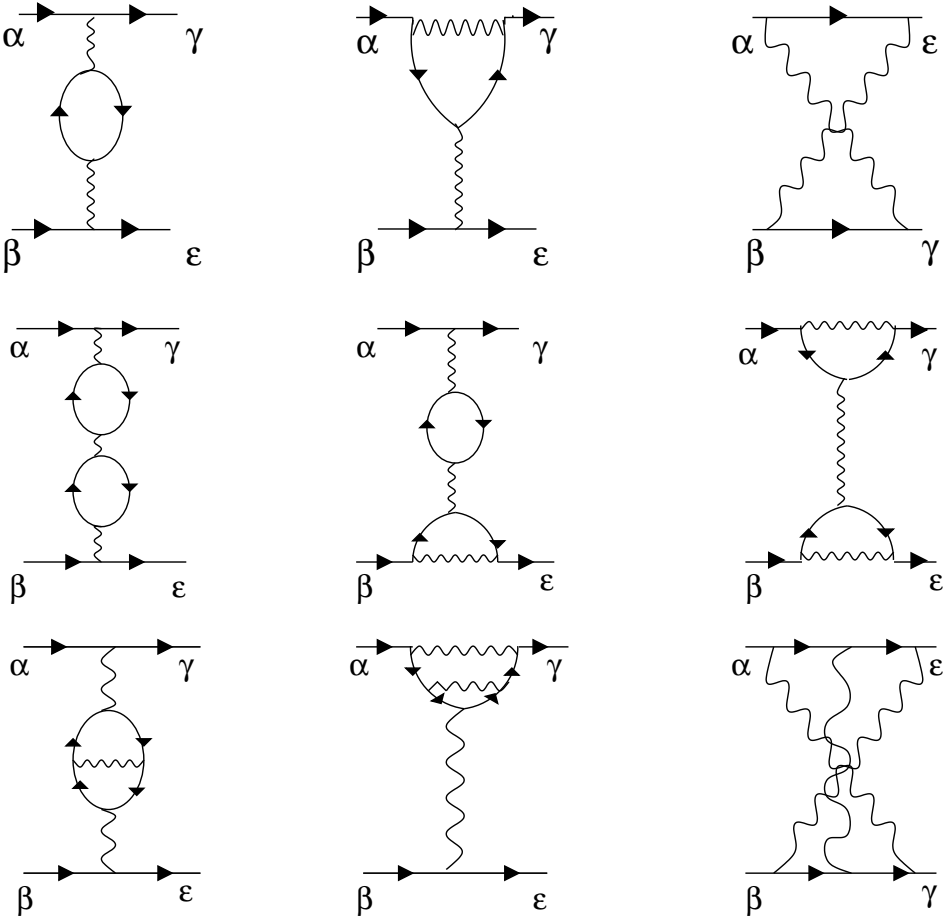}
\end{center}
\caption{Vertex diagrams with maximum number of particle-hole
bubbles to third in the interaction. Additional diagrams, obtained
from those in the second column by a permutation
$\alpha\to\beta,\gamma\to\epsilon$, are not shown.}
\label{fig:vertex2}
\end{figure}

The diagrams for $\Gamma $ with the maximum number of
particle-hole bubbles form familiar ladder series (see
Fig.~\ref{fig:vertex2}), when $\Gamma $ is anti-symmetrized with
respect to a permutation of either initial or final states.
However, the procedure of finding an overall prefactor at order
$\nu $ is somewhat involved \cite{agd}, as it requires counting
the number of diagrams at the same order in a conventional
diagrammatic technique operating with a non-symmetrized vertex,
$\bar{\Gamma}$. We choose to sum the diagrams for a
non-symmetrized vertex to all orders first, and then
anti-symmetrize the result. The second- and third-order diagrams
for
$\bar{%
\Gamma}$  are shown in Fig.~\ref
{fig:vertex2}. A general procedure of summing such diagrams to all orders
is
described in Appendix \ref{sec:appendixA}. It leads to the following
result
for $\bar{\Gamma}$%
\begin{equation}
\bar{\Gamma}_{\alpha \beta ,\gamma \varepsilon
}(p_{1},p_{2};p_{1}-q,p_{2}+q)={\bar \Gamma} (q) = - U \left[\delta _{\alpha \gamma }\delta
_{\beta \varepsilon } \left( \frac{1}{2}+\mathcal{G}_{\rho }\right) +
\sigma
_{\alpha \gamma }^{a}\sigma _{\beta \varepsilon }^{a} \left( \frac{1}{2}+%
\mathcal{G}_{\sigma }\right)\right],  \label{b4}
\end{equation}
where $\sigma _{\alpha \beta }^{a}$ are Pauli matrices ($a=x,y,z)$, and
\begin{equation}
\mathcal{G}_{\rho }\equiv \frac{1}{2}\frac{1}{1-U\Pi \left(
q\right) };\text{
}\mathcal{G}_{\sigma }=-\frac{1}{2}\frac{1}{1+U\Pi \left( q\right) }
\label{f8}
\end{equation}
are the (dimensionless) charge- and spin vertices, respectively. An
anti-symmetrized vertex is obtained from $\bar{\Gamma}$ by the following
procedure
\begin{equation}
\Gamma _{\alpha \beta ;\gamma \varepsilon }(q)=%
\bar{\Gamma}_{\alpha \beta ;\gamma \varepsilon
}(q)-\bar{\Gamma}_{\alpha \beta ;\gamma \varepsilon
\gamma }(q).  \label{f9}
\end{equation}
For the case of $U>0$, which we are interested in, the retarded
charge
vertex, $\mathcal{G}%
_{\rho }^{R}$, has a pole determined from the equation $1-U\Pi
^{R}(\mathbf{q}%
,\Omega )=0$. A two-particle excitation corresponding to the pole in $%
\mathcal{G}_{\rho }$ is a zero-sound collective mode. Since $\Pi^R
$ is real for $\Omega ^{2}>v_{F}^{2}Q^{2}$, and can be arbitrarily
large (and positive) when $|\Omega|$ approaches $v_{F}Q$, the
zero-sound pole exists already for an arbitrarily small $U$. Near
the pole, $\mathcal{G}_{\rho }^{R}$ is of the form
\begin{equation}
\mathcal{G}_{\rho }^{R}=\frac{u^{2}v_{F}^{2}Q^{2}}{\left( \Omega
+i0^{+}\right) ^{2}-c^{2}Q^{2}},  \label{c2}
\end{equation}
where $c$ is the zero-sound velocity
\begin{equation}
c=v_{F}\sqrt{1+u^{2}/\left( 1+2u\right) } \approx v_{F} (1 + u^2/2) > v_{F}.
\end{equation}
We see that  zero-sound velocity $c$ differs from $v_{F}$ only by
a $u^{2}$-term. This means that the zero-sound mode $|\Omega| =cQ$
is just above the upper boundary of the particle-hole continuum,
which, for small $Q,$ is given by
$%
v_{F}Q.$ We also see from  Eq.~(\ref{c2}) that the quanta of zero sound
are
\emph{not }%
free bosons as the residue of the zero-sound pole in  Eq.~(\ref{c2}) is
proportional to $Q^{2}.$
The spin vertex ($\mathcal{G}_{\sigma }^{R})$ also has a pole, but it
is located on the imaginary axis. Consequently, the corresponding
collective
mode is over-damped.
As an independent check, we verified that  the diagrams in Fig.~\ref{fig:vertex2} sum
up
to zero for the case of spinless fermions. This result is a manifestation
of
the Pauli principle: spinless fermions do not interact via contact forces
as
the Pauli principle forbids them to be at the same point in space.

\subsubsection{Dyson equation}

The self-energy due to forward scattering is related to the vertex
function via the
Dyson equation \cite{statphys}
\begin{equation}
\Sigma _{\rm{F},\alpha \beta }(p)=\delta _{\alpha \beta
}\int_{q}UG(p-q)-\int_{p',p"}U\Gamma _{\gamma \alpha ;\gamma \beta
}(p,p'+p''-p;p',p'')G(p')G(p'')G(p'+p''-p),
\label{dyson}
\end{equation}
where
\begin{equation}
\int_{k}\dots \equiv T\sum_{\omega_{m}}\int d^{2}k/\left( 2\pi \right)
^{2}\dots \label{notation}
\end{equation}
In principle, the Green's functions in the Dyson equation are the
exact ones. However, it can be verified that self-energy
insertions into the diagrams diverging near the mass shell do not give
rise to additional mass-shell singularities. As we
 keep $u$ small, regular corrections are thus irrelevant,
 and we can safely use
bare $G^{\prime }$s instead of the exact ones in  Eq.~(\ref{dyson}).
Substituting  Eq.~(\ref{b4}) into  Eq.~(\ref{dyson}), we obtain $\Sigma
_{\rm{F},\alpha
\beta }=\delta _{\alpha \beta }\Sigma _{\rm{F}}$ where
\begin{equation}
\Sigma _{\rm{F}}\left( p\right) =\int_{q}\left[ U+U^{2}\Pi
(q)+\frac{1}{2}\frac{%
U^{3}\Pi ^{2}\left( q\right) }{(1-U\Pi \left(
q\right) )}-\frac{3}{2}\frac{%
U^{3}\Pi ^{2}(q)}{(1+U\Pi (q))}\right] G\left( p-q\right) .  \label{b5}
\end{equation}
Technical details of the derivation leading to  Eq.~(\ref{b5}) are
presented in Appendix \ref{sec:appendixA}. Expanding
 Eq.~(\ref{b5}) to third order in $U$, we reproduce the results of
the conventional perturbation theory, Eqs.~(\ref{c1b}) and
(\ref{ims3}). We remind that in the perturbation theory
\textrm{Im}$\Sigma^{R} $ diverges upon approaching the mass
shell-- logarithmically to second order in $U$, and as
$1/\sqrt{\omega -\epsilon
_{k}%
}$ to third order. It is convenient to rearrange the terms in
(\ref{b5}) and decompose $\Sigma _{\rm{F}}$ into three parts
making use of the charge- and spin vertices, introduced in
 Eq.~(\ref{f8}), as
\begin{subequations}
\begin{eqnarray}
\Sigma _{\rm{F}}\left( p\right) &=&\Sigma _{\rho }\left( p\right) +\Sigma
_{\sigma }\left( p\right) +\Sigma _{\text{ex}};  \label{b7} \\
\Sigma _{\rho }\left( p\right) &=&U\int_{q}\mathcal{G}_{\rho }\left(
q\right) G\left( p-q\right) ;  \label{b8} \\
\Sigma _{\sigma }\left( p\right) &=&3U\int_{q}\mathcal{G}_{\sigma }\left(
q\right) G\left( p-q\right) ;  \label{b9} \\
\Sigma _{\text{ex}} &=&\int_{q}\left[ 2U-U^{2}\Pi (q)\right] G\left(
p-q\right) .  \label{b9_1}
\end{eqnarray}
Terms $\Sigma _{\rho }$ and $\Sigma _{\sigma }$ correspond to the
interaction in the charge- and spin channels, respectively, and are summed
to all orders in $U.$ The remainder, $\Sigma _{\text{ex}},$
contains extra contributions of the first and second orders in
$U,$ not included in the first two terms,
 reproduces the second-order result  Eq.~(\ref{c1b}).

Before we proceed further, a comment is in order. Our results for
the vertex and self-energy formally coincide with those found in
the paramagnon (spin-fluctuation) model \cite{brinkman} (except
for the remainder term, $\Sigma
_{\text{%
ex}}$, in  Eq.~(\ref{b7}) which was neglected in
Ref.~\cite{brinkman}). However, our results have been obtained in a
more general approach. In the paramagnon model, the self-energy is
given only by diagrams of the type (b1) and (b5) in
Fig.~\ref{fig:selfenergy}, \emph{i.e.}, it involves only RPA
diagram in the charge-channel and
 ladder diagrams in the spin channel. We included
\emph{all } diagrams with the maximum number of bubbles \ and
found that the overall combinatorial coefficients at each order
are such that the summation to all orders results in two
independent geometric series--one for the charge channel, and the
other for the spin channel. It does not mean, however, that we
have obtained exact results for $\Gamma $ and $\Sigma $. Indeed,
we considered only forward scattering, kept $u$ small, and
neglected all diagrams that constitute regular series in $u$ and
does not give rise to proliferating mass-shell singularities in
$\Sigma $. From this perspective, the controlling parameter for
our approximation is not the coupling constant $u$ itself but a
combined parameter $u^{2}|\omega| /|\Delta| $ which measures the
proximity to the mass shell. We sum up the series in
$u^{2}|\omega| /|\Delta| $, and neglect regular corrections in $u$
at every order .
\subsection{Imaginary part of the self-energy to all orders in the
interaction} \label{sec:imsigma_all} We now evaluate the
forward-scattering part of \textrm{Im}$\Sigma ^{R}$ near the mass
shell. The imaginary part of the retarded self-energy comes from
two sources: from the particle-hole continuum $\left( |\Omega|
<v_{F}Q\right) , $ where \textrm{Im}$\Pi ^{R}\neq
0$%
, and from the collective mode at $|\Omega| =cQ$, where
$\mathcal{G}_{\rho }^{R}$ has a pole. The spin-channel part of the
self-energy, \textrm{Im}$\Sigma _{\sigma }^{R}$, comes only from
the continuum, whereas the charge-channel part contains
contributions from both the continuum and collective mode.
Accordingly, the imaginary part of the total self-energy can be
represented as
\end{subequations}
\begin{equation}
\text{\textrm{Im}}\Sigma _{\rm{F}}^{R}=\text{\textrm{Im}}\Sigma
_{\rm{PH}}^{R}+%
\text{\textrm{Im}}\Sigma^{R} _{\rm{ZS}}+\text{\textrm{Im}}\Sigma^{R}
_{\text{ex}},
\label{sum}\end{equation}
where($\dots $)$_{\rm{PH}}$ and ($\dots $)$_{\rm{ZS}}$ stand for the
particle-hole and zero-sound contributions, respectively:
\begin{equation}
\text{\textrm{Im}}\Sigma _{\rm{PH}}^{R}=\left( \text{\textrm{Im}}\Sigma
_{\rho }^{R}+\text{\textrm{Im}}\Sigma _{\sigma }^{R}\right) _{\rm{PH}},~~%
\text{\textrm{Im}}\Sigma^{R} _{\rm{ZS}}=\left( \text{\textrm{Im}}\Sigma
_{\rho
}^{R}\right) _{\rm{ZS}}.
\end{equation}
\subsubsection{remainder term $\Sigma _{\text{ex}}$}
Comparing  Eq.~(\ref{b9_1}) and  Eq.~(\ref{a1}), we see that the
remainder term in decomposition  Eq.~(\ref{b7}), \textrm{Im}$\Sigma^{R}
_{\text{ex}}$, is opposite in sign and equal in magnitude to the
second-order forward-scattering contribution to the self-energy,
 Eq.~(\ref{c1b}):
\begin{equation}
\text{\textrm{Im}}\Sigma^{R} _{\text{ex}}=-\frac{u^{2}}{8\pi }\frac{\omega
^{2}}{%
E_{F}}\ln \frac{E_{F}}{|\Delta |}.  \label{h2}
\end{equation}
\subsubsection{particle-hole contribution}
\begin{figure}[tbp]
\begin{center}
\epsfxsize=0.5 \columnwidth \epsffile{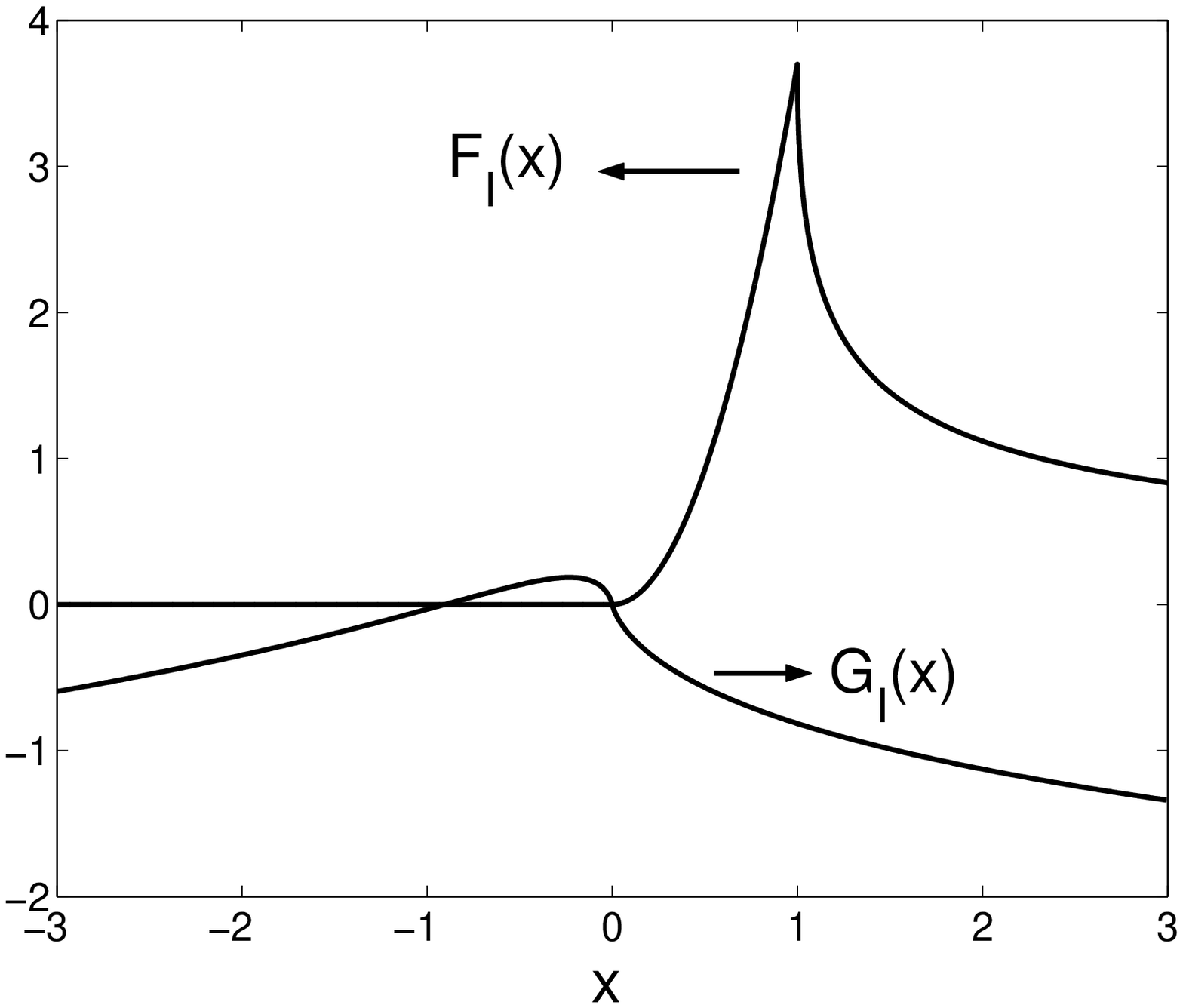}
\end{center}
\caption{Scaling functions $G_{I}\left( x\right) $ [ Eq.~(\ref{may_5_3})]
and $%
F_{I}\left( x\right) $ [ Eq.~(\ref{j1})].}
\label{fig:plot}
\end{figure}

Term $\text{\textrm{Im}}\Sigma _{\rm{PH}}^{R}$ contains the imaginary
parts of the retarded vertices in the charge- and spin channels:
\begin{equation}
\text{\textrm{Im}}\mathcal{G}_{\rho ,\sigma
}^{R}=\frac{1}{2}~\frac{U\text{%
\textrm{Im}}\Pi ^{R}}{\left( 1\mp U\text{\textrm{Re}}\Pi ^{R}\right)
^{2}+\left( U\text{\textrm{Im}}\Pi ^{R}\right) ^{2}}.  \label{h22}
\end{equation}
The first term in the denominator of  Eq.~(\ref{h22}) can be
replaced by unity because for $q\leq 2k_{F}$, \textrm{Re}$\Pi
^{R}=-m/2\pi
$, \emph{i.e.}, $-U%
\text{\textrm{Re}}\Pi ^{R}=u\ll 1$. Substituting then  Eq.~(\ref{h22}) into
 Eq.~(\ref{b8}) and  Eq.~(\ref{b9}), we obtain
\begin{equation}
\text{\textrm{Im}}\Sigma _{\rm{PH}}^{R}=-2U^{2}\int_{-\omega }^{0}\frac{%
d\Omega }{\pi }\int \frac{d^{2}Q}{(2\pi
)^{2}}\text{\textrm{Im}}G^{R}(\omega
+\Omega ,\mathbf{k}+\mathbf{Q})\frac{\text{\textrm{Im}}\Pi ^{R}}{1+\left(
U%
\text{\textrm{Im}}\Pi ^{R}\right) ^{2}}.  \label{may_5_1}
\end{equation}
Substituting \textrm{Im}$\Pi ^{R}$ from  Eq.~(\ref{sqrt1}) and
\textrm{Im}$G^{R}$
from  Eq.~(\ref{sqrt2}) into  Eq.~(\ref{may_5_1}) and keeping only the  forward
scattering contribution, we obtain after some algebra
\begin{equation}
\text{\textrm{Im}}\Sigma _{\rm{PH}}^{R}=\frac{u^{2}}{4\pi }~\frac{\omega
^{2}}{E_{F}}\left[ \ln \frac{E_{F}}{u^{2} |\omega| }+G_{I}\left(
\frac{2\Delta
}{u^{2}|\omega |}\right) \right] ,  \label{may_5_2}
\end{equation}
where
\begin{equation}
G_{I}(x)=2\ln 2-1/2+\ln
|x|^{-1}-2~\mathrm{Re}\int_{0}^{1}zdz\frac{1}{\sqrt{%
1-x/z}}~\ln \frac{1+\sqrt{1-x/z}}{1-\sqrt{1-x/z}}{.}  \label{may_5_3}
\end{equation}
Subscript $I$ in $G_{I}$ implies that this is a scaling function
for the imaginary part of the self-energy. We subtracted off a
constant term in $G_I$ so that $G_I(0)=0$. This is equivalent to
neglecting a regular, $\omega ^{2}$-contribution to
\textrm{Im}$\Sigma _{\rm{PH}}^{R}$. A plot of $G_{I}(x)$ is
presented in Fig.~\ref{fig:plot}. At large and positive $x$, the
integral term in scaling function $G_{I}(x)$ falls of as
$x^{-1/2}$, whereas for large and negative $x,$ it falls off
as $%
\left( -x\right)^{-1}$. In either of these limits, $G_{I}\approx \ln
|x|^{-1} $ and, consequently,
\begin{equation}
\text{\textrm{Im}}\Sigma _{\rm{PH}}^{R}=\frac{u^{2}}{4\pi }~\frac{\omega
^{2}}{E_{F}}\ln \frac{E_{F}}{|\Delta |}{.}  \label{may_5_4_1}
\end{equation}
Further expansion in powers of $1/x$ yields
\begin{equation}
\text{\textrm{Im}}\Sigma
_{\text{3,PH}}^{R}=-2\frac{\sqrt{2}u^{3}}{20}\frac{%
\omega ^{2}}{E_{F}}\sqrt{\Big|\frac{\omega }{\Delta }\Big|}.
\label{may_5_7}
\end{equation}

In the opposite limit of small $x$, function $G_I(x)$ vanishes as $x\ln
|x|$%
. As a result, net $\text{\textrm{Im}}\Sigma _{\rm{PH}}^{R}$
remains finite at $\Delta =0$, and  for $x\ll 1$ (i.e., for
$\Delta\ll u^2 \omega)$ it behaves as
\begin{equation}
\text{\textrm{Im}}\Sigma _{\rm{PH}}^{R}=\frac{u^{2}}{4\pi }~\frac{\omega
^{2}}{E_{F}}~\left[\ln \frac{E_{F}}{u^{2}|\omega
|}+\frac{2\Delta}{u^2\omega}
\ln \frac{|\Delta|}{u^2 |\omega|}\right].  \label{may_5_6}
\end{equation}
Comparing the limiting forms of  Eq.~(\ref{may_5_4_1}) and  Eq.~(\ref{may_5_6}), we
see
that higher order terms in $u$ simply cut the logarithmic divergence in $%
\text{\textrm{Im}}\Sigma _{\rm{PH}}^{R}$ for $|\Delta |<
u^{2}|\omega |$. To logarithmic accuracy, one can then approximate
$\text{\textrm{Im}}\Sigma
_{%
\rm{PH}}^{R}$ by
\begin{equation}
\text{\textrm{Im}}\Sigma _{\rm{PH}}^{R}=\frac{u^{2}}{4\pi }\frac{\omega
^{2}}{E_{F}}\ln \frac{E_{F}}{|w|},  \label{h3}
\end{equation}
where $w\equiv \max \left\{ \Delta ,u^{2}|\omega |\right\} $. The
appearance of $u$ under the logarithm in  Eq.~(\ref{h3}) is a
reminder that  Eq.~(\ref{h3}) includes all orders of the
perturbation theory. Indeed, a crossover between
Eqs.~(\ref{may_5_4_1}) and  (\ref{may_5_6}) occurs at a scale
$|\Delta| \simeq u^{2}|\omega| $, which is not accessible within
the perturbation theory.

\subsubsection{zero-sound contribution}
For the collective-mode contribution to the self-energy in the
vicinity of the mass shell, \emph{i.e.}, for $|\Delta| \ll
|\omega|$, we obtain from
 Eq.~(%
\ref{b8}) and  Eq.~(\ref{h22})
\begin{equation}
\mathrm{Im}\Sigma^{R} _{\rm{ZS}}=\frac{u^{2}U}{4\pi }\text{%
\textrm{Re}}\left[ \int_{0}^{|\omega|
/v_{F}}\frac{QdQ}{\sqrt{\left(2\Delta /v_{F}Q\right) \text{sgn}
\omega - u^2}}\right] =\frac{u^{2}}{4\pi }~\frac{\omega
^{2}}{E_{F}}F_{I}\left( \frac{2\Delta }{u^{2}\omega }\right) .
\label{d2}
\end{equation}
For $x>0,$ scaling function $F_{I}\left( x\right) $ is given by
\begin{equation}
F_{I}\left( x\right) \equiv 2\pi \int_{0}^{\min
\{1,\sqrt{x}\}}dy\frac{y^{4}%
}{\sqrt{x-y^{2}}}=\left\{
\begin{array}{cl}
3\pi ^{2}x^{2}/8,\;\mathrm{for}\;x<1; &  \\
\frac{\pi }{2}\left[ \frac{3x^{2}}{2}\sin
^{-1}\frac{1}{\sqrt{x}}-\sqrt{x-1}%
\left( \frac{3}{2}x+1\right) \right] ,\;\mathrm{for}\;x>1. & \label{j1}
\end{array}
\right.
\end{equation}
For negative $x$, $F_{I}(x)=0$. At $x=1$, $F_I(x)$ is continuous but its derivative is singular: $dF_I(x)/dx \propto
1/\sqrt{x-1}$ for $x>1$. A plot of $F_{I}(x)$ is shown in
Fig.~\ref {fig:plot}. For $x\gg 1$, \emph{i.e}., for $u\rightarrow
0$,
\begin{equation}
F_{I}(x)\approx 2\pi /5\sqrt{x}.
\end{equation}
In this limit,  $F_{I}(x)\propto u$ and thus $ \mathrm{Im}%
\Sigma^{R} _{\rm{ZS}}\propto u^{3},$ as is to be expected, as the
zero-sound propagator, $U{\cal G}_{\rho}$, is of third order in
$U$ near the pole. Substituting this limiting form into
 Eq.~(\ref{d2}),
 we obtain for $\omega/\Delta >0$
\begin{equation}
 \mathrm{Im}\Sigma ^{R} _{\rm{ZS}}=\frac{\sqrt{2}u^{3}}{20}\frac{%
\omega ^{2}}{E_{F}}\sqrt{\frac{\omega }{\Delta }}.  \label{may_5_10}
\end{equation}
Combining  Eq.~(\ref{may_5_10}) and the third-order particle-hole
contribution as given in   Eq.~(\ref{may_5_7}), we reproduce the result of the third-order
perturbation theory,  Eq.~(\ref{ims3}). Expanding $\left( \mathrm{Im}\Sigma
_{\rho }\right)^{R}_{\mathrm{ZS}}$ in powers of $u$ further$,$ we indeed
reproduce the structure of higher order terms in the perturbation theory,
 Eq.~(\ref{imsn}). All these terms diverge when $\Delta \rightarrow 0$.
However, the full result shows that the perturbative expansion in $u$ for
the zero-sound contribution is valid only for $|\Delta |\gg u^{2}|\omega
|$.
At $|\Delta |=u^{2}|\omega |/2,$ \emph{i.e.}, at $x=1$, $\mathrm{Im}%
\Sigma^{R} _{\mathrm{ZS}}$ has a maximum. Upon further approach to
the mass-shell, $\mathrm{Im}\Sigma^{R} _{\mathrm{ZS}}$ decreases
as $\Delta^2$ and eventually vanishes on the mass shell
($\Delta=0$).
Vanishing of $\mathrm{Im} \Sigma^{R} _{\mathrm{ZS}}$ on the mass
shell is due to a Cherenkov-type restriction: because the
zero-sound velocity $c > v_F$, an on-shell fermion cannot emit a
zero-sound boson, hence the fermion's lifetime becomes infinite.

 Observe that $\mathrm{Im}\Sigma^{R} _{\mathrm{ZS}}$ is asymmetric with respect
to a change in sign of $\Delta$: $\mathrm{Im}\Sigma^{R}
_{\mathrm{ZS}}\neq 0$ only if $\Delta$ and $\omega$ are of the
same sign. This asymmetry follows simply from the energy and
momentum conservation. For example, a fermion of energy $\omega>0$
above the Fermi level can emit a soft zero-sound boson of frequency
$0\leq \Omega\leq\omega$ provided that
\begin{equation}
\omega-\epsilon_k=\Omega-v_FQ\cos\theta=\Omega\left[1-\left(v_F/c\right)\cos\theta\right],
\end{equation}
which, for $c\approx v_F\left(1+u^2/2\right)$, is equivalent to
\begin{equation}
0\leq u^2\Omega/2\leq\Delta\leq 2\Omega\leq 2\omega.
\label{ineq}\end{equation} Thus, emission is possible only if both
$\omega$ and $\Delta$ are positive. A similar consideration for
the case when a fermion of energy $\omega<0$ below the Fermi level
absorbs a zero-sound boson of frequency $\Omega$ in the interval
$(-|\omega|,0)$ shows that absorption is possible only for
$\Delta<0$.   Eq.~(\ref{ineq}) also clarifies the meaning of a
characteristic scale $\Delta^*=\omega u^2/2$. For $\Delta
>\Delta^*$ emission of bosons with any frequency in the interval
$0\leq\Omega\leq\omega$ is possible. In particular, a fermion can
emit only one boson of frequency $\Omega=\omega$ and ``land'' on
the Fermi level. For $\Delta<\Delta^*$, {\em i.e.}, when a fermion
is close to the Fermi level, emission of bosons with frequency
$\Omega>2\Delta/u^2$ is impossible, and the fermion relaxes to the
Fermi level via emitting a large number of low-frequency bosons.
As a result, the relaxation slows down which corresponds to a
decrease in $\mathrm{Im}\Sigma^{R} _{\mathrm{ZS}}$ for
$\Delta<\Delta^*$.

 Combining the results for $\Sigma^R_{{\rm  ex}}, \Sigma^R_{{\rm  PH}}$ and $\Sigma^R_{{\rm ZS}}$,
 we see  that the
summation of the power-law divergent diagrams for the self-energy
leads to a non-trivial result: the total self-energy due to
forward scattering undergoes a non-monotonic variation near the
mass-shell. All power-law divergences of the form $u^{n}\Delta
^{1-n/2}$ are now eliminated. However, we still have a
logarithmically divergent term \textrm{Im}$\Sigma^{R}
_{\text{ex}}$, given by  Eq.~(\ref{h2}). As this term does not
contain higher than the second order in $u$, its divergence  can
be cut only by a finite curvature of the fermion dispersion (see
Sec.~\ref
{sec:curv}).

Notice also that the collective-mode contribution to $\mathrm{Im}%
\Sigma _{\rm{F}}^{R}$ is smaller than the rest of the contributions by a
large
logarithm. Indeed, the maximum value of $F_{I}(x)$ in  Eq.~(\ref{d2}) is of
order one, so that $\mathrm{Im}\Sigma _{\mathrm{ZS}}^{R}\lesssim
u^{2}\omega
^{2}/E_{F}$, whereas
\begin{equation}
\mathrm{Im}\Sigma _{\rm{B}}^{R}\simeq \mathrm{Im}\Sigma _{\rm{PH}%
}^{R}\simeq \mathrm{Im}\Sigma _{\text{ex}}^{R} \simeq\left( u^{2}\omega
^{2}/E_{F}\right) \ln E_{F}/\left| \omega \right| \gg u^{2}\omega
^{2}/E_{F}.
\end{equation}
Still,  $\mathrm{Im}\Sigma
_{%
\mathrm{ZS}}^{R}$ exhibits a non-monotonic and rapid variation near the
mass
shell at $\Delta \simeq u^{2}\omega \ll \omega $, whereas other
contributions are either constant or vary only smoothly on this
scale. This
feature will have consequences for the spectral function,
discussed in Sec.~\ref{sec:spectral}.

\subsubsection{modifications due to a finite curvature of the fermion
dispersion}
\label{sec:curv}
As we have already mentioned in Sec.~\ref{sec:imsigma}, the logarithmic
mass-shell singularity in \textrm{%
Im}$\Sigma _{\rm{F}}^{R}$ at the second order can be eliminated by
accounting for the finite curvature of the dispersion.
Technically, this amounts to retaining the quadratic-in-$q$ term
in the expansion of $\epsilon _{\mathbf{k+Q}}$ in $\mathbf{Q}$. A
straightforward analysis shows~\cite{chm} that the logarithmic
singularity in the second-order diagram is cut at a certain
distance to the mass shell $|\Delta |\simeq \Delta _{c} $, where,
we remind, $\Delta _{c}\equiv\omega ^{2}/W$,
$W=k_{F}^{2}/(2m_{c})$, and $1/m_{c}$ is the curvature. For the
quadratic dispersion $\epsilon _{k}=k^{2}/(2m)$, $m_{c}=m$, hence
$W=E_{F}$. For a non-quadratic dispersion, $W$ and $E_{F}$ are not
equivalent, but, at least for any power-law spectrum, they are of
the same order. Therefore we will not distinguish between $W$ and
$E_{F}$ in the rest of the paper. Cutting the log-singularity in
\textrm{Im}$\Sigma _{2,\rm{F}}^{R}$ at $\Delta _{c} $, we obtain
\begin{equation}
\text{\textrm{Im}}\Sigma _{2,\rm{F}}^{R}\left( \Delta =0,\omega
\right) =\frac{%
u^{2}}{4\pi }\frac{\omega ^{2}}{E_{F}}\ln \frac{E_{F}}{|\omega |}.
\label{i3}
\end{equation}
The net second-order self-energy, \emph{i.e.}, the sum of
backscattering and forward-scattering contributions, is then given
by
\begin{equation}
\text{\textrm{Im}}\Sigma _{2}^{R}\left( \omega
\right) =\text{\textrm{Im}}%
\Sigma^{R} _{\rm{B}}\left( \omega \right) +\text{\textrm{Im}}\Sigma
_{2,\rm{F}}^{R}\left(
\omega ,\Delta =0\right) =2\text{\textrm{Im}}\Sigma^{R} _{\rm{B}}\left(
\omega
\right) =\frac{u^{2}}{2\pi }\frac{\omega ^{2}}{E_{F}}\ln \frac{E_{F}}{%
|\omega |}.
\end{equation}
The elimination of $\ln |\Delta| $ singularity at the second order
does not eliminate the need for re-summation of the perturbation
theory, since higher-order terms diverge as powers of $|\Delta|
^{-1}$. Indeed, cutting the singularities at $|\Delta| =\Delta
_{c}$ in the general expression  Eq.~(\ref{b1}) for \textrm{Im}$\Sigma
_{\rm{F}}^{R}(\omega )$, we obtain
\begin{equation}
\text{\textrm{Im}}\Sigma _{\rm{F}}^{R}(\omega )=\frac{u^{2}}{4\pi
}\frac{\omega
^{2}}{E_{F}}\left[ \ln \frac{E_{F}}{|\omega |}+\sum_{n=1}^{\infty
}C_{n}\left( \frac{\omega _{c}}{\omega }\right) ^{n/2}\right] ,
\label{b1_1}
\end{equation}
where
\begin{equation}
\omega _{c}\equiv u^{2}E_{F} \simeq \frac{\omega^2}{u^2} \Delta_c.
\label{h4}
\end{equation}
Obviously, the series for $\text{\textrm{Im}}\Sigma _{\rm{F}}^{R}$
does not converge for $|\omega| \lesssim \omega _{c}$,
\emph{i.e.}, one still needs to re-sum the perturbation theory. We
already know, however, that all power-law
divergences are eliminated after such  re-summation even for infinite $%
\omega _{c}$. Finite curvature is not going to modify the results for
$\Delta\gg\Delta_c$. For arbitrary $\Delta$, inclusion of the curvature
will modify scaling functions $F_{I}$ and $G_{I}$, which will now depend
on
two variables: $F_{I}(2\Delta /(u^{2}\omega ),\Delta/\Delta_{c})$ and $%
G_{I}(2\Delta /(u^{2}\omega ),\Delta/\Delta_{c})$.
We have not attempted to
determine the most general form of these functions. However, we can make
certain
conclusions about their behavior near the mass shell.
Indeed, as power-law singularities are cut at $\Delta_c$, a
particular contribution to the self-energy for
$\Delta\ll\Delta_c$ is obtained by taking  an explicit result for this
contribution
for $\Delta\gg\Delta_c$ and replacing $\Delta$ by $\Delta_c$.
For example, the particle-hole contribution, given by  Eq.~(\ref{may_5_2})
for $\Delta\gg\Delta_c$, takes the following form  for
$\Delta\ll\Delta_c$:
\begin{equation}
\text{\textrm{Im}}\Sigma _{\rm{PH}}^{R}=\frac{u^{2}}{4\pi }~\frac{\omega
^{2}}{E_{F}}\left[ \ln \frac{E_{F}}{u^{2}|\omega|}+G_{I}\left(
\frac{2|\omega|}{\omega_c}\right) \right].   \label{may_5_2_1}
\end{equation}
 Function $G_I(x)$ in this form is still given by  Eq.~(\ref{may_5_3}). We
recall that
$G_I(x)$ behaves as $\ln|x|^{-1}$ and $x\ln|x|$ for $|x|\gg 1$ and $|x|\ll
1$,
correspondingly. Using the small-$x$ asymptotic form of $G_I(x)$, we find
that  the second term
in  Eq.~(\ref{may_5_2_1}) is much smaller than the first one
for $|\omega|\ll \omega_c$. Therefore, $\mathrm{Im}\Sigma^{R}_{\rm{PH}}$
in this limit is given by
\begin{equation}
\mathrm{Im}\Sigma _{\rm{PH}}^{R}(\omega)=\frac{u^{2}}{4\pi }~\frac{\omega
^{2}}{E_{F}}\ln \frac{E_{F}}{u^{2}|\omega|},\quad{\rm
for}\quad\omega\ll\omega_c.
\label{june7_1}
\end{equation}
The opposite limit of $|\omega|\gg\omega_c$ (large $x$) exists
only for a finite curvature. In this limit,
$\mathrm{Im}\Sigma^{R}_{\rm{PH}}$ reduces to
\begin{equation}
\mathrm{Im}\Sigma _{\rm{PH}}^{R}(\omega)=\frac{u^{2}}{2\pi }~\frac{\omega
^{2}}{E_{F}}\ln \frac{E_{F}}{|\omega|},\quad{\rm
for}\quad\omega\gg\omega_c.
\label{june7_2}
\end{equation}
A similar procedure is applied to the contribution from
\textrm{Im}$\Sigma^R_{\text{ex}}$.
Away from the mass-shell, \textrm{Im}$\Sigma^R _{\text{ex}}$ is
given by  Eq.~(\ref{h2}). A finite curvature cuts the infrared logarithmic
divergence in the same way as in the second-order diagram. As a result, we
obtain on the mass shell
\begin{equation}
\mathrm{Im}\Sigma^{R}_{\text{ex}}(\omega)=-\frac{u^{2}}{4\pi }\frac{\omega
^{2}}{%
E_{F}}\ln \frac{E_{F}}{|\omega |}.  \label{h2_1}
\end{equation}
To logarithmic accuracy, a general form of $\mathrm{Im}\Sigma^R
_{\mathrm{ex}}$
can be written as
\begin{equation}
\mathrm{Im}\Sigma _{\text{ex}}^{R}(\omega ,\Delta) =-\frac{u^{2}}{8\pi
}\frac{%
\omega ^{2}}{E_{F}}\ln \frac{E_{F}}{\max \{|\Delta |,\Delta _{c}\}}.
\label{h2_2_2}
\end{equation}
Finally, the scaling function for the zero-sound contribution
[$F_I(x)$ from  Eq.~(\ref{j1})] is small as a power-law of either
$x$ (for small $x$) or $x^{-1}$ (for large $x$). Therefore,
for both $|\omega|\ll \omega_c$ and $|\omega|\gg\omega_c$ regimes, the
zero-sound contribution $\text{Im}\Sigma^R_{{\rm ZS}}$ can be
neglected compared to $\mathrm{Im}\Sigma^R_{{\rm PH}}
+\mathrm{Im}\Sigma^R_{{\rm ex}}$.

Combining the formulas for the mass-shell forms
of $\mathrm{Im}\Sigma^{R}_{{\rm PH}}$ [Eqs.~(\ref{june7_1}) and
(\ref{june7_2})]
and $\mathrm{Im}\Sigma^{R}_{{\rm ex}}$ [ Eq.~(\ref{h2_1})], we arrive at
\begin{eqnarray}
\mathrm{Im}\Sigma_{\rm{F}}^R=\frac{u^2}{4\pi}\frac{\omega^2}{E_F}\times\left\{
\begin{array}{cl}
|\ln u^2|,\quad{\rm for}\quad|\omega|\ll\omega_c;\\
\ln E_F/|\omega|,\quad{\rm
for}\quad|\omega|\gg\omega_c.\label{imsigma_f_in}
\end{array}
\right.
\end{eqnarray}
Notice that for $|\omega|\ll\omega_c$, there is no $\omega$-
dependence in the logarithm, \emph{i.e}., the frequency dependence of
\textrm{Im}$\Sigma _{\rm{F}}^{R}$ is perfectly regular in this range of
$\omega$. Notice also that \textrm{%
Im}$\Sigma _{\rm{F}}^{R}(\omega )$ remains positive for all
frequencies, as it should in order for the quasi-particles to be
stable.

\subsubsection{final result for imaginary part of the self-energy on the mass shell}
\paragraph{{\bf contact potential}} The net self-energy is a sum of forward
scattering and backscattering contributions. On the mass shell
($\Delta =0$),
  forward- and backscattering contributions to
\textrm{Im}$\Sigma ^{R}$ are equal to each other  for $|\omega
|\gg \omega _{c}$ [see Eqs.~(\ref{c1c}) and (\ref{imsigma_f_in})],
whereas
 in the opposite limit of  $|\omega |\ll \omega _{c},$
the forward-scattering part [ Eq.~(\ref{imsigma_f_in})] is smaller by a
large
logarithmic factor than the backscattering one. The leading-order result for the
on-shell \textrm{Im}$\Sigma ^{R}$ can then be written as
\begin{equation}
\mathrm{Im}\Sigma ^{R}\left( \omega \right) =\mathrm{Im}\Sigma
_{\rm{F}}^{R}\left( \omega \right) +\mathrm{Im}\Sigma _{\rm{B}}^{R}\left(
\omega
\right) =\frac{u^{2}}{2\pi }\frac{\omega ^{2}}{E_{F}}\ln \frac{E_{F}}{%
|\omega |}\Phi_0\left( \frac{\left| \omega \right|
}{u^{2}E_{F}}\right) , \label{imsigmamain_1}
\end{equation}
where
\begin{equation}
\Phi_0\left( x\right) =\left\{
\begin{array}{cl}
1,\text{ for }x\gg 1; &  \\
1/2,\text{ for }x\ll 1. &
\end{array}
\right.   \label{fi_1}
\end{equation}
As we see, the non-perturbative effect in $\mathrm{Im}\Sigma
^{R}$ on the mass shell is rather benign: all we have is a smooth
crossover function
interpolating between two different values of the numerical prefactor in a
familiar $\omega ^{2}\ln \left| \omega \right| $-dependence \cite{2D}.
Away from the mass shell (at $\Delta \neq 0$), the non-perturbative effect
in $%
\mathrm{Im}\Sigma ^{R}$ is much more pronounced, and a non-monotonic
behavior of the zero-sound term (%
\ref{d2}, \ref{j1}) gives rise to a
 non-monotonic variation of  $\mathrm{Im}\Sigma ^{R}$ near $\Delta =0$.
We will return to this issue in Sec.~\ref{sec:spectral}, where
we discuss the spectral function.

\paragraph{{\bf finite range potential}}
For a finite-range potential, a factor of $u^2$ in the
backscattering contribution is \cite{chm}
 replaced by $u^2_0 + u^2_{2k_F} - u_0 u_{2k_F}$,
where \begin{equation} u_{0}\equiv mU\left( 0\right) /2\pi,\;
u_{2k_F}\equiv mU(2k_F)/2\pi .\label{diffu}
\end{equation}
 In the forward-scattering contribution, $u$ is just replaced by
$u_0$. As a result, the net imaginary part of the self-energy on
the mass shell becomes
\begin{equation}
\mathrm{Im}\Sigma ^{R}\left( \omega \right) =\mathrm{Im}\Sigma
_{\rm{F}}^{R}\left( \omega \right) +\mathrm{Im}\Sigma _{\rm{B}}^{R}\left(
\omega
\right) =\frac{u^2_0}{2\pi}~\frac{\omega ^{2}}{E_{F}}\ln \frac{E_{F}}{%
|\omega |}\Phi \left( \frac{\left| \omega \right| }{
u_{\max}^{2}E_{F}}\right) , \label{imsigmamain}
\end{equation}
where $u_{\max} \equiv \max\{u_0, u_{2k_F}\}$, and
\begin{equation}
\Phi \left( x\right) =\left\{
\begin{array}{cl}
1 + (2 u_0)^{-1}u_{2k_F} (u_{2k_F} - u_0),\text{ for }x\gg 1; &  \\
1/2 +(2 u_0)^{-1}u_{2k_F} (u_{2k_F} - u_0),\text{ for }x\ll 1. &
\end{array}
\right.   \label{fi}
\end{equation}

\subsection{Real part of the self-energy}
\label{sec:resigma}

Next, we consider what happens to \textrm{Re}$\Sigma ^{R}(\omega
)$ near the mass shell. For definiteness, we consider $\omega
>0$
but $\Delta =\omega -\epsilon _{k}$ can be of any sign. The real part of
the
self-energy can be obtained either by a Kramers-Kr{\"o}nig transformation
of
\textrm{Im}$\Sigma^{R} $ or directly, by evaluating the self-energy in
Matsubara
frequencies and analytically continuing it to real frequencies.

\subsubsection{backscattering}

First, we present the result for the total backscattering
contribution to the self-energy. (By ``total'', we mean the sum of
$g_{2}-$ and $2k_{F}-$ contributions). A Kramers-Kr{\"o}nig
transformation of  Eq.~(\ref{c1c}) yields, on the mass shell,
\begin{equation}
\mathrm{Re}\Sigma _{\rm{B}}^{R}\left( \omega \right) =\frac{2}{\pi
}\mathcal{P}%
\int \frac{\mathrm{Im}\Sigma _{\rm{B}}^{R}\left( E,\epsilon _{k}=\omega
\right) }{%
E-\omega }=-\frac{u^{2}}{8}\frac{\omega \left| \omega \right| }{E_{F}}.
\label{g1}
\end{equation}
A non-analytic, $\omega |\omega |$-behavior of $\mathrm{Re}\Sigma
_{\rm{B}}^{R}\left( \omega \right) $ is obviously related to a
non-analytic, $\omega
^{2}\ln |\omega |$-behavior of \textrm{Im}$\Sigma
_{\rm{B}}^{R}(\omega,k)$.

\subsubsection{forward scattering}

The real part of the self-energy consists of three contributions:
from the remainder term ($\Sigma^R _{\text{ex}}$), from the
particle-hole continuum
($\Sigma^R _{\rm{PH}}$), and the from the collective mode ($\Sigma^R _{\rm{ZS%
}}$).

\paragraph{\textbf{remainder}} Performing
Kramers-Kr{\"o}nig
transformation of \textrm{Im}$\Sigma _{\text{ex}}^{R}$ (given by
(\ref{h2})), we  find
\begin{eqnarray}
\text{\textrm{Re}}\Sigma _{\text{ex}}^{R}(\omega ,k) &=&-\frac{u^{2}}{8\pi
^{2}E_{F}}~\mathcal{P}\int_{-\infty }^{\infty }dz\frac{z^{2}}{z-\omega
}~\ln
\frac{E_{F}}{|z-\epsilon _{k}|}  \notag \\
&=&-\frac{u^{2}}{8\pi ^{2}E_{F}}\mathcal{P}\text{ }\lim_{E_{F}\rightarrow
\infty }\int_{-E_{F}}^{E_{F}}dx\frac{\left( x+\omega -\Delta
\right) ^{2}}{%
x-\Delta }\ln \frac{E_{F}}{|x|}  \notag \\
&=&-\frac{u^{2}}{2\pi ^{2}}\left( \omega -\frac{1}{2}\Delta
\right) +\frac{%
u^{2}}{16}~\frac{\omega |\omega |}{E_{F}}~\text{\textrm{sgn}}\Delta +%
\mathcal{O}\left( E_{F}^{-2}\right) +\dots  \label{KK_2}
\end{eqnarray}
The first term in  Eq.~(\ref{KK_2}) is responsible for the renormalization
of
the effective mass and $Z-$ factor, and we neglect it. The second term has
the right --$\omega |\omega |$-- frequency dependence, but its value on
the
mass shell depends on how we take the limit $\Delta \rightarrow 0$. This
ambiguity is due to the logarithmic
 singularity in \textrm{Im}$\Sigma_{\text{ex}%
}^{R}$ [cf.  Eq.~(\ref{h2})]. To eliminate this ambiguity, one has to
re-evaluate the integral using the full form of \textrm{Im}$\Sigma
_{\text{ex%
}}^{R}$, obtained by keeping the curvature finite. This form is
given by
 Eq.~(%
\ref{h2_1}) and is independent of $\Delta $ for $\Delta \rightarrow 0.$
Performing a Kramers-Kr{\"o}nig transformation of  Eq.~(\ref{h2_1}), we
obtain
\begin{equation}
\text{\textrm{Re}}\Sigma _{\text{ex}}^{R}=\frac{u^{2}}{8}\frac{\omega
|\omega |}{E_{F}} + O(\Delta^2 \log \Delta).  \label{h2_3}
\end{equation}
Alternatively, one could just notice that
\begin{equation}
\ln |\omega |=\text{\textrm{Im}}\left( \frac{i}{2}\ln \left[
-(\omega +i0^{+})^{2}\right] \right).  \label{identity}
\end{equation}
Substituting  Eq.~(\ref{identity}) into  Eq.~(\ref{h2_1}), one obtains the full
(complex) $\Sigma^R_{\text{ex}}$:
\begin{equation}
\Sigma _{\text{ex}}^{R}=-\frac{u^{2}}{8\pi }\frac{\omega ^{2}}{E_{F}}i\ln
\left[ -\frac{E_{F}^{2}}{(\omega +i0^{+})^{2}}\right]
.  \label{fullsigmaex}
\end{equation}
Taking the real part of  Eq.~(\ref{fullsigmaex}), we indeed reproduce  Eq.~(\ref
{h2_3}).

\begin{figure}[tbp]
\begin{center}
\epsfxsize=0.5 \columnwidth
\epsffile{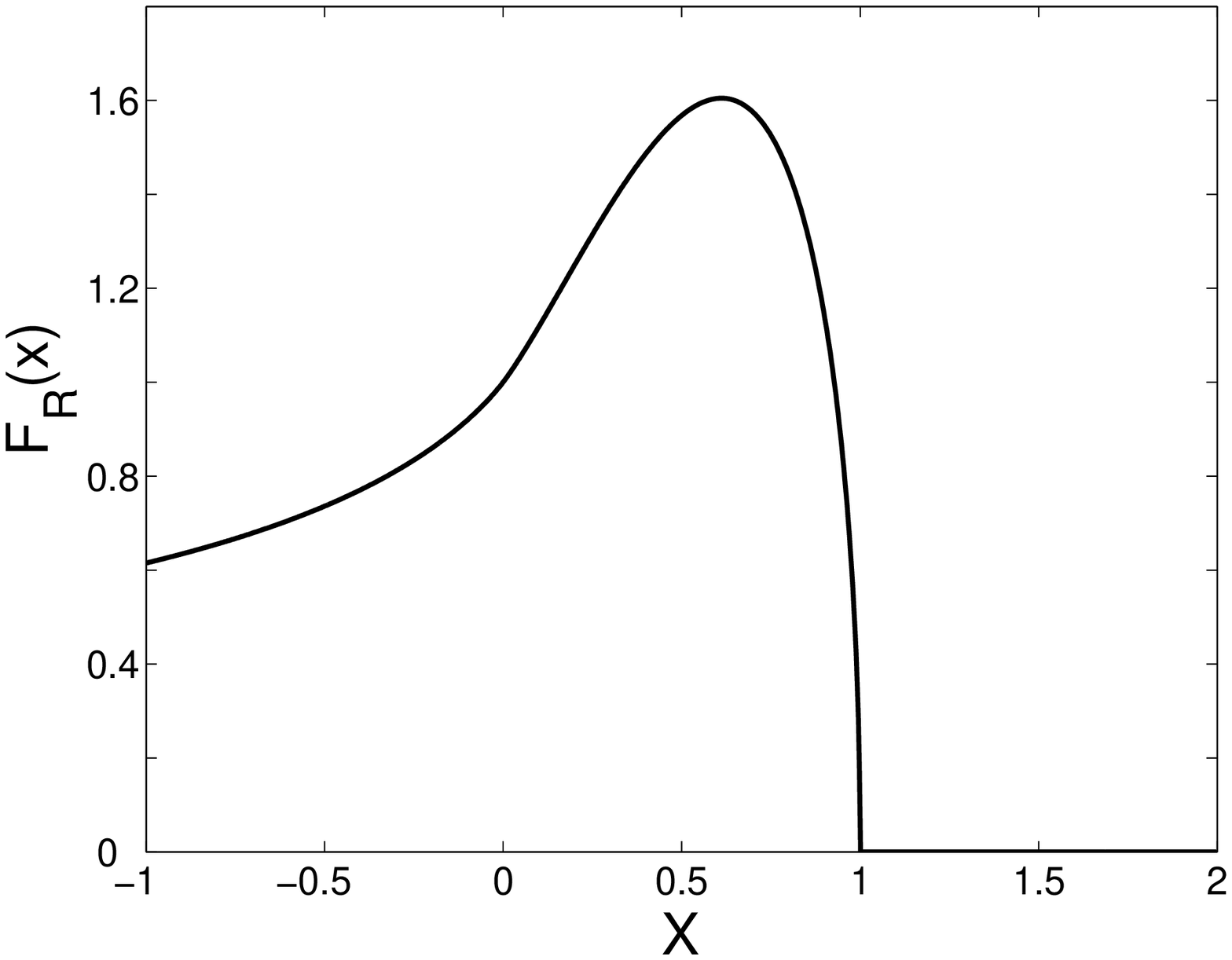}
\end{center}
\caption{Scaling function $F_{R}\left( x\right) $ [ Eq.~(\ref{feb6_7})].}
\label{fig:plotFR}
\end{figure}

\paragraph{\textbf{particle-hole contribution}}
The same reasoning can be applied to the particle-hole contribution,
$\Sigma
_{\rm{PH}}^{R}$. The imaginary part of $\Sigma
_{\rm{PH}}^{R}$  near the mass
shell is given by  Eq.~(\ref{may_5_6}). Using relation
(\ref{identity}) again,
we restore the full $\Sigma _{\rm{PH}}^{R}$ as
\begin{equation}
\Sigma _{\rm{PH}}^{R}=i\frac{u^{2}}{8\pi }\frac{\omega ^{2}}{E_{F}}\ln %
\left[ -\frac{E_{F}^{2}}{u^{2}(\omega +i0^{+})^{2}}\right] +i\frac{\left|
\omega \right| \Delta }{4\pi E_{F}}\ln \left[ -\frac{\Delta ^{2}}{%
u^{4}(\omega +i0^{+})^{2}}\right] ,  \label{fullsigmaph}
\end{equation}
where we have also kept a first sub-leading term in $\Delta .$ The
real part of  Eq.~(\ref{fullsigmaph}) is given by
\begin{equation}
\text{\textrm{Re}}\Sigma _{\rm{PH}}^{R}=-\frac{u^{2}}{8}\frac{\omega
|\omega |}{E_{F}}-\frac{|\omega |\Delta }{4E_{F}}.  \label{h2_4}
\end{equation}
Adding up Eqs.~(\ref{h2_3}) and (\ref{h2_4}), we see that the leading $%
\omega \left| \omega \right|$-terms cancel each other, whereas the
rest vanishes linearly on the mass shell:
\begin{equation}
\text{\textrm{Re}}\Sigma
_{\rm{PH}}^{R}+\text{\textrm{Re}}\Sigma^{R}_{\text{ex%
}}=-\frac{|\omega |\Delta }{4E_{F}}.  \label{h2_5}
\end{equation}
Absence of a non-analytic, $\omega |\omega |$-term in \textrm{Re}$%
\Sigma _{\rm{PH}}^{R}$ + \textrm{Re}$\Sigma^R _{\text{ex}}$ is consistent
with our earlier observation that on the mass shell $\text{\textrm{Im}}%
\Sigma _{\rm{PH}}^{R}+\text{\textrm{Im}}\Sigma^{R} _{\text{ex}}$
is an analytic function of frequency [it scales as $\omega^2
u^{2}\ln u$, see
(%
\ref{imsigma_f_in})]. Notice also that  Eq.~(\ref{h2_5}) is
independent of $u$ . On its own, such a term in the self-energy
will give rise to the linear-in-$\omega $ and
$u$%
-independent correction to the density of states. We will see, however,
that
this term will be cancelled out by the contribution from the zero-sound
collective mode, so that the full density of states remains analytic in $%
\omega .$
\paragraph{\textbf{zero-sound contribution}}
Next, we consider the contribution from the zero-sound collective
mode. The real part of $\Sigma _{\rm{ZS}}^{R}$ can be obtained
either by a Kramers-Kr{\"o}nig transformation of  Eq.~(\ref{d2}), or
directly from  Eq.~(\ref{b8}), by expanding $1-U\Pi (q)$ near the pole
and performing the frequency and
angular%
\textbf{\ }integrations. Either way, we obtain for $\Delta \ll \omega $
\begin{equation}
\mathrm{Re}\Sigma _{\rm{ZS}}^{R}=\frac{u^{2}U}{4\pi }\text{\textrm{Re}}%
\left[ \int_{0}^{\omega /v_{F}}\frac{QdQ}{\sqrt{u^{2}-2\Delta /(v_{F}q)}}%
\right] .  \label{feb6_5}
\end{equation}
Evaluating the integral, we obtain
\begin{equation}
\mathrm{Re}\Sigma^{R} _{\rm{ZS}}=\frac{u^{2}}{8}~\frac{\omega
^{2}}{E_{F}}%
~F_{R}\left( \frac{2\Delta }{u^{2}\omega }\right) ,  \label{feb6_6}
\end{equation}
where
\begin{equation}
F_{R}(x)=\text{\textrm{Re}}\left[ (1+\frac{3}{2}x)\sqrt{1-x}+\frac{3}{2}%
x^{2}\ln \frac{1+\sqrt{1-x}}{\sqrt{-x}}\right] .  \label{feb6_7}
\end{equation}
Subindex $R$ implies that this is the scaling function for \textrm{Re}$%
\Sigma _{\rm{ZS}}^{R}$. A plot of $F_{R}(x)$ is presented in Fig.~\ref
{fig:plotFR}. \
Contrary to $F_{I}(x)$ [ Eq.~(\ref{j1})],  scaling function
$F_{R}(x)$ is not one-sided, \emph{i.e.}, it is nonzero for both
positive and negative $x$. However, it is clear from the plot that
this function is asymmetric with respect to $x$.
 In fact, $F_{R}(x)=0$ for $x>1.$
For large and negative $x$,
\begin{equation}
F_{R}(x)\approx 4\mathrm{Re}[1/(5\sqrt{-x})],
\end{equation}
which means that
\begin{equation}
 \mathrm{Re}\Sigma^{R} _{\rm{ZS}}\propto u^{3}\omega
^{2}~(\omega /-\Delta )^{1/2}
\end{equation}
for $\Delta <0.$ This is consistent with the large $x$ behavior of
$%
F_{I}(x)\approx \mathrm{Im}[2\pi/(5\sqrt{-x})]$. Obviously, for
large $x$, full $\Sigma_{\rm{ZS}}^R$ can be written as
\begin{equation}
\Sigma _{\rm{ZS}}^{R}=\frac{u^{3}\sqrt{2}}{20}~\frac{\omega |\omega |}{%
E_{F}}~\left( -\frac{\omega }{\Delta }\right) ^{1/2}.
\end{equation}
We see that away from the mass shell, \emph{i.e.}, at $|\Delta |\gg
\omega $, the collective-mode component of \textrm{Re}$\Sigma
^{R}(k,\omega ) $ scales as $u^{3}\omega ^{2}$, which is smaller
by a factor of $u$ compared to  the $u^{2}\omega ^{2}$ -contribution to
the self-energy from backscattering ( Eq.~(\ref{g1})). This
smallness is another consequence of the fact that quanta of zero
sound are not free bosons: the residue of the corresponding
propagator scales as $u^{2}\omega ^{2}$ and is thus small. For
$x\rightarrow 0,$
\begin{equation}
F_{R}\left( x\right) =1+x+\frac{3}{4}x^{2}\ln x^{-1}+\dots .  \label{frx0}
\end{equation}
On the mass shell, \emph{i.e., }for $x=0$, the function $F_{R}$ approaches a
finite value $F_{R}=1$, so that
\begin{equation}
\mathrm{Re}\Sigma _{\rm{ZS}}^{R}|_{\Delta =0}=\frac{u^{2}}{8}~\frac{\omega
|\omega |}{E_{F}}.  \label{feb6_8}
\end{equation}
We see that the real part of the self-energy due to the interaction with
the
collective mode is strongly enhanced near the mass shell, such that at $%
\Delta =0$ one power of the small parameter $u$ is eliminated, and $
\mathrm{Re}\Sigma^{R} _{\rm{ZS}}$ \textit{becomes of the same order as
the self-energy due to backscattering}. This is one of the central results
of this paper.
Keeping the linear-in-$x$ term in  Eq.~(\ref{frx0}) results in a
linear-in-$%
\Delta $ correction to the self-energy
\begin{equation}
\mathrm{Re}\Sigma _{\rm{ZS}}^{R}=\frac{u^{2}}{8}~\frac{\omega |\omega |}{%
E_{F}}+\frac{|\omega |\Delta }{4E_{F}}.  \label{feb6_9}
\end{equation}
Adding up this result with  Eq.~(\ref{h2_5}), we find for the total
contribution to the on-shell $\mathrm{Re}\Sigma^{R} $ from forward
scattering
\begin{equation}
\text{\textrm{Re}}\Sigma _{\rm{F}}^{R}=\text{\textrm{Re}}\Sigma _{\rm{PH}%
}^{R}+\text{\textrm{Re}}\Sigma^{R}_{\text{ex}}+\mathrm{Re}\Sigma
_{\rm{ZS}%
}^{R}=\frac{u^{2}}{8}~\frac{\omega |\omega
|}{E_{F}}.  \label{resigmaffull}
\end{equation}

\subsubsection{final result for the real part of the self-energy}
\paragraph{\textbf{contact-potential}}
Combining the backscattering and forward scattering contributions
to the self-energy for the contact potential [Eqs.~(\ref{g1}) and
(\ref {resigmaffull}), respectively], we find that non-analytic
($\omega \left| \omega \right| $ and $\omega \Delta$) terms cancel
out, and the net self-energy vanishes on the mass shell:
\begin{equation}
\text{\textrm{Re}}\Sigma ^{R}=\ \text{\textrm{Re}}\Sigma _{\rm{B}}^{R}+%
\text{\textrm{Re}}\Sigma _{\rm{F}}^{R}= O \left(u^2 \Delta ^{2}\ln \Delta
\right).  \label{resigmaf}
\end{equation}
This is another central result of the paper. It means that a
non-perturbative contribution of the zero-sound mode totally changes the
 result of the second order perturbation theory, where to order $u^2$
 we had  \textrm{Re}$\Sigma ^{R}\propto
u^{2}\omega \left| \omega \right| .$

\paragraph{\textbf{finite-range potential}}
For a finite-range potential, the backscattering part of the
self-energy changes to \cite{chm}
\begin{equation}
\text{\textrm{Re}}\Sigma _{\rm{B}}^{R}=-\frac{\omega \left| \omega \right|
}{%
8E_{F}}~\left( u_{0}^{2}+u_{2k_{F}}^{2}-u_{0}u_{2k_{F}}\right),
\label{la-la}
\end{equation}
where $u_{0}$ and $u_{2k_{F}}$ are given by  Eq.~(\ref{diffu}). The
forward-scattering contribution comes only with $u_{0}^{2}$ and is
obtained from  Eq.~(\ref{resigmaffull}) by replacing $u\rightarrow
u_{0}.$ A cancellation between backward and forward-scattering
parts of \textrm{Re}$\Sigma _{{}}^{R}$ is no longer in place, and
the  net \textrm{Re}$\Sigma _{{}}^{R}$ is given by
\begin{equation}
\text{\textrm{Re}}\Sigma ^{R}=\frac{\omega \left| \omega \right|
}{8E_{F}}%
~u_{2k_{F}}\left( u_{0}-u_{2k_{F}}\right) .
\end{equation}
For $u_0\neq u_{2k_F}$, it is a non-analytic function of $\omega $ on the
mass shell.

Finally, in the ZS contributions to the self-energy,
Eq.(\ref{d2}), $u$ is replaced by $u_0$ as this contribution comes
only from forward scattering.

\subsection{Spectral function}
\label{sec:spectral}
\subsubsection{short-range potential}
A non-monotonic variation in the zero-sound part of the
self-energy is manifested in a specific feature in the spectral
function
\begin{eqnarray}
A\left( \omega ,k\right)  &=&-\frac{1}{\pi
}\text{\textrm{Im}}G^{R}\left(
\omega ,k\right)   \notag \\
&=&\frac{1}{\pi }\frac{\text{\textrm{Im}}\Sigma ^{R}\left( \omega
,k\right) }{\left[ \Delta +\text{\textrm{Re}}\Sigma ^{R}\left(
\omega ,k\right) \right] ^{2}+\left[ \text{\textrm{Im}}\Sigma
^{R}\left( \omega ,k\right) \right]
^{2}%
}.  \label{A}
\end{eqnarray}
Having in mind a potential comparison with the experiment, we
present a detailed discussion of $A(\omega,k)$ in this Section.

\begin{figure}[tbp]
\begin{center}
\epsfxsize=0.5 \columnwidth
\epsffile{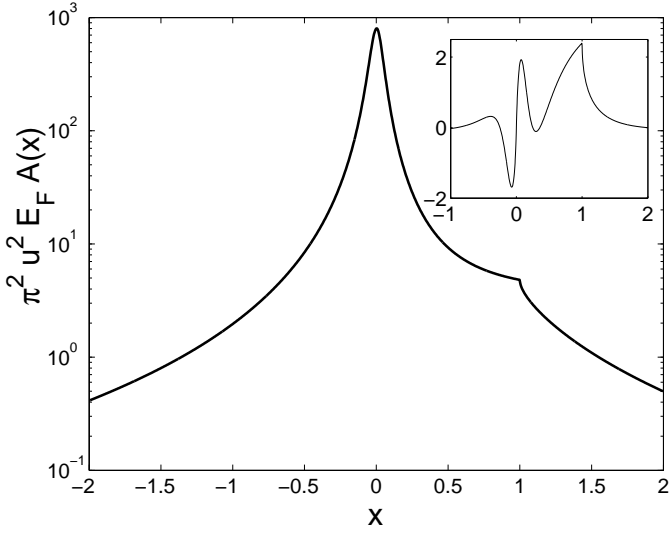}
\end{center}
\caption{Main panel: a $\protect\log$-plot of the spectral
function $A(\omega,k)$ [ Eq.~(\ref {A_full})] in units of
$1/\protect\pi^2u^2E_F$ as a function of
$x=2(\omega-\epsilon_k)/u^2\protect\omega$ for
 $L_\omega=2$ and $\gamma=0.05$.
A kink at $x=1$ is due to the interaction of fermions with the
 zero-sound mode. Inset: part of the spectral function $A_1(\omega,k)$
 [ Eq.~(\ref{AZS})] in the same units
function of $x$
  for $\protect\gamma=0.25$.  A maximum in $A_1$ at $x=1$ gives rise to a
kink in total $A$ (main panel). } \label{fig:spectral}
\end{figure}

A variation of \textrm{Re}$\Sigma ^{R}\left( \omega ,k\right)$
 has only a little effect on the shape of the
spectral function, and we verified that it can be safely ignored. The
effect
 of \textrm{Im}$\Sigma ^{R}\left( \omega ,k\right)$ is much stronger.
 As we have shown in Sec.~\ref{sec:imsigma_all},
\textrm{Im}$\Sigma ^{R}\left( \omega ,k\right)$  is a sum  of four
contributions
\begin{equation}
\text{\textrm{Im}}\Sigma ^{R}\left( \omega ,k\right) =\text{\textrm{Im}}%
\Sigma _{\rm{B}}^{R}+\text{\textrm{Im}}\Sigma
_{\text{ex}}^{R}+\text{\textrm{Im}}%
\Sigma _{\rm{PH}}^{R}+\text{\textrm{Im}}\Sigma _{\rm{ZS}}^{R}.
\label{june5_1}
\end{equation}
  The particle-hole
 and zero-sound contributions contain scaling functions $G_I$ and $F_I$
 which evolve as a function of $\Delta=\omega-\epsilon_k$ on a scale
$\Delta \simeq \Delta^{\ast} \equiv u^{2}\omega /2\ll\omega$. The
backscattering part, on the other hand,  evolves only on much
larger scale:
 $\Delta \simeq
\omega $ (cf. Appendix A).
Therefore, one can safely put $\Delta=0$ in $\mathrm{Im}\Sigma^R
_{\rm{B}}$, \emph{i.e.},
use its mass-shell value given by  Eq.~(\ref{c1c}).
Finally, $\mathrm{Im}\Sigma^R _{\rm{ex}}$
crosses over between the forms given by Eqs.~(\ref{h2}) and (\ref{h2_1})
at  $\Delta \simeq \Delta _{c}=\omega ^{2}/E_{F}.$
Two situations are then possible,
depending on the relation between $\Delta _{c}$ and $\Delta ^{\ast }.$
If $\Delta _{c}\gg \Delta^{\ast}$ or, equivalently,
$\omega \gg
\omega _{c}=u^{2}E_{F},$ the variation of Im$\Sigma^R _{\rm{ZS}}$ and
Im$\Sigma^R_{\rm{PH}}$ occurs in the range where
 Im$\Sigma^R _{\text{ex}}$ can be approximated by its
small-$\Delta $ form [ Eq.~(\ref{h2_1})], which is independent of $\Delta
.$
The sum of \textrm{Im}$\Sigma ^R_{\text{ex}}$ and \textrm{Im%
}$\Sigma^R _{\rm{B}}$ then vanishes, so that
$\mathrm{Im}\Sigma^R(\omega,k) = \mathrm{Im}\Sigma^R_{\rm{PH}} +
\mathrm{Im}\Sigma^R_{\rm{ZS}}$. This sum can be further decomposed
as
\begin{equation}
\mathrm{Im}\Sigma^R(\omega,k)=\Gamma_0+\Gamma_1 (x).
\label{spec_sigma}\end{equation} where $x=2\Delta /u^{2}\omega =
\Delta/\Delta^{\ast}$, $\Gamma_0$ is a
 $\Delta -$ independent
part of the self-energy [the first term in the particle-hole
contribution,  Eq.~(\ref{may_5_2})]:
\begin{equation}
\Gamma _{0}=\frac{u^{2}}{4\pi }\frac{\omega ^{2}}{E_{F}}\ln \frac{E_{F}}{%
u^{2}\left| \omega \right| },\label{gamma_0}
\end{equation}
and $\Gamma_1 (x)$ is a sum of the scaling terms in
\textrm{Im}$\Sigma^R_{\rm{PH}}$ and \textrm{Im}$\Sigma
^R_{\rm{ZS}}$ [Eqs.~(\ref{may_5_2}) and (\ref{d2}), respectively]
\begin{equation}
\Gamma _{1} (x) =\frac{u^{2}}{4\pi }\frac{\omega ^{2}}{E_{F}}\left[
G_{I}\left(
x\right) +F_{I}\left( x\right) \right].\label{gamma_1}
\end{equation}
Notice that $F_I (0) =0$ and $G_I (0) =0$.
Substituting  Eq.~(\ref{spec_sigma})-(\ref{gamma_1}) into  Eq.~(\ref{A}) and
neglecting $\mathrm{Re}\Sigma^R$, we obtain a scaling form
of the spectral function
\begin{equation}
A(\omega,k)=\frac{1}{\pi^2 u^2 E_F}\frac{L_{\omega}+G_I(x)+F_I(x)}
{x^2+\gamma^2\left[L_{\omega}+G_I(x)+F_I(x)\right]^2},
\label{A_full}
\end{equation}
where
\begin{equation}
\gamma \equiv \frac{\left| \omega \right| }{2\pi E_{F}} \ll 1,
\end{equation}
and $L_\omega \equiv\ln\left(E_F/u^2|\omega|\right)$ is a large
factor. We consider a setup when $\omega$ is fixed and the
spectral function is measured as a function of the momentum. This
is equivalent to varying $x$ at fixed $\omega$ in
 Eq.~(\ref{A_full}). In photoemission measurements, this setup
produces what is known as a ``momentum distribution curve'' (MDC).
A plot of $A$ [ Eq.~(\ref{A_full})] as a function of $x$ is shown in
the main panel of
 Fig.~\ref{fig:spectral}. For solely illustrative purposes, we have chosen $
 L_\omega = 2$
and $\gamma = 0.05$. We see that the spectral function contains
 not only a narrow  quasi-particle peak at $x=0$ (i.e, $\omega = \epsilon_k$)
 but also a well-pronounced kink at $x=1$.
To understand the reasons for the kink in $A(\omega,k)$, we notice that
at typical $\Delta \simeq \Delta^{\ast}$,  $\Gamma _{1}$ is of order $%
\omega ^{2}/E_{F}$, which is smaller than $\Gamma _{0}$ by  a large
$L_\omega$.
Therefore, we can expand
 the spectral function in $\Gamma _{1}$ and
  represent $A(\omega,k)$ as a sum of two contributions
\begin{subequations}
\begin{eqnarray}
A(\omega,k)&=&A_0(\omega,k)+A_1(\omega,k)  \label{AA} \\
A_0(\omega,k)&=&\frac{1}{\pi^2u^2E_F}\frac{L_\omega} {%
x^2+\gamma_1^2}  \label{A0} \\
A_{1}\left( \omega ,k\right)&=&\frac{1}{\pi^2u^{2}E_{F}}\left[ G_{I}\left(
x\right) +F_{I}\left( x\right) \right] \frac{x^{2}-\gamma_1 ^{2}}{\left(
x^{2}+\gamma_1 ^{2}\right) ^{2}}.  \label{AZS}
\end{eqnarray}
\end{subequations}
where $\gamma_1 = \gamma L_\omega$.
The first term  Eq.~(\ref{A0}) describes
 a regular quasi-particle peak at $\Delta=0
$ of width $\gamma_1$.
 The scaling behavior of the self-energy
 shows up in the second term,  Eq.~(\ref{AZS}).
Since $\gamma_1 \ll 1$,  damping  affects the behavior of
$A_{1}\left( \omega ,k\right) $ only at very small $x$: $x\simeq
\gamma_1 \ll 1$.
 For these $x$, $A_{1}$ exhibits a rapid non-monotonic variation,
but it  is overshadowed
by the rapid variation of $A_0$. For $|x|\gg\gamma_1$, $A_0$ is smooth,
whereas $A_1$ is determined by the sum of the scaling functions, which is
non-monotonic in $x$
\begin{equation}
A_{1}\left( \omega ,k\right) \approx
\frac{1}{u^{2}E_{F}}~\frac{G_{I}\left(
x\right) +F_{I}\left( x\right) }{x^{2}}.  \label{AZS_1}
\end{equation}
Comparing now the behavior of two scaling functions, $G_I$ and $F_I$,
 we see from Fig.~\ref{fig:plot} that $G_I(x)$ varies smoothly at $x\simeq
1$,
  whereas $F_I (x)$ has a sharp peak at $x=1$.
We remind (cf. discussion in Sec.~\ref{sec:imsigma_all}) that this
sharp peak
 is associated with  the fact that, for $|\Delta|>\Delta^*$, a fermion with energy $\omega>0$ above the Fermi level
 can emit ZS bosons with any
 frequency in the interval $0<\Omega<\omega$ (or absorb bosons in
 the interval $0<\Omega<-\omega$ for $\omega<0$), whereas for $|\Delta|<\Delta^*$, a Cherenkov-type
 restriction makes it impossible to emit and absorb
 bosons with frequencies
 above $|\Delta|/(1-v_F/c)\approx
 2|\Delta |/u^2$.
 The  sharp peak in $F_I (x)$  gives rise to a peak in
  $A_1\left( \omega ,k\right)$  at $x =1$, see inset in
Fig.~\ref{fig:spectral}. The peak in $A_1$
 gives rise to a kink in the full $A(\omega,k)$ at $x=1$.
We emphasize that  the kink originates from the zero-sound
contribution
 to the self-energy, \emph{i.e.}, it reflects an essentially non-perturbative effect.

 The second situation occurs when $\Delta
_{c}\gg \Delta ^{\ast},$ \emph{ i.e.}, $\omega \ll \omega
_{c}=u^{2}E_{F}.$ In this case,
 for $x \simeq 1$,  Im$\Sigma^{R}_{\text{ex}} $
can be replaced by its large-$\Delta $ form
[ Eq.~(\ref{h2})]. Re-expressing
Im$\Sigma^R_{\text ex}$ in terms of the dimensionless variable $x$ we obtain
\begin{equation}
\text{Im}\Sigma^R_{\text{ex}}=-\frac{u^{2}}{8\pi }\frac{\omega
^{2}}{E_{F}}%
\left( \ln \frac{E_{F}}{u^{2}\left| \omega \right| }-\ln |x|\right) .
\label{june4_1}
\end{equation}
Decomposing again Im$\Sigma ^{R}$ into $x -$ independent and
$x -$ dependent parts, we obtain instead of Eqs.~(\ref{AA}-\ref{AZS})
\begin{subequations}
\begin{eqnarray}
A(\omega,k)&=&A_0(\omega,k)+A_1(\omega,k)  \label{AA_1} \\
A_0(\omega,k)&=&\frac{1}{2\pi^2u^2E_F}\frac{\ln E_F^3/u^2|\omega|^3}{%
x^2+\gamma_2^2}  \label{A0_1} \\
A_{1}\left( \omega ,k\right) &=&\frac{1}{2\pi^2u^{2}E_{F}}\left[
G_{I}\left(
x\right) +F_{I}\left( x\right) +\frac{1}{2}\ln |x|\right]
\frac{x^{2}-\gamma
^{2}_2}{\left( x^{2}+\gamma^{2}_2\right) ^{2}},  \label{A2}
\end{eqnarray}
\end{subequations}
where
\begin{equation}
\gamma_2 \equiv \frac{\left| \omega \right| }{4\pi E_{F}}\ln
\frac{E_{F}^{3}}{%
u^{2}\left| \omega \right| ^{3}}.
\end{equation}
The behavior of $A\left(\omega ,k\right) $ in this case is a bit
more involved than for  $\Delta_c < \Delta^{\ast}$; nevertheless,
the a general structure is the same as before: the spectral
function has both a quasi-particle peak at $x =0$ and a kink at
$x=1$, due to the contribution from the zero-sound mode.

\subsubsection{Coulomb potential}
A kink in the spectral function due to the interaction of fermions
with the collective mode is not a special feature of the model
with a short-range repulsion  but a general phenomenon. To
illustrate this point, we consider a 2D system with the Coulomb
interaction, when the collective mode is a plasmon with dispersion
$\Omega _{0}(Q)=\left( e^{2}mv_{F}^{2}Q\right) ^{1/2}.$ An
on-shell electron can emit plasmons only if its energy exceeds a
certain critical value: $|\omega |=|\epsilon _{k}|\geq \omega
_{\mathrm{pl}}=\sqrt{2}r_{s}E_{F}$ \cite{quinn}, where $r_{s}$ is
the usual ideal-gas parameter for a charged system, which is
assumed to be small. Although formally there is an interval of
energies in between $\omega _{\mathrm{pl}}$ and $E_{F},$ in
practice it cannot be very large. In what follows, we will
consider only the low-energy limit: $\left| \omega \right|
,|\epsilon _{k}|\ll \omega _{\mathrm{pl}}.$ In this case, emission
of plasmons by electrons is possible only away from the mass shell
($\omega \neq \epsilon _{k}$), and the effect we are interested in
is a kink in the spectral function rather than the lifetime of an
electron.

Near the plasmon pole, the Coulomb potential reduces to
\begin{equation}
V\left( \Omega ,Q\right) =\frac{2\pi e^{4}mv_{F}^{2}}{\left(
\Omega +i\delta \right) ^{2}-\Omega _{0}^{2}\left( Q\right) }
\end{equation}
and, correspondingly, the imaginary part of the self-energy is given by $%
\left( \omega >0\right) $%
\begin{equation}
\text{Im}\Sigma ^{R}\left( \omega ,\epsilon _{k}\right) =2\pi
e^{4}mv_{F}^{2}\int_{0}^{\omega }d\Omega \int d^{2}Q\delta \left(
\omega-\epsilon_k -\Omega -v_{F}Q\cos \theta \right) \delta \left(
\Omega ^{2}-\Omega _{0}^{2}\left( Q\right) \right) .
\label{sigma_c}\end{equation} The second $\delta -$ function
forces the boson momentum  $Q$ to be small:  $Q=\left| \Omega
\right| ^{2}/me^{2}v_{F}^{2}\simeq (|\Omega |/v_{F})\left( \left|
\Omega \right| /\omega _{\mathrm{pl}}\right) \ll \left| \Omega
\right| /v_{F}.$ Therefore, one can neglect the $Q$-dependent term
in the argument of the first $\delta -$ function. Performing an
elementary integration and considering the case of $\omega <0$ in
the same
way, we obtain for Im$\Sigma ^{R}\left( \omega ,\epsilon _{k}\right) :$%
\begin{eqnarray}
\text{Im}\Sigma ^{R}\left( \omega ,\epsilon _{k}\right) =\left\{
\begin{array}{cl}\pi \left(\epsilon_k-\omega \right)^{2}/E_{F},\text{ }\min \{0,\omega \}\leq
\epsilon_k \leq \max \left\{ 0,\omega
\right\}; \nonumber\\
0,\text{ otherwise.}
\end{array}\right.
\label{kink_c}\end{eqnarray} For fixed $\omega$, $\text{Im}\Sigma
^{R}$ has a kink at $\epsilon_k=0$, \emph {i.e.}, at the Fermi
surface, where $\text{Im}\Sigma^R $  vanishes discontinuously (see
comment \cite{kink}). (Keeping the $Q$-dependence in the first
$\delta$-function in Eq.~(\ref{sigma_c}), one sees that in fact the
kink and zero of $\text{Im}\Sigma ^{R}$ are separated by a small
energy scale $\omega ^{2}/\omega _{\mathrm{pl}}$.)  Notice that
the electron charge dropped out of the result. It can be shown
that the spectral function of a bi-layer system with two plasmon
modes--with $\sqrt{Q}$ and acoustic dispersions--behaves in a
similar way but we defer a detailed discussion of this case to
later occassion.

A kink in the spectral function could, in principle,  be detected
in photoemission experiments on layered materials \cite{photo} or
in a momentum-conserving tunneling between two parallel layers of
2D gases \cite{eisenstein}.

\subsubsection{absence of a non-analytic correction to the tunneling
density of states} Both the particle-hole and zero-sound
contributions to the real part of the self-energy contain a
specific term, which is proportional to the product $\left| \omega
\right| \Delta $ and is independent of the interaction [cf.
Eqs.~(\ref{h2_4}) and (\ref{feb6_9})]. Each of these terms on its
own would give rise to a
linear-in-$%
|\omega |$ and $u$-independent correction to the density of
states. Indeed, a term in the self-energy of the form
\begin{equation}
s\left| \omega \right| \Delta /E_{F},  \label{term}
\end{equation}
where $s$ is a numerical  coefficient, gives rise to a
non-analytic frequency dependence of  the renormalization factor:
\begin{equation}
Z(\omega )=\left( 1+s\frac{\left| \omega \right| }{E_{F}}\right)
^{-1}\approx 1-s\frac{\left| \omega \right| }{E_{F}}.
\label{feb6_10}
\end{equation}
[We defined $Z(\omega )$ in such a way that $G^{R}(k,\omega
)=Z(\omega )/(\omega -\epsilon _{k}+i0^{+})$.] The linear
-in-$|\omega|$-term in $Z(\omega )$ results in a linear frequency
dependence of the tunneling density of states
\begin{equation}
N(\omega )=-\frac{2}{\pi }\int \frac{d^{2}k}{\left( 2\pi
\right) ^{2}}\text{%
\textrm{Im}}G^{R}\left( \omega ,k\right) =\frac{m}{\pi }~Z(\omega
)\approx \left( m/\pi \right) \left( 1-s\frac{\left| \omega
\right| }{E_{F}}\right) . \label{tunnel}
\end{equation}
 However, we see that the $\Delta|\omega|$-terms in Eqs.~(\ref{h2_5}) and (\ref{feb6_9})
 cancel out in full \textrm{Re}$\Sigma ^{R}$ [ Eq.~(%
\ref{resigmaffull})], so that $s=0$.  Therefore, to order $u^2$,
 $N(\omega )$ is \emph{analytic} in $\omega.$
 This result is  valid for any finite-range potential as the
cancellation of $\left| \omega \right| \Delta -$ terms occurs
between the forward-scattering contributions to the self-energy,
all of which contain the same coupling $u_{0}.$
 Our result that there is no linear-in
 -$|\omega |$ correction to the DOS  is in agreement with
Ref.~\cite{anton}, where  the tunneling density of states was
obtained for the case of a multi-layer system with the Coulomb
interaction. Inter-layer screening gives rise to an acoustic
branch of the plasmon spectrum which is an analog of the ZS mode
of our model. Notice that an $|\omega|$-correction to the density
of states does exist for a single layer with the Coulomb potential
\cite{reizer,anton}.

\section{Specific heat}
\label{sec:sh}

As we pointed out in the Introduction, the $\omega |\omega|$%
-non-analyticity in the second-order self-energy gives rise to a
non-analytic, $u^{2}T$-correction to the ratio $C\left( T\right)
/T$ \cite {chm}. However, it was shown in Sec.~\ref{sec:resigma}
that, upon re-summation, higher-order forward-scattering
contributions to $\mathrm{Re}\Sigma^{R}$ also becomes of the order
$U^2$ near the mass shell and modify the second order result. For
a contact interaction, the non-perturbative contribution even
cancels the second-order $u^2 \omega |\omega|$-term in
$\mathrm{Re}\Sigma^{R}$.
 The question addressed in this Section is whether the forward-scattering
 component of the self-energy modifies the non-analytic term
 in  the specific heat.  We  show in several ways that this does not happen,
  \emph{i.e.}, the enhancement of the forward scattering self-energy near the mass shell
 does not affect the specific heat.

We also go beyond the weak-coupling limit in this Section, and
consider the non-analytic behavior of the specific heat in a
generic Fermi liquid.

\subsection{Specific heat via self-energy}
\label{sec:cviasigma}

A relation between the entropy (and thus the specific heat) and an
exact fermion  Green's function can be found in Ref.~\cite{agd}.
 However, this relation is justified only for the Fermi-liquid,
 linear-in-$T$ part of $C(T)$.
 In order to find a sub-leading, non-analytic contribution to $C(T)$,
 one needs to re-examine the
 assumptions, made in Ref.~\cite{agd}, and to establish
a correct relation between $C(T)$ and the self-energy beyond the
leading order in $T$.

The relation between the thermodynamic potential and the Green's
function reads \cite{agd}
\begin{equation}
\Xi = 2 T \sum_{\omega_m} \int \frac{d^2k}{(2\pi)^2} \ln{G (
\omega_m,k,T=0)}, \label{ya_2}
\end{equation}
where $G(\omega_m,k,T=0)$ is the Green's function evaluated at
discrete Matsubara frequencies but with no additional
$T$-dependence. Converting the Matsubara sum into the contour
integral and using a familiar thermodynamic relation
\begin{equation}
C\left( T\right) =-T\frac{\partial ^{2}\Xi }{\partial T^{2}},
\label{comega}
\end{equation}
one obtains
\begin{equation}
C(T)/T=-\frac{2T}{\pi }~\frac{\partial }{\partial T}\left[ \frac{1}{T}%
~\int \frac{d^{2}k}{(2\pi )^{2}}~\int_{-\infty }^{\infty }d\omega
\omega \frac{\partial n_{0}}{\partial \omega }\arg G^{R}(\omega
,k) \right]. \label{feb5_2}
\end{equation}
(Notice that there is no need to distinguish between $C_{P}$ and
$C_{V}$ here, as we are interested in the $T^{2}-$ term in
$C\left( T\right) $, whereas the difference between $C_{P}$ and
$C_{V}$ is of the order $T^{3}$ \cite{statphys}.) Both the
retarded and advanced Green's functions in  Eq.~(\ref{feb5_2}) are
evaluated
 at $T=0$, thus the derivative in the r.h.s. of  Eq.~(\ref{feb5_2}) affects only the Fermi function
 [the temperature derivative of $n_0$ was converted into the frequency dependence by using
 a familiar identity: $\partial n_0/\partial T = - (\omega/T)
\partial n_0/\partial \omega$].

 Eq.~(\ref{feb5_2}) correctly describes the regular, linear-in-$T$
part of the specific heat. Indeed, substituting the analytic, FL
form of the self-energy, Eq.(\ref{sigma_an}), into
Eq.~(\ref{feb5_2}), one finds that $C(T)$ is given by the
Fermi-gas result [ Eq.~(\ref{FG})] but with a renormalized mass,
which is composed from coefficients $a$ and $b$ in
Eq.~(\ref{sigma_an}).

It was  conjectured in Ref.~\cite{agd} that  Eq.~(\ref{feb5_2})
describes not only the leading but also the sub-leading terms in
$C(T)$. However, this conjecture is questionable, as the accuracy
of the low-temperature expansion used in the derivation of
 Eq.~(\ref{feb5_2}) is not specified. In other words, it is not obvious
that if one retains contain higher powers of $\omega$ in $\Sigma$,
one should not at the same time take into account an explicit
temperature dependence of $\Sigma$. Of particular concern are the
situations when the self-energy depends on $T$ via a scaling
function of variable $\omega/T$ (this happens in our case; see
below). As typical $\omega$ are of order $T$, the argument of the
scaling function is of order unity, thus neglecting the
$T$-dependence is not justified. Moreover, it was shown in
Ref.\cite{amit} that the zero-temperature and
temperature-dependent parts of the self-energy give comparable
contributions to the non-analytic, $T^3\ln T$ part of $C(T)$ in
3D.

We will still be considering the case of a weak interaction, when
$|\Sigma^R|\ll |\omega|$.  In this case,  Eq.~(\ref{feb5_2}) can be
simplified further by expanding the logs of Green's function in
the self-energy,
which results in
\begin{equation}
C(T) = C_{{\rm  FG}} (T) + \delta
C(T),\label{sep6_1}\end{equation}
 where $C_{{\rm  FG}} (T)
$ is the specific heat for free fermions in 2D [ Eq.~({\ref{FG})]
and $\delta C(T)$ is given by
\begin{equation}
\delta C(T)/T=\frac{2}{\pi }~\frac{\partial }{\partial T}\left[
\frac{1}{T} ~\int \frac{d^{2}k}{(2\pi
)^{2}}~\int_{-\infty }^{\infty }d\omega \omega \frac{\partial
n_0}{\partial \omega } \text{\textrm{Im}} \left[\Sigma ^{R}(\omega
,k,T=0)G_{0}^{R}(\omega,k)\right]\right]. \label{feb5_3T0}
\end{equation}
 There are
two contributions to $\delta C(T)$--one from
\textrm{Re%
}$\Sigma ^{R}$ and another from \textrm{Im}$\Sigma ^{R}$--which we
label as $C_{1}(T)$ and $C_{2}(T)$, correspondingly:
\begin{subequations}
\begin{eqnarray}
\delta C\left( T\right) &=& C_{1}\left( T\right) +C_{2}\left( T\right) ;
\label{ct} \\
C_{1}(T)/T &=&- 2~\frac{\partial }{\partial T}\left[
\frac{1}{T}~\int \frac{d^{2}k}{(2\pi
)^{2}}~\int_{-\infty }^{\infty }d\omega \omega \frac{\partial
n_0}{\partial \omega }~\delta \left( \omega -\epsilon _{k}\right)
\text{\textrm{Re}}\Sigma ^{R}(\omega ,k,T=0) \right];  \label{ct1}
\\
C_{2}(T)/T &=& \frac{2}{\pi }~\frac{\partial }{\partial T}\left[
\frac{1}{T}%
~\int \frac{d^{2}k}{(2\pi )^{2}}~\int_{-\infty
}^{\infty
}d\omega \omega \frac{\partial n_0}{\partial \omega }\mathcal{P}~\frac{1}{%
\omega -\epsilon _{k}}\text{\textrm{Im}}\Sigma ^{R}(\omega
,k,T=0)\right] . \label{ct2}
\end{eqnarray}
\end{subequations}
In the expression for $C_1 (T)$, we have used the fact that $\text{\textrm{Im}} G_{0}^{R}(\omega
-k)=-\pi \delta (\omega -\epsilon _{k})$. The delta-function in
(\ref{ct1}) implies that $C_1 (T)$ is determined by
\textrm{Re}$\Sigma ^{R} $ only on the mass shell, where $\omega =
\epsilon_k$. The second term $C_2 (T)$ contains the integral of
$\mathrm{Im} \Sigma^R (\omega,k)$. If \textrm{Im}$\Sigma
^{R}(\omega ,k)$ depends on $\omega $ but not $k$, which is the
case, \emph{e.g.}, for the electron-phonon
interaction~\cite{eli_1},
 the momentum integral in  Eq.~(\ref{ct2}) vanishes once one
approximates the density of states by a constant, so that $C_{2}(T)$ drops
out. Indeed,
\begin{equation}
C_{2}\left( T\right) \propto \int d\omega \omega \frac{\partial
n_0}{\partial \omega }\mathrm{Im}\Sigma ^{R}(\omega
)\mathcal{P}\int_{-\infty }^{\infty }d\epsilon _{k}\frac{1}{\omega
-\epsilon _{k}}=0\mathrm{.} \label{c2zero}
\end{equation} However, for a general case, when $\Sigma $ depends on both
$\omega $ and $k$, there are no \emph{a priori} reasons for
$C_{2}(T)$ to vanish, and thus the imaginary part of the
self-energy contributes to the specific heat as well. In what
follows, we will omit the analytic terms in $\Sigma^R$ which just
renormalizes the coefficient of the linear $T$-dependence in
 Eq.~(\ref{FG}), and consider only the non-analytic contributions to
$C_1$ and $C_2$.

Next, we compare the result for the correction to the specific
heat given by Eqs.~(\ref{ct}-\ref{ct2}) to the one obtained in a
different way, namely, employing the Luttinger-Ward formula for
the thermodynamic potential $\Xi$:
\begin{equation}
\Xi -\Xi _{0}= - 2 T \sum_{\omega_m} \int \frac{d^2k}{4\pi^2}
\left[ \ln \left( G_{0}G^{-1}\right) -\Sigma G+\sum_{\nu
}\frac{1}{2\nu }\Sigma_{\nu }G\right].  \label{o1}
\end{equation}
Here, $\Xi _{0}$ is the thermodynamic potential of the free Fermi
gas per unit area, $G_{0}=\left( i\omega _{m}-\epsilon _{k}\right)
^{-1}$, $G=\left( i\omega _{m}-\epsilon _{k}+\Sigma \right)
^{-1}$, $\Sigma $ is the exact (to all orders in the interaction)
self-energy,  and $\Sigma _{\nu }$ is the skeleton self-energy of
order $\nu$.
 Both the skeleton and full self-energy, related via
\begin{equation}
\Sigma=\sum_{\nu}\Sigma_{\nu}, \end{equation}
 are evaluated at {\it
finite} $T$. The diagrams for $\Sigma_{\nu}$ are obtained from
those  in Fig.~\ref{fig:selfenergy} by replacing the bare Green's
function and interaction lines by the exact ones. Expanding both $G$
and $\Sigma_{\nu}$ in  Eq.~(\ref{o1}) back in $\Sigma$, one
generates a perturbative expansion for $\Xi$. To second order,
diagrams generated by the first two terms in  Eq.~(\ref{o1})
correspond to self-energy insertions into a free thermodynamic
potential (circle). Such diagrams just renormalize the prefactor
of the leading, linear-in-$T$ part of $C(T)$. The non-analytic
contributions come from the third (skeleton) term. To second order
in the interaction, this contribution is
\begin{equation}
\delta\Xi =-\frac{1}{2}~ T \sum_{\omega_m} \int
\frac{d^2k}{4\pi^2} ~\Sigma(\omega_m,k,T)  G_0 (k, \omega_m),
\label{o_1}
\end{equation}
where $\Sigma(\omega_m,k,T)$ is (a non-analytic part of) the
second-order self-energy. Converting the Matsubara sum to an
integral over real frequencies and using relation  Eq.~(\ref{comega})
between $\Xi$ and $C(T)$, one obtains
\begin{equation}
\delta C\left( T\right) /T=-\frac{1}{2\pi }~\frac{\partial^2
}{\partial
T^2}\left[~\int \frac{d^{2}k}{(2\pi )^{2}}%
~\int_{-\infty}^{\infty } d\omega ~\textrm{Im}
\left\{G_0^{R}(\omega ,k) \Sigma ^{R}(\omega ,k, T)\right\}~\left(
n_0 (\omega)-\frac{1}{2}\right)\right]. \label{feb5_3_11}
\end{equation}
We emphasize that in this approach
 $\Sigma^R(k, \omega_m, T)$ is evaluated at finite
temperature.

 Generally speaking,  Eq.~(\ref{feb5_3_11}) and
Eqs.~(\ref{ct}-\ref{ct2}) give different results
 for $\delta C$. Indeed, let us assume for a moment
 that $\Sigma^R$ depends only on frequency but not on $k$ and $T$.
Then $\delta C(T)$ from  Eq.~(\ref{ct1}) and from
 Eq.~(\ref{feb5_3_11}) differ by a factor of four.
 The derivation based on the Luttinger-Ward formula is free from
assumptions on what constitutes the main source of the
$T$-dependence in $\delta C(T)$. In fact,  Eq.~(\ref{feb5_3_11}) is
valid for {\it any} temperature albeit for weak interactions.
 The safe way to proceed therefore is to use
 Eq.~(\ref{feb5_3_11}) but not  Eq.~(\ref{feb5_3T0}). It appears, though
that  for our case,
 there exists a deeper relation between the two formulas.
Namely, the two expressions yield identical results for $\delta
C(T)$, provided that one replaces $\Sigma (\omega, k, T=0)$  by
the temperature-dependent self-energy in  Eq.~(\ref{feb5_3T0}),
{\it i.e.}
\begin{equation}
(\ref{feb5_3T0})\to \delta C(T)/T=\frac{2}{\pi }~\frac{\partial
}{\partial T}\left[ \frac{1}{T} ~\int \frac{d^{2}k}{(2\pi )^{2}}~\int_{-\infty }^{\infty }d\omega
\omega \frac{\partial n_0}{\partial \omega }\mathrm
{Im}\left[\Sigma ^{R}(\omega
,k,T)G_{0}^{R}(\omega,k)\right]\right]. \label{feb5_3T}
\end{equation}  This is how
$\delta C(T)$ was calculated in Ref.~\cite{chm}. The overall
factor of four difference between   Eq.~(\ref{feb5_3_11}) and  Eq.~(\ref{feb5_3T0}) is eliminated by two reasons. First, the
imaginary part of the self-energy to order $u^2$ does depend on
$\epsilon_k$ albeit only logarithmically:
$\mathrm{Im}\Sigma^R(\omega,k) \propto \omega^2 \ln|\omega+
\epsilon_k|$. Then  Eq.~(\ref{ct2}) gives the same contribution as
 Eq.~(\ref{ct1}), The details of this calculation are presented in
Appendix ~\ref{app:extra_1}.
 An additional factor  of two appears because the derivative over $T$
 in  Eq.~(\ref{feb5_3T}) now acts not only on $n_0$ but also on $\Sigma^R$.
In
Appendix~\ref{app:extra_1}, we show that these two
 terms contribute equally to $\delta C(T)$; hence, an additional
 factor of two.

To summarize,  Eq.~(\ref{feb5_3_11}) gives a correct result for a
non-analytic term in $C(T)$ to second order in the interaction
without any assumptions or constraints. At the same time,
 Eq.~(\ref{feb5_3T0}) gives the correct result provided that the
self-energy in  Eq.~(\ref{feb5_3T0})  is
 evaluated at finite $T$ rather than
at $T=0$, as specified by  Eq.~(\ref{feb5_3T}). This is a
consequence
 of the $\omega/T$ scaling in the non-analytic part of $\Sigma (\omega, T)$.
 We did not study, however, whether or not this statement is specific to our weak-coupling
 case  or has
a wider range of applicability. Having this precaution in mind,
we will be using  Eq.~(\ref{feb5_3T}) in the following analysis.

It is convenient now to separate the self-energy into the
zero-sound part and the rest, which includes the backscattering-
and PH-contributions from spin- and charge channels, as well as
the remainder term, $\Sigma_{{\rm ex}}$. Such a separation is
convenient because the imaginary part of $\text{Im}\Sigma^R$ has a
substantial $k$-dependence and thus, according to the discussion
in the previous Section, gives a contribution to the specific
heat. On the other hand, the imaginary part of the rest of the
self-energy depends on $k$ only logarithmically, and will be shown
not to contribute to $C(T)$.
\subsubsection{non-zero-sound contribution to $C(T)$}
We begin with the part of the self-energy that contains all
contributions but the zero-sound one:
\[{\tilde \Sigma}\equiv\Sigma-\Sigma_{{\rm ZS}}=\Sigma_B+\Sigma_{{\rm  PH}} + \Sigma_{{\rm ex}}.\] Consider first the
contribution to the specific heat from the real part of the
self-energy, $C_1 (T)$,  Eq.~(\ref{ct1}). A sum of the two
contributions,  \textrm{Re}$\Sigma_{{\rm  PH}}$ +
\textrm{Re}$\Sigma_{{\rm ex}}$, vanishes on
 the mass shell and therefore does not contribute to $C_1 (T)$. A
non-analytic part of \textrm{Re}$\Sigma^{R}_{\rm{B}}$ on the mass
shell and at $T=0$ is given by  Eq.~(\ref{g1}). At finite
temperatures,  \textrm{Re}$\Sigma^{R}_{\rm{B}}$ has been
calculated in ~\cite{chm}; the result of this calculation is that
\textrm{Re}$\Sigma^{R}_{\rm{B}}(k,\omega, T)$ differs from
\textrm{Re}$\Sigma^{R}_{\rm{B}}(k,\omega, T=0 )$ by a
multiplicative factor which is a scaling  function of  $\omega/T$:
\begin{eqnarray}
\textrm{Re}\Sigma^{R}_{\rm{B}}(k,\omega,
T)&=&\textrm{Re}\Sigma^{R}_{\rm{B}}(k,\omega, T=0
)g(\omega/T);\nonumber\\
 g(x)&=&1+\frac{4}{x^{2}}\left[ \frac{\pi
^{2}}{12}+\mbox{Li}_{2}\left( -e^{-|x|}\right) \right],
\label{2.111}
\end{eqnarray}
where $\mbox{Li}_{2}(x)$ is a polylogarithmic function.

Substituting  Eq.~(\ref{2.111}) into  Eq.~(\ref{ct1})
 we obtain~\cite{chm}
\begin{equation}
C_1\left( T\right) /T=-~\frac{9\zeta (3)}{\pi
^{2}}u^{2}%
C_{{\rm  FG}}/E_F,  \label{aa1}
\end{equation}
where $C_{{\rm  FG}}$ is given by  Eq.~(\ref{FG}). As  is
expected, the non-analytic $\omega \left| \omega
\right|$-dependence of \textrm{Re}$\Sigma^R _{\rm{B}}$
 gives rise to a non-analytic contribution to the specific heat $C_{1}\left( T\right) /T\propto u^{2}T$.

 To logarithmic accuracy,
\textrm{Im}${\tilde \Sigma} (\omega,k) \propto \omega^2 \ln
|\omega|$
 does not depend on $k$, hence, according to  Eq.~(\ref{c2zero}), $C_2 (T)$ vanishes.
To demonstrate unambiguously that $C_{2}\left( T\right) $
vanishes, one has to go a bit deeper and analyze
\textrm{Im}$\Sigma^{R}(k,\omega )$ beyond the logarithmic
accuracy, focusing specifically on the momentum dependence under
the logarithm. In Appendix~\ref{app:extra_1}, we show that the
contributions to
 $C_{2}\left( T\right)$ from $\mathrm{Im}\Sigma^R_B$ and \textrm{Im}$\Sigma_{{\rm  PH}}$ + \textrm{Im}$\Sigma_{{\rm  ex}}$
  cancel each other, \emph{i.e.}, there is indeed no
 contribution to the specific heat from \textrm{Im}${\tilde\Sigma^R}$.

\subsubsection{zero-sound mode contribution to $C(T)$}
\label{sec:zstoct}

We now
use the results for \textrm{Re}$\Sigma^R
_{\rm{ZS}%
} $ and \textrm{Im}$\Sigma^{R}_{\rm{ZS}}$ from
Sec.~\ref{sec:sigma} and calculate the contribution to the
specific heat from the zero-sound mode. We show that the
contributions from
 \textrm{Re}$\Sigma^R
_{\rm{ZS}}$ and  \textrm{Im}$\Sigma^R_{\rm{ZS}}$ cancel each
other, {\it i.e.}, that the zero-sound mode does not contribute to
$C(T)$ despite the fact that \textrm{Re}$\Sigma^R _{\rm{ZS}}$ is
enhanced near the mass shell. As we only need to prove the
cancellation, we just use zero-temperature forms of
$\Sigma^R_{{\rm ZS}}$; as we explained in the previous Section,
the difference between the results for the specific heat found
using the zero- or finite-temperature forms of the self-energy, is
just an overall numerical factor.

Substituting \textrm{Re}$\Sigma^R _{\rm{ZS}}$ from  Eq.~(\ref{feb6_8}) into
 Eq.~(\ref{ct1}), we obtain
\begin{equation}
C_{1}(T)/T=\frac{4}{\pi }Nu^{2}~\frac{T}{v_{F}^{2}},  \label{feb7_1}
\end{equation}
where
\begin{equation}
N\equiv \int_{0}^{\infty }\frac{dxx^{3}}{\cosh ^{2}x}=\frac{9}{8}\zeta
\left( 3\right) .
\end{equation}
Substituting next   Eq.~(\ref{d2}) for
\textrm{Im}$\Sigma^{R}_{\rm{ZS}}(k,\omega )$ into
 Eq.~(%
\ref{ct2}), we obtain
\begin{equation}
C_{2}(T)/T=-\frac{1}{\pi }\frac{\partial }{\partial T}\left(
\frac{Z}{T^{2}}%
\right) ,  \label{feb7_2}
\end{equation}
where
\begin{equation}
Z(T)=\frac{1}{2m}~u^{3}{\cal P}\int_{0}^{\infty }\frac{d\omega
\omega
}{%
\cosh ^{2}\frac{\omega }{2T}}\int \frac{d^{2}k}{\omega -\epsilon
_{k}}~\text{\textrm{%
Im}}\left[ \int_{0}^{\epsilon
_{k}/v_{F}}\frac{QdQ}{\sqrt{u^{2}-2(\omega -\epsilon
_{k})/(v_{F}Q)}}\right].  \label{feb7_3}
\end{equation}
Re-scaling the variables and replacing $\int d^{2}k$ by $(m2\pi
)\int d\epsilon_k$, we obtain from  Eq.~(\ref{feb7_3})
\begin{equation}
Z(T)=\frac{u^{3}}{4\pi }~\int_{0}^{\infty }\frac{d\omega \omega
}{\cosh ^{2}\omega /(2T)}J(\omega ),  \label{feb7_4}
\end{equation}
where
\begin{equation}
J(\omega )={\cal P}\int_{-\infty }^{\infty
}\frac{dP}{P}\text{\textrm{Im}}\left[ \int_{0}^{\frac{\omega
}{v_{F}}-\frac{P}{2}}~\frac{Q^{3/2}dQ}{\sqrt{u^{2}Q-P}%
}\right] .~  \label{feb7_5}
\end{equation}
The integration region over momenta $P$ and $Q$ are defined by the
following conditions: $P>u^{2}Q$ and $\omega >v_{F}P/2$.
Evaluating the integrals in
(%
\ref{feb7_5}) over this region, we find
\begin{equation}
J(\omega )=\frac{\pi }{2u}~\left( \frac{\omega }{v_{F}}\right) ^{2}.
\label{feb7_6}
\end{equation}
Substituting this result into  Eq.~(\ref{feb7_4}), and then into
(\ref{feb7_2}), we  obtain
\begin{equation}
C_{2}(T)/T=-\frac{u^{2}}{8\pi v_{F}^{2}}\frac{\partial }{\partial
T}%
\left[ \frac{1}{T^{2}}\int_{0}^{\infty }\frac{\omega ^{3}d\omega }{\cosh
^{2}\omega /(2T)}\right] =-\frac{4}{\pi }~u^{2}N~\frac{T}{v_{F}^{2}}~.
\label{feb7_7}
\end{equation}
Comparing Eqs.~(\ref{feb7_1}) and  (\ref{feb7_7}), we find that
these two contributions to the specific heat cancel each other,
\emph{i.e.}, \textit{there is no non-analytic contribution to the
specific heat from the zero-sound collective mode to second-order
in the interaction},
despite the non-perturbative enhancement of $\Sigma _{\rm{ZS}%
}^{R}$ near the mass shell. An absence of the
 zero-sound contribution to the specific heat is
 another central result of the
paper.

The final result for a weak, contact interaction is then given
just by the perturbative contribution,  Eq.~(\ref{aa1}):
\begin{equation}
C_1\left( T\right) /T=-~\frac{9\zeta (3)}{\pi ^{2}}u^{2} C_{{\rm
FG}}/E_F.  \label{aa1_1}
\end{equation}
\subsubsection{finite-range potential}
For a finite-range potential, the forward-scattering part of the
self-energy involves only $u_{0}.$ Hence, all cancellations
discussed in the preceding Sections are still in place. In
particular, the sum of \textrm{Re}$\Sigma
_{%
\text{ex}}^{R}$ and \textrm{Re}$\Sigma _{\rm{PH}}^{R}$ still vanishes on
the mass shell and the contributions from real and imaginary parts of $%
\Sigma _{\rm{ZS}}^{R}$ to $C\left( T\right) $ still cancel each other. In
addition, the momentum integral  in  Eq.~(\ref{ct2}) of each of the three
terms $\mathrm{Im}\Sigma^{R} _{\rm{B}},$ \textrm{Im}$\Sigma^{R}
_{\rm{PH}},$ and
\textrm{Im}$\Sigma^{R}_{\text{ex}}$ still vanishes. As a result, the
specific
heat is again determined by the real part of the self-energy from
backscattering. The self-energy due to backscattering for a generic $U(Q)$
is given by  Eq.~(\ref{la-la}), hence $\delta C\left( T\right) $ becomes
\begin{equation}
\delta C\left( T\right) /T=-~\frac{9\zeta (3)}{\pi ^{2}}
\left(u_{0}^{2}+u_{2k_{F}}^{2}-u_{0}u_{2k_{F}}\right)C_{{\rm
FG}}/E_F. \label{genu}
\end{equation}
This expression is the final result for the non-analytic correction to the
specific heat to order $u^2$.
  For $u_{0}=u_{2k_{F}}=u$, it reduces to the
contact-potential result,  Eq.~(\ref{aa1}).

\subsection{Specific heat via the thermodynamic potential}
\label{sec:sh_omega_m}

The calculation of the specific heat via the self-energy presented
in the previous Section is quite involved, as it requires a
detailed knowledge of $\Sigma \left( \omega, k\right) $. To verify
the main result of the previous Section--that there is no
contribution to the specific heat from the collective mode--and to
understand it from a different perspective, we employ an alternate
approach. In particular, we obtain the specific heat by
finding the thermodynamic potential, $\Xi$, directly, and then
using relation  Eq.~(\ref{comega}). In this approach, fermions are
integrated out from the very beginning, and the intricate details
of their self-energy are not important. Another advantage of
working with the thermodynamic potential is that the entire
calculation can be performed in Matsubara frequencies.
\begin{figure}[tbp]
\begin{center}
\epsfxsize=0.7\columnwidth \epsffile{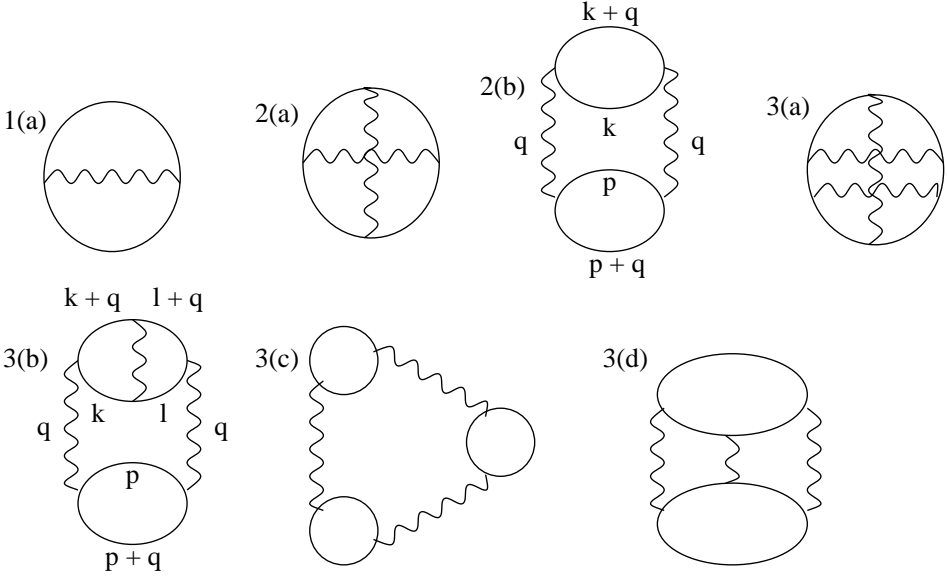}
\end{center}
\caption{Diagrams for the thermodynamic potential containing
maximum number of particle-hole bubbles. For the Coulomb potential,
diagrams
1(a), 2(b) and 3(c) represent  ring diagram series to third order in
the  interaction}
\label{fig:omega}
\end{figure}

\subsubsection{Luttinger-Ward expansion}

To generate a perturbative expansion of $\Xi $, we follow the
Luttinger-Ward approach \cite{luttinger,wasserman}, in which $\Xi$
is expressed in terms of the exact Green's functions and the skeleton
self-energies, as specified in  Eq.~(\ref{o1}). To begin with, we
consider the contact-interaction case. The diagrams for the
self-energy describing the interaction of fermions with collective
modes have been discussed in
Sec.~\ref{sec:sigma}%
. Expanding the result for the self-energy,  Eq.~(\ref{b5}), back in powers
of
interaction $U$ \ and substituting the resulting series for $\Sigma $ and
$%
\Sigma _{\nu }$ into  Eq.~(\ref{o1}), we generate the series for
the thermodynamic potential. The non-trivial  diagrams for $\Xi $ up to
third order in $U$ are shown in Fig.~\ref{fig:omega}. Explicitly,
\begin{eqnarray}
\Xi &=& \Xi _{0}+\int_{q}\left[ -U\Pi _{m}-\frac{1}{2}\left( U\Pi
_{m}\right) ^{2}+\frac{1}{3}\left( U\Pi _{m}\right)
^{3}-\frac{1}{2}\left( U\Pi _{m}\right) ^{4}+\frac{1}{5}\left(
U\Pi _{m}\right) ^{5}+\dots \right]\nonumber\\
 & =& \int_q \left[-2
U\Pi _{m} + \frac{1}{2} \left( U\Pi _{m}\right)^2 + \frac{1}{2}
\ln (1 - U\Pi _{m}) + \frac{3}{2} \ln (1 +  U\Pi _{m})\right],
\label{e1}
\end{eqnarray}
where $\Pi _{m}(q)$ is the polarization bubble in Matsubara
frequencies. Again,
we will need only the asymptotic forms of $\Pi _{m}(q)$ for $Q$ near $0$
and
near $2k_{F}.$ Analytically continuing the corresponding retarded
expressions [Eqs.~(\ref{feb6_1}) and  (\ref{apr20_1})] ] to Matsubara
frequencies $\Omega_m=2\pi mT$, we obtain
\begin{equation}
\Pi _{m}(\Omega _{m},Q)=-\frac{m}{2\pi }\left[ 1-\frac{|\Omega
_{m}|}{\sqrt{%
\Omega _{m}^{2}+(v_{F}Q)^{2}}}\right] ,\text{ }Q\rightarrow
0;  \label{pi20}
\end{equation}
and
\begin{equation}
\Pi _{m}(\Omega _{m},Q)=-\frac{m}{2\pi }~\left[ 1-\left( \frac{Q-2k_{F}}{%
2k_{F}}+\sqrt{\left( \frac{Q-2k_{F}}{2k_{F}}\right) ^{2}+\left(
\frac{\Omega
_{m}}{2k_{F}v_{F}}\right) ^{2}}\right) ^{1/2}\right] ,Q\approx 2k_{F}.
\label{apr20_3}
\end{equation}
One observes immediately that the series in  Eq.~(\ref{e1})
converges for small $U$ because $\Pi _{m}(\Omega _{m},Q)$ is
regular for any $\Omega _{m}$ and $Q$ [in contrast with $\Pi
^{R}(\Omega ,Q)$ which is singular at the boundary of the
particle-hole continuum $\Omega ^{2}=(v_{F}Q)^{2}$]. Every order
in
$%
U $ then gives a finite contribution to the thermodynamic
potential which can be calculated separately from other orders.
Therefore, there is no need for re-summation of the perturbation
theory for $\Xi $ for a weak interaction. This tells us that the
$\mathcal{O}(U^{2})$ term in the specific heat cannot have any
non-perturbative contributions, \emph{i.e.}, to second order in
$U$, $\Xi$ is given just by
\begin{equation}
\Xi-\Xi_0 =\Xi _{2}=-\frac{1}{2}\int_{q}\left( U\Pi _{m}\right)
^{2}. \label{apr20_2}
\end{equation}
This explains the  absence of  collective-mode contribution to
$C(T)$, which we have demonstrated explicitly in Sec.~\ref{sec:zstoct}.
\subsubsection{Evaluation of the thermodynamic potential in Matsubara
 frequencies}
 \label{sec:matsubara}

Next, we show how the non-analytic $T^{2}$ correction to the
specific heat emerges in the Matsubara formalism. The $T^{2}$-term
in $C(T)$ comes from a non-analytic, $T^{3}$-piece in $\Xi $, and
we will be searching for this term in  Eq.~(\ref{apr20_2}). We
first show that the $Q-$integral of $\Pi _{m}^{2}$ taken over
momenta near $Q=0$ and $Q=2k_{F}$ contains a non-analytic, $\Omega
_{m}^{2}\ln |\Omega _{m}|$ part. Indeed, squaring the small-$Q$
form of $\Pi _{m}$ [ Eq.~(\ref{pi20})] and substituting the result
into  Eq.~(\ref{apr20_2}), we find
\begin{equation}
\int \frac{dQQ}{2\pi }\Pi _{m}^{2}(\Omega _{m},Q)\rightarrow
\frac{m^{2}}{%
(2\pi )^{3}}~\Omega _{m}^{2}~\int_{0}^{\simeq k_{F}}\frac{dQQ}{\Omega
_{m}^{2}+(v_{F}Q)^{2}}=\frac{m^{2}}{(2\pi )^{3}v_{F}^{2}}~\Omega
_{m}^{2}\ln
\frac{E_{F}}{|\Omega _{m}|}.  \label{apr20_5}
\end{equation}
Similarly, the square of the second term in the bubble near
$2k_{F}$ [Eq.~(\ref {apr20_3})], yields another $\Omega _{m}^{2}\ln
|\Omega _{m}|$-term with the same prefactor, as from the region of
small $Q$. To see this, we substitute the square of
 Eq.~(\ref{apr20_3})
into  Eq.~(\ref{apr20_2}), re-define the  integration variable as $%
x=(Q-2k_{F})/(2k_{F}),$ and retain only the square of the second
term in $\Pi _{m}^{2}$:
\begin{equation}
\int_{Q\approx 2k_{F}}\frac{QdQ}{2\pi }~\Pi _{m}^{2}(\Omega
_{m},Q)\Longrightarrow \frac{m^{2}}{(2\pi )^{3}}(2k_{F})^{2}\int_{-1}^{1}dx%
\left[ x+\sqrt{x^{2}+\left( \frac{\Omega _{m}}{2k_{F}v_{F}}\right) ^{2}}%
\right] .
\end{equation}
The precise limits of the integration over $x$ are not
important. Expanding
in frequency, we obtain
\begin{eqnarray}
\int_{Q\approx 2k_{F}}\frac{dQQ}{2\pi }~\Pi _{m}^{2}(\Omega _{m},Q)
&\Longrightarrow &\frac{m^{2}}{(2\pi
)^{3}}(2k_{F})^{2}\int_{-1}^{1}dx\left(
x+|x|+\frac{\Omega _{m}^{2}}{4k_{F}v_{F}^{2}|x|}+...\right) \\
&\Longrightarrow &\frac{m^{2}}{(2\pi )^{3}v_{F}^{2}}~\Omega _{m}^{2}\ln
\frac{E_{F}}{|\Omega _{m}|}.  \label{apr20_4}
\end{eqnarray}
Comparing  Eq.~(\ref{apr20_4}) and  Eq.~(\ref{apr20_5}), we see that
the non-analytic contributions from $Q=0$ and $2k_{F}$ are equal.
Substituting the sum of the two contributions into
 Eq.~(\ref{apr20_2}), we obtain
\begin{equation}
\mathrm{\Xi }_{2}=-\frac{1}{2}U^{2}\int_{q}\Pi _{m}^{2}=-\frac{u^{2}T}{\pi
v_{F^{{}}}^{2}}\sum_{\Omega _{m}=0}^{E_{F}}\Omega _{m}^{2}\ln
\frac{E_{F}}{%
\Omega _{m}}=-\frac{4\pi u^{2}T^{3}}{v_{F}^{2}}S\left( M\right) ,
\label{e2}
\end{equation}
where
\begin{equation}
S\left( M\right) \equiv \sum_{m=0}^{M}m^{2}\ln \frac{M}{m}=\frac{1}{6}%
M\left( M+1\right) \left( 2M+1\right) \ln M-\sum_{m=1}^{M-1}\left(
m+1\right) ^{2}\ln \left( m+1\right)  \label{e22}
\end{equation}
and $M=\left[ E_F/2\pi T\right] \gg 1.$ The choice of $E_F$ as an
upper limit in  Eq.~(\ref{e2}) is completely arbitrary; since
are looking for a universal contribution to the thermodynamic
potential, the choice of  cutoff is not important. Next, we use
the Euler-Maclaurin formula :
\begin{equation}
\sum_{m=1}^{M-1}f(m)=\int_{0}^{M}f(x)dx-\frac{1}{2}\left[ f(M)+f(0)\right]
+\sum_{p=1}^{N}B_{2p}\frac{f^{(2p-1)}(M)-f^{(2p-1)}(0)}{(2p)!},
\label{apr20_14}
\end{equation}
 where $B_{k}$ are the Bernoulli coefficients and $f^{(n)}$ is the $n$-th
derivative of $f$. Applying this formula to  Eq.~(\ref{e22}), we see
that the derivatives $f^{(2k-1)}(M)$ for $k\geq 2$ form a series
in $1/M$ for large $M$, whereas $f^{(1)}\left( M\right) $ and
$%
f^{(2k-1)}(0)$ give $M-$independent contributions for $k\geq 1$.
Combining the result of the Euler-Maclaurin expansion with the
first term in  Eq.~(\ref {e22}) and taking the limit of
 $M\rightarrow \infty ,$ we arrive at
\begin{equation}
\sum_{m=0}^{M}m^{2}\ln \frac{M}{m}=\frac{1}{9}M^{3}-\frac{1}{12}M-\alpha
+%
\frac{1}{360M}+\dots  \label{e3}
\end{equation}
The $M$-independent term, $\alpha$, is represented by the
following series
\begin{eqnarray}
\alpha  &=&\frac{1}{9}-\sum_{p=1}^{\infty }\frac{1}{\left( 2p\right) !}%
~A_{2p}\frac{d^{2p-1}}{dx^{2p-1}}[x^{2}\ln
x ]|_{x=1}  \notag \\
&=&\frac{1}{9}-\frac{1}{12}+\frac{1}{360}-\frac{1}{7560}+\cdots
=0.0304\dots \label{e33}
\end{eqnarray}
Although we have not been able to prove this analytically, we
observe that, to very high accuracy,
\begin{equation}
\alpha=\frac{%
\zeta (3)}{4\pi ^{2}}. \label{alpha}\end{equation} The same
constant is obtained when calculating the specific heat in real
frequencies, when the Matsubara sums are converted into ingegrals
(see Appendix \ref{sec:omega_re_freq}). We will thus treat
relation (\ref{alpha}) as an exact one.

 Terms of order $M^{3}$, $M$, $1/M$, etc. in
$S\left( M\right) $ generate regular--$T^{0}$, $T^{2}$, $T^{4}$,
etc.--corrections to $\Xi _{2}$, whereas a constant term ($-\alpha
$) gives a universal, non-analytic
 $T^{3}-$ contribution
to $\Xi _{2}$. This contribution is precisely what we need.
Retaining only this term in  Eq.~(\ref{e2}) and substituting the
result into  Eq.~(\ref{comega}), we obtain the same correction to
the specific heat as the one found by expressing the specific heat
via the self-energy,  Eq.~(\ref{aa1_1}).

For a finite-range interaction, a slight modification of the
analysis presented in this Section leads to the result identical
to that in  Eq.~(\ref{genu}).

 For  the sake of completeness, in Appendix~\ref{sec:omega_re_freq}
 we evaluate
the specific heat by computing the thermodynamic potential in real
frequencies and show explicitly how the contribution from the
zero-sound is cancelled out.

\subsection{specific heat in a generic Fermi liquid}
\label{sec:generic}

Now we are in a position to discuss a more general question--what
happens to the $T^2$ term in the specific heat if the interaction is
not weak. First, we discuss a model case of contact interaction of
arbitrary strength and then move on to the case of a generic Fermi
liquid.

\subsubsection{contact interaction}
To second order in contact interaction $u$, relevant diagrams for
the thermodynamic potential contain the square of the polarization
bubble [cf. Fig.~\ref{fig:omega}, 2(a) and 2(b)]. Since $Q=0$ and
$Q=2k_F$ contributions to the thermodynamic potential are
identical for this case, we evaluate the $Q=0$-contribution first,
and then just double the result at the end. In
Sec.~\ref{sec:matsubara}, we have shown that the non-analytic,
$T^3$ contribution to the thermodynamic potential comes from the
square of the dynamic part of $\Pi_m$. According to  Eq.~(\ref{e1}),
higher orders in $u$ generate higher powers of $\Pi_m$. Another
effect of higher orders is that self-energy insertions
 result in replacing the bare Green's functions by exact one. As
 the main contribution to the $T^3$-term in $\Xi$ comes from the
 states near the Fermi surface, one can approximate exact $G$ by
 its expression near the pole
 \begin{equation}
 G(\omega_m,k)=\frac{Z}{i\omega_m-\epsilon^*_k},
 \end{equation}
 where $Z$ is the renormalization factor,
 $\epsilon^*_k=v^*_F(k-k_F)$, $v^*_F=k_F/m^*$, and $m^*$ is the
 renormalized mass. Parameters $Z$ and $m^*$ are some functions of
 the bare interaction $u$, whose forms, in general, are not known.
 This amounts to replacing the prefactor and the Fermi velocity in
  Eq.~(\ref{pi20})
 \begin{equation}
 \Pi_m(\Omega_m,Q)\to
 \Pi_m^*(\Omega,Q)=-\frac{m^*Z^2}{2\pi}\left[1-
\frac{|\Omega_m|}{\sqrt{\Omega_m^2+(v^*_FQ)^2}}\right].
 \end{equation}
 A term of order $n$ in series  Eq.~(\ref{e1}) contains $(\Pi_m^*)^n$
and  thus generates a binomial expansion in powers of ${\cal
D}^*=|\Omega_m|/\sqrt{\Omega^2+(v^*_FQ)^2}$
\begin{equation}
(\Pi_m^*)^n=(-)^n(Z^2m^*/2\pi)^n\sum_{l=0}^n (-)^lC^l_n{\cal
D^*}^l \label{binom},
\end{equation}
where $C^l_n$ is the binomial coefficient. It is easy to make sure
that only the term with $l=2$ in  Eq.~(\ref{binom}) yields
$\Omega_m^2{\ln}|\Omega_m|$ upon the momentum integration, whereas
all other terms yield just $\Omega_m^2$. As  was shown in
Sec.~\ref{sec:matsubara}, the frequency sum
\begin{equation}
T\sum_{\Omega _{m}=0}^{E_{F}}\Omega _{m}^{2} {\ln}|\Omega_m|
\end{equation}
gives a universal $T^3$ contribution to the thermodynamic
potential, responsible for the $T^2$ term in the specific heat (see
(\ref{e2})). At the same time, the frequency sum
\begin{equation}
T\sum_{\Omega _{m}=0}^{E_{F}}\Omega _{m}^{2}
\end{equation}
contributes only analytic--$T^2, T^4$, etc.--terms to $\Xi $, but
no $T^{3}-$ term. The problem therefore reduces to collecting the
combinatorial coefficients of ${\cal D^*}^2$ terms
 at each order and re-summing the perturbation series.

Expanding each term  in  Eq.~(\ref{e1}) to order ${\cal D^*}^2$ and
 using
\begin{equation}
\sum_{k=1}^\infty k x^{k-1}   = \left(\frac{1}{1-x}\right)^2,
\end{equation}
 we find that the $T^2$- term in the specific heat is given by
\begin{equation}
\delta C\left( T\right)/T =-\frac{9\zeta (3)}{\pi ^{2}} u_{{\rm
eff}}^{2}C^*_{\rm{FG}}/E_F^*. \label{e4_4}
\end{equation}
Here $C^*_{\rm FG}=\pi m^*T/3$, $E_F^*=k_Fv_F^*/2$, and
\begin{equation}
u_{{\rm
eff}}^{2}=(u^*)^2\left[1+2u^*+6(u^*)^{2}+4(u^*)^{3}+\dots\right]=(u^*)^{2}\left[
\frac{3}{2} \left(\frac{1}{1 -u^*}\right)^2 + \frac{1}{2}
\left(\frac{1}{1 +u^*}\right)^2 -1 \right], \label{ueff}
\end{equation}
where
\begin{equation}
u^*\equiv Z^2m^*U/2\pi. \label{ustar}\end{equation} This result is
valid for $0\leq u^* < 1$.  The divergence of $u_{{\rm eff}}^{2}$
at $u^*=1$, resulting from the spin-channel, signals an
instability towards a magnetically-ordered state. A plot of
$u_{{\rm eff}}^{2}\left( u^*\right) $ is presented in
Fig.~\ref{fig:ueff}.
\begin{figure}[tbp]
\begin{center}
\epsfxsize=0.4\columnwidth \epsffile{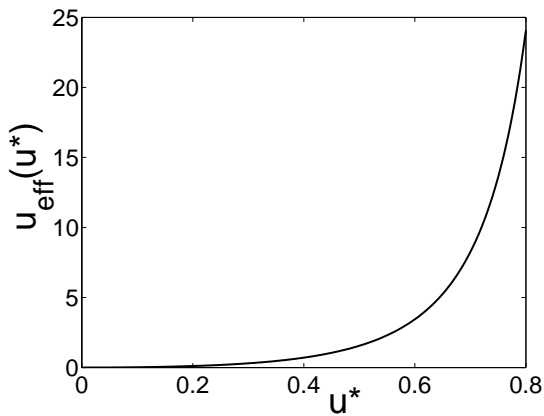}
\end{center}
\caption{Effective coupling for a $T^2$-term in the specific heat
$u_{{\rm eff}}^{2}$ [ Eq.~(\ref {ueff})] as a function of the
renormalized interaction $u^*$, Eq.~(\ref{ustar}).}
\label{fig:ueff}
\end{figure}

\subsubsection{generic interaction}
The result for a contact interaction,  Eq.~(\ref{e4_4}),
 is of a rather limited use,
as, in general, the Fourier transform of the interaction does
depend on the momentum transfer. To
 obtain a form of $\delta C(T)$, valid for a  generic Fermi liquid, we first
go back to the second-order diagrams for the thermodynamic
potential, and identify the structure of the vertices contributing
to the non-analytic part of $C(T)$.

We consider first a small-$Q$ contribution to the diagram 2(b) in
Fig.~\ref{fig:omega}.
 It is proportional to $U^2 (0)~ T \sum_\Omega \int d^2 Q \Pi^2 (\Omega_m,Q)$,
 where the integration is restricted to small $Q$. Each of the two polarization bubbles
 is obtained by the integration over internal fermion momenta,
 ${\bf k}$ in the upper bubble and ${\bf p}$ in the lower one.
 At first glance, ${\bf k}$ and ${\bf p}$
are completely uncorrelated,
 as the integrations over ${\bf k}$ and ${\bf p}$ are independent of each
other. This would imply that the total momenta for the two
vertices in diagram 2(b) are arbitrary.
 In general, this is indeed true. However, a
 non-analytic, $T^3$-term in the thermodynamic potential arises only
 from a product of non-analytic, $|
 \Omega_m|/Q$ parts of the two polarization bubbles, integrated over $Q$.
 In Appendix~\ref{app:extra}, it is shown that the $|\Omega_m|/Q$-term
  in $\Pi_m(\Omega_m,Q)$ comes from an integral $\int d^2k G\left(\omega_m+\Omega_m,{\bf k}+{\bf
  Q}\right)G(\omega_m,k)$ from only over those  regions of $k$ where
${\bf k}$ is nearly
 orthogonal to ${\bf Q}$ (and the same for ${\bf p}$ in the other bubble).
 Since both ${\bf k}$ and ${\bf p}$ are almost orthogonal to ${\bf Q}$,
 they are either nearly parallel or nearly antiparallel to each
 other.
 In Appendix~\ref{app:extra}, it is also shown that the
contribution from the nearly parallel ${\bf k}$ and ${\bf p}$
vanishes, {\it i.e.}, the non-analytic , ``$Q=0$'' contribution to
$\Xi$ involves only a vertex with a small momentum transfer $Q$
{\it and} small total momentum ${\bf k} + {\bf p}$ (
``backscattering" vertex). The same consideration holds for the
$U^2 (2k_F)$-term from  diagram 2(b) in Fig.~\ref{fig:omega} and for
 the $U(0) U(2k_F)$-term from diagram 2(a) in Fig.~\ref{fig:omega}. In both cases, the
 $T^3$-term in $\Xi$
  comes only from those momentum range, where
 ${\bf k}$ and ${\bf p}$ are nearly antiparallel.
  We thus see that the $T^2$-term in $C(T)$
involves only vertices of the type
 $({\bf k},-{\bf k};{\bf k},{\bf -k})$ and
$({\bf k},-{\bf k};-{\bf k},{\bf k})$.

Consider now what happens when we add higher-order terms in the
interaction. They lead to two types of corrections: self-energy
insertions into the fermion lines in the two bubbles and
corrections to the vertices. The self-energy corrections are of
the Fermi-liquid type: they account for the appearance of the
quasi-particle $Z$-factors, and for the replacement of the bare
fermion mass by the effective one. Vertex corrections lead to a
variety of diagrams. A typical
 $n$-th order diagram contains $n$ bubbles. [We remind that these bubbles are either explicit, as in diagrams 2(b) and 3(c),
  or are obtained after integrating  over the fermion variables, as in the rest of the diagrams. Some diagrams, \emph{e.g.},
  3(b),
  contains bubbles both explicitly and implicitly.] To obtain a $T^2$
contribution to $C(T)$, one needs to take the dynamic,
$|\Omega_m|/Q$-parts of the two out of $n$ bubbles, and set
$\Omega =0, Q \rightarrow 0$ in the remaining $n-2$ ones, as any
extra factor
  of $\Omega/Q$, as well as any extra factors of $Q$
 eliminates the logarithmic divergence of the momentum integral.
  The dynamic terms can come either
 from the two bubbles,  already present in the skeleton second-order diagrams, or from the
   bubbles associated with the vertex corrections.  It
is intuitively plausible that, once the two dynamic bubbles are
chosen at the $n-$th order, while the rest of diagram is evaluated
at $\Omega_m =0$,  thus  constituting  the
 $n-$th order correction to the \emph{static} vertex. In other words, it is plausible
 that  the non-analytic,  $T^3$-term in the thermodynamic potential
 can be expressed as the two second-order \emph{skeleton} diagrams, in which the wavy lines are replaced
 by exact static vertices, $\Gamma^k ({\bf k}, -{\bf k}; {\bf k}, -{\bf k})$ and $\Gamma^k ({\bf k},
 -{\bf k};-{\bf k},{\bf k})$, and bare fermion lines are replaced by solid ones.

This conjecture, however, needs to be verified, as  diagrams
 for the thermodynamic potential contain combinatoric factors.
 {\it A priori}, it is not clear whether these factors, combined with those associated with the selection
  of the two dynamic bubbles, would give just the right combinatoric factors to form the perturbative series
   for the static vertices. In order to verify this conjecture, we
   explicitly evaluated
  the $T^3$-term in the thermodynamic potential to the third order in
 $U(Q)$, and compared the result with that obtained by inserting renormalized static vertices into
 two-bubble skeleton diagrams.
This derivation is  presented in Appendix~\ref{app:extra}. We
find the two expressions, obtained directly and via skeleton
diagrams, are identical. We did not attempt to prove that this
equivalence
 holds to all orders in $U(Q)$, but the agreement between the two third-order
 results is a promising sign. In what follows, we assume that this agreement
  holds to all orders in $U(Q)$, i.e., the non-analytic part of the thermodynamic potential is
 a product of  two (renormalized) dynamic bubbles and two exact static
 vertices,
 $\Gamma^k ({\bf k}, -{\bf k}; {\bf k}, -{\bf k})$ and $\Gamma^k ({\bf k}, -{\bf k};-{\bf k},{\bf k})$.
To first order in $U$, these vertices reduce just to $U(0)$ and
$U(2k_F)$, respectively.

Vertices $\Gamma^k ({\bf k}, -{\bf k}; {\bf k}, -{\bf k})$ and
$\Gamma^k ({\bf k}, -{\bf k};-{\bf k},{\bf k})$ are exact in a
sense that they
 include {\it all} static corrections, coming from
 the states both away from \emph{and} near the Fermi surface. The latter produce powers of
 the static bubble, $\Pi_m (\Omega_m =0,Q\rightarrow 0) = -Z^2m^*/2\pi$, which, we remind,
 comes from the states in a narrow range near the Fermi surface~\cite{statphys}.
In other words, the vertices include
 all corrections, except those coming from  the dynamic part of the
 polarization bubble. In conventional notations~\cite{agd,statphys},
 vertices $\Gamma ({\bf k}, -{\bf k}; {\bf k}, -{\bf k})$ and $\Gamma ({\bf k}, -{\bf k};-{\bf k},{\bf k})$ are related to $\Gamma^k (\theta)$,
 defined by  Eq.~(\ref{gammak}). As the incoming momenta are nearly
 anti-parallel to each other, angle $\theta$ in  Eq.~(\ref{gammak})
 can be put equal to $\pi$.

Vertex $\Gamma^k (\pi)$, as a tensor in the spin space, can be
represented as
\begin{equation}
\Gamma^k_{\alpha \beta ,\gamma \delta }(\pi )= \Gamma^k ({\bf k},
-{\bf k};{\bf k},-{\bf k}) \delta _{\alpha \gamma }\delta _{\beta
\delta } - \Gamma^k ({\bf k},-{\bf k};-{\bf k},{\bf k}) \delta
_{\alpha \delta }\delta _{\beta \gamma }. \label{jul3_2}
\end{equation}
Quasi-particle $Z$-factors, resulting  from the self-energy
insertions into the fermion lines of the  bubbles, can be
incorporated into a relation between $\Gamma^k$ and
 the quasi-particle scattering amplitude~\cite{statphys}
\begin{equation}
f_{\alpha \beta ,\gamma \delta }(\pi)=Z^{2}\Gamma _{\alpha \beta
,\gamma \delta }^{k}(\pi ). \label{ya4}
\end{equation}
Next, representing the scattering amplitude in terms of its
charge- and spin components
\begin{eqnarray} f_{\alpha\gamma,\beta\delta} ({\bf k}, -{\bf k})
&=& f_{\alpha\gamma,\beta\delta} (\pi) =\frac{\pi }{m^*}
\left[f_c(\pi) \delta_{\alpha \gamma} \delta_{\beta \delta} + f_s
(\pi) {\bf \sigma}_{\alpha \gamma}
{\bf \sigma}_{\beta \delta}\right] \nonumber \\
&&= \frac{\pi}{m^*} \left[\left\{f_c (\pi) - f_s (\pi)\right\}
\delta_{\alpha \gamma} \delta_{\beta \delta} +2 f_s (\pi)
\delta_{\alpha \delta} \delta_{\beta \gamma}\right],
\label{jul3_1}
\end{eqnarray}
and comparing Eq.~(\ref{jul3_1}) with  Eq.~(\ref{jul3_2}), we obtain
\begin{equation}
Z^{2} \Gamma^k(k,-k; k,-k) =\frac{\pi }{m^*}\left[ f_c (\pi )-f_s
(\pi )\right] ,~~Z^{2} \Gamma^k(k,-k;-k,k) =-2\frac{\pi}{m^*}f_s
(\pi ) . \label{jul3_3}
\end{equation}
Substituting  $Z^2 \Gamma^k ({\bf k},-{\bf k},{\bf k},-{\bf k})$
instead of $U (0)$ and $Z^2 \Gamma ({\bf k},-{\bf k};-{\bf k},{\bf
k})$ instead of $U(2k_F)$ into
 Eq.~(\ref{genu}), we obtain the final form of the non-analytic part
 of the specific heat
 in a generic Fermi liquid
\begin{equation}
\delta C(T)/T =-\frac{3\zeta (3)}{2\pi\left(v^*_F\right)^2}
~\left[ f^2_c (\pi )+3f^2_s (\pi )\right]T . \label{jul3_4}
\end{equation}
On the other hand, $\delta C(T)$ does not have a simple closed
form in terms of the Landau interaction function, $F \left( \theta
\right)$. Indeed, the Landau function is related to vertex
$\Gamma^\omega$, defined in  Eq.~(\ref{gammaomega}), rather than to
$\Gamma^k$:
\begin{eqnarray}
F_{\alpha \beta ,\gamma \delta }(\theta)=Z^{2}\Gamma _{\alpha
\beta ,\gamma \delta }^{\omega}(\theta )=\frac{\pi}{m^*}
\left[\left\{\gamma_c (\pi) - \gamma_s (\pi)\right\}
\delta_{\alpha \gamma} \delta_{\beta \delta} +2 \gamma_s (\pi)
\delta_{\alpha \delta} \delta_{\beta \gamma}\right]. \label{ya44}
\end{eqnarray}
A perturbative expansion for $\Gamma^\omega (\pi)$ includes static
vertex corrections from the states away from the Fermi surface but
not in its  vicinity, \emph{i.e.}, it neglects the  vertex
corrections associated with $\Pi (\Omega =0, Q \rightarrow 0)$.
Whereas there is a simple relation between the partial components
of functions $\Gamma^k (\theta)$ and $\Gamma^\omega(\theta)$
\cite{statphys}, in order to relate these two functions at a given
angle, \emph{e.g.},  $\theta =\pi $, one has to invoke an infinite
number of partial components. In this respect, the universal
sub-leading term in the specific heat is different from the
leading, Fermi-liquid term, $C(T)/T=$const, which is  expressed in
terms of $\langle F_c(\theta)\cos\theta\rangle$, where $F_c$ is
the charge component of the Landau function.

To emphasize an absence of a simple relation between
$\Gamma^{k}(\theta)$ and $\Gamma^{\omega}(\theta)$, we present the
relation between the charge- and spin components of scattering
amplitude, $f_a (\pi)$, and those of the vertex,
 $\gamma_a (\pi)$ (a = c,s), to
third order in $U(Q)$. Using results
 from Appendix~\ref{app:extra}, we obtain
\begin{eqnarray}
f_c (\pi) &=& \gamma_c (\pi) - \left( \gamma_c (\pi) -
 \gamma_s (\pi)\right) \left( \gamma_c (\pi)-  \gamma_s (\pi)
 + 2 \langle \gamma_s (\theta)\rangle \right) -  \langle \gamma_s (\theta)  \gamma_s
 (\pi -\theta)\rangle; \nonumber \\
f_s (\pi) &=& \gamma_s (\pi) -  \langle \gamma_s (\theta) \gamma_s
(\pi -\theta)\rangle, \label{10_9_1}
\end{eqnarray}
where $\langle ...\rangle$ stands for averaging over $\theta$. To
the same accuracy, the terms of order
 $(\Gamma^\omega)^2$ can be expressed via the bare interaction potential $U(Q = 2 k_F \sin{ \theta/2})$ using
 the first-order relations:  $\Gamma^\omega_c (\theta) = U(0) -
 U(\theta)/2$ and
$\Gamma^\omega_s (\theta) = -(1/2) U(\theta)$. The relations
between $\Gamma^\omega$ and $U(Q)$ to second order are presented
in Appendix~\ref{app:extra}.

The rather complicated relations between $\Gamma^k_a (\pi)$ and
$\Gamma^\omega_a (\pi)$
 are simplified if  bare interaction $U(Q)$ is
 strongly peaked at $Q=0$, so that $\Gamma_c^\omega (\pi)$ is much larger than
$\Gamma_s^\omega (\theta)$  for a generic $\theta$, including
$\theta = \pi$. In this limit, perturbative series for the
relation between $\Gamma_c^k$ and $\Gamma_c^\omega$ can be summed
up exactly, and $\Gamma_c^k (\pi)$ can be expressed in terms of
$\Gamma_c^\omega (\pi)$ as
\begin{equation}
f_c (\pi) = \frac{\gamma_c (\pi)}{1 +  \gamma_c (\pi)}
\label{10_9_2}
\end{equation}
In addition, if the condition $\gamma_s (\theta)$ is met then,, the contribution
to the specific heat from $\gamma_c(\pi)$ dominates, and the
singular term in the specific heat becomes
\begin{equation}
\delta C\left( T\right) /T=- ~\frac{3m \zeta (3)}{4\pi}~\left(
\frac{\gamma_c (\pi)}{1 + \gamma_c (\pi)}\right)^2
\frac{T}{E_F}.\label{genu_111}
\end{equation}

\subsubsection{Non-analytic term in the specific heat for the Coulomb interaction}
\label{sec:coulomb}

The limiting case of  $\gamma_c (\pi) \rightarrow \infty$ in
 Eq.~(\ref{genu_111}) corresponds to the Coulomb interaction in the
weak coupling limit,
 when the dimension-less gas parameter,
 $r_s\equiv\sqrt{2}me^2/k_F$,
 is small. [We remind that  Eq.~(\ref{genu_111}) is valid only in the
 limit $\gamma_s(\theta)\ll 1$.]
[Recall that $\Gamma_c^\omega (\theta)$ is expressed via $U(0)$
and $U(\theta)$,  whereas $\Gamma_s^\omega(\theta)$ is expressed
only via $U(\theta)$ and, for generic $\theta$, is proportional to
$r_s$.]
 We see that in this limit
$\Gamma_c^\omega (\pi)$ drops out from  Eq.~(\ref{genu_111}), and the
singular term in the specific heat becomes
\begin{equation}
\delta C\left( T\right) /T=- ~\frac{3 m\zeta
(3)}{4\pi}~\frac{T}{E_{F}}. \label{genu_1111}
\end{equation}
For the Coulomb interaction, the  $T^{2}-$ term in the specific
heat is universal, {\it i.e.}, it is independent of $r_s$ at small
$r_s$. This result is in full agreement with
Ref.~\cite{aleiner}.

For the sake of completeness, we explicitly calculate the specific heat for
 the Coulomb potential by  summing-up the RPA (ring) sequence
 of  diagrams for the
thermodynamic potential (diagrams 1(a), 2(b), 3(c) ... in
 Fig.~\ref{fig:omega}). A sum of these diagrams
results in a familiar expression for $\Xi $
\begin{equation}
\Xi =-\frac{T}{2}\sum_{i\Omega _{m}}\int \frac{d^{2}Q}{(2\pi )^{2}}\ln
\frac{%
1}{1-2V(Q)\Pi _{m}(\Omega _{m},Q)},
\end{equation}
where $V(Q)=2\pi e^{2}/Q$ and a factor of two in front of $\Pi
_{m}$ comes from the spin summation. Replacing the Matsubara sum
by a contour integral and using  Eq.~(\ref{comega}), we obtain the
following expression for the correction to the specific heat:
\begin{equation}
\delta C\left( T\right) /T=\frac{\partial }{\partial T}\frac{1}{8\pi
^{2}T^{2}}\int_{0}^{\infty }d\Omega \frac{\Omega }{\sinh ^{2}\Omega /2T}%
\int_{0}^{\infty }dQQ\arg \frac{1}{1-2V\left( Q\right) \Pi ^{R}\left(
\Omega,Q \right) }.
\end{equation}
The pole of the effective interaction at $\Omega ^{2}=\Omega
_{p}^{2}\left(
Q\right) \equiv v_{F}^{2}\kappa Q/2$ corresponds to a 2D plasmon mode,
where
$\kappa \equiv 2me^{2}$ is the screening wave-vector. Near the plasmon
pole,
\begin{equation}
\frac{1}{1-2V\left( Q\right) \Pi ^{R}\left( \Omega,Q \right) }\approx
\frac{%
\Omega _{p}^{2}\left( Q\right) }{\left( \Omega +i0^{+}\right) ^{2}-\Omega
_{p}^{2}\left( Q\right) }
\end{equation}
the argument of the effective interaction changes by $-\pi $ when
$Q$ intersects the plasmon curve. The plasmon's contribution to
$\delta C\left( T\right) $ comes from the region
$%
2\Omega ^{2}/v_{F}^{2}\kappa <Q<\Omega /v_{F}.$ As typical $\Omega \simeq
T$
, one can neglect the contribution from the lower limit of $Q$ for $T\ll
\kappa v_{F}.$ We then obtain a universal and interaction-independent
plasmon contribution to $\delta C(T)$
\begin{equation}
C_{\text{PL}}\left( T\right) /T=-\frac{9\zeta \left( 3\right) }{2\pi
^{2}}%
\frac{C_{{\rm  FG}}}{E_{F}}.  \label{f4}
\end{equation}
For the contribution from the particle-hole region, we find
\begin{equation}
C_{\rm{PH}}\left( T\right) /T=-\frac{\partial }{\partial T}\frac{1}{8\pi
^{2}T^{2}}\int_{0}^{\infty }d\Omega \frac{\Omega }{\sinh ^{2}\Omega /2T}%
\int_{\Omega /v_{F}}dQQ\tan ^{-1}W\left( Q,\Omega \right) ,  \label{f5}
\end{equation}
where
\begin{equation}
W\left( Q,\Omega \right) \equiv \frac{\kappa \Omega }{\sqrt{%
v_{F}^{2}Q^{2}-\Omega ^{2}}~\left( Q+\kappa \right) }.  \label{f2}
\end{equation}
The integral over $Q$ in   Eq.~(\ref{f5}) diverges logarithmically
at the upper limit, where $\tan ^{-1}W\left( Q,\Omega \right)
\approx W\left( Q,\Omega \right) $. The divergence is cut at
$Q\simeq k_{F}$ as   Eq.~(\ref{f2}) is only valid for small $Q$.
Subtracting off and adding $W\left( Q,\Omega \right)$ in
Eq.~(\ref{f5}), we split
$C_{\rm{PH}%
}\left( T\right) $ into two parts as
\begin{subequations}
\begin{eqnarray}
C_{\rm{PH}}\left( T\right) &=&C_{\rm{PH}}^{\left( 1\right) }\left(
T\right) +C_{\rm{PH}}^{\left( 2\right) }\left( T\right) ;  \label{x1a} \\
C_{\rm{PH}}^{(1)} \left( T\right) /T &=&-\frac{\partial }{%
\partial T}\frac{1}{8\pi ^{2}T^{2}}\int_{0}^{\infty }d\Omega \frac{\Omega
}{%
\sinh ^{2}\Omega /2T}\int_{\Omega /v_{F}}^{\infty }dQQ\left[ \tan
^{-1}W\left( Q,\Omega \right) -W\left( Q,\Omega \right) \right] ;
\label{x1ab} \\
C_{\rm{PH}}^{\left( 2\right) }\left( T\right) /T &=&-\frac{\partial }{%
\partial T}\frac{1}{8\pi ^{2}T^{2}}\int_{0}^{\infty }d\Omega \frac{\Omega
}{%
\sinh ^{2}\Omega /2T}\int_{\Omega /v_{F}}^{k_{F}}dQQW\left( Q,\Omega
\right)
.  \label{x1c}
\end{eqnarray}
\end{subequations}
The integral over $Q$ for $C_{\rm{PH}}^{\left( 1\right) }\left(
T\right) $ is convergent at the upper limit, and the integral can
be extended to infinity--thus this contribution is universal. One
can readily make sure that typical $Q\simeq \Omega /v_{F}\simeq
T/v_{F}\ll \kappa $, so that $\kappa $ drops out from function
$W\left( Q,\Omega \right)$ in   Eq.~(\ref{f2}). Evaluating the
integral, we obtain
\begin{equation}
C_{\rm{PH}}^{\left( 1\right) }\left( T\right)/T =\frac{9\zeta
\left( 3\right) }{4\pi ^{2}} \frac{C_{{\rm  FG}}}{E_{F}},
\end{equation}
which differs by a factor of $\left( -1/2\right) $ from the
plasmon contribution  Eq.~(\ref{f4}). The sum of the two contributions
is
\begin{equation}
\delta C\left( T\right) /T=C_{\text{PL}}\left( T\right) /T+C_{\rm{PH}%
}^{\left( 1\right) }\left( T\right) /T=-\frac{9 m \zeta \left(
3\right) }{4\pi}\frac{T}{E_{F}},
\end{equation}
which coincides with   Eq.~(\ref{genu_1111}).

Finally, the second term in   Eq.~(\ref{x1a}), $C_{\rm{PH}}^{\left(
2\right) }\left( T\right) ,$ gives a regular, linear-in-$T,$
correction to $C\left( T\right) .$ To logarithmic accuracy,
\begin{equation}
\int_{\Omega /v_{F}}^{k_{F}}dQQW\left( Q,\Omega \right) \approx \left(
\kappa \Omega /v_{F}\right) \int_{\kappa }^{k_{F}}dQ/Q=\left( \kappa
\Omega
/v_{F}\right) \ln k_{F}/\kappa,
\end{equation}
and
\begin{equation}
C_{\rm{PH}}^{\left( 2\right) }\left( T\right) =-C_{{\rm  FG}}\left( T\right) ~%
\frac{\sqrt{2}}{\pi }r_{s}|\ln r_{s}|.  \label{f1}
\end{equation}
We remind that our treatment is applicable for $r_{s}\ll 1.$
Notice that the $r_{s}|\ln r_{s}|$ correction to $C\left( T\right)
/T$ can be interpreted as the result of the mass renormalization
\begin{equation}
\delta m/m=-\frac{\sqrt{2}}{\pi }r_{s}\ln r_{s}^{-1}.
\end{equation}
This last expression coincides with that obtained in Ref. \cite
{dassarma} by evaluating the low-energy asymptotic form of
the self-energy.

\section{conclusions}
\label{sec:conclusions} This paper presents a detailed
perturbation theory for interacting fermions in 2D and  analyzes
non-analytic corrections to the Fermi-liquid behavior beyond the
second-order in interaction. We derived a full expression for
 the fermion self-energy near the mass shell, valid to an infinite order in a weak short-range interaction.
Recent study \cite{chm} found, that to second order in $U$, the
imaginary part of the self-energy diverges as $\ln |\Delta |$ upon
approaching the mass shell, where $\Delta =\omega -\epsilon
_{k}=0$. Following this lead, we demonstrated that beyond second
order, divergences become of the power-law form in $|\Delta |$,
and get stronger with increasing order of the perturbation theory.
We identified the divergent contribution as originating from the
interaction between fermions and zero-sound collective
excitations. In the perturbation theory, the collective mode
coincides with the upper edge of the particle-hole continuum. This
degeneracy causes divergences at any finite order of the
perturbation theory.  We demonstrated that a re-summation of the
power-law divergent terms to all orders in the interaction
eliminates the power-law divergences in  the self-energy, as the
zero-sound mode splits off the continuum.
  The still remaining logarithmic divergences near the
mass shell  are eliminated by a finite curvature of the fermion
spectrum. A fully renormalized self-energy is then completely free
from divergences. We found that for a contact interaction, the
real part of the self-energy  vanishes on the Fermi surface, while
the imaginary part of $\Sigma $ behaves as $\omega ^{2}\ln
|{\omega }|$. Near the mass shell, both \textrm{Re}$\Sigma^R $ and
\textrm{Im}$\Sigma^{R} $ evolve rapidly as scaling functions of
variables
$%
\Delta /(u^{2}\omega )$ and $\Delta E_{F}/\omega ^{2}$. The first scaling
variable emerges after the re-summation of power-law divergent diagrams,
whereas the second one describes the effect of a finite curvature of the
Fermi surface.
We demonstrated that the interaction of fermions with the zero-sound mode
 gives rise to a kink
 in the spectral function near $\Delta = u^2 \omega/2$.
 This prediction is amenable to a direct check in
photoemission measurements or in momentum-conserved tunneling
between two layers of the 2D electron gases.

In the second part of the paper, we discussed the non-analytic
part of the specific heat: $\delta C(T) \propto T^2$. We found
that the collective-mode contribution to the fermion self-energy
 does not affect the specific heat, i.e., a non-analytic
term in $C(T)$ is determined only by the perturbative part of the
self-energy. This result was also verified by calculating the
thermodynamic potential directly, in both real- and Matsubara
frequency formalisms. We also considered the $T^2$-term in the
specific heat for a generic Fermi liquid. We showed that it can be
expressed in a simple way
 via the spin and charge components of the quasi-particle scattering amplitude
  at angle $\theta = \pi$ between the incoming momenta. On the other hand, $\delta C(T)$
 cannot be expressed compactly in terms of the Landau interaction function. Finally, we found  for the Coulomb
interaction, not amenable to a direct perturbative treatment due
to infrared singularities,
 a non-analytic $T^2$-term in $C(T)$ is universal and
independent of the gas parameter, $r_s$, for small $r_s$.

\subsection{acknowledgments}
We acknowledge stimulating discussions with I. Aleiner, B.
Altshuler, A. Andreev, F. Essler,  W. Metzner, A. Millis, and A.
Nersesyan. The research has been supported by NSF DMR 0240238 (A.
V. Ch.), NSF DMR-0308377 (D. L. M.), and NSF DMR-0237296 (L. I.
G.). The work of D. L. M. and L. I. G. during their stay at the
Argonne National Laboratory was supported by the US DOE \ Office
of \ Science under contract No. W-31-109-ENG-38.

\appendix
\section{Backscattering contribution to the self-energy}
\label{sec:appendix_se} In this Appendix, we present the
calculation of a non-analytic part \ of the self-energy, resulting
from backscattering processes at $T=0$. This part of the
self-energy contains two contributions:
from processes with small momentum transfers $\vec{k}_{1}\approx \vec{k}%
_{1}^{\prime }$ , $\vec{k}_{2}\approx \vec{k}_{2}^{\prime }$ [see
Fig.~\ref
{fig:proc}(b)] and from processes with momentum transfers near $2k_{F}$: $%
\vec{k}_{1}\approx -\vec{k}_{1}^{\prime },$ $\vec{k}_{2}\approx -\vec{k}%
_{2}^{\prime }$ [see Fig.~\ref{fig:proc}(c)] ($g_{2}$- and $2k_{F}$%
-processes, correspondingly). The contribution to the self-energy from the $%
g_{2}$-process was obtained in \cite{chm}, and here we just cite
the result. In Matsubara frequencies,
\begin{equation}
\Sigma _{g_{2}}(\omega _{m},k)=-i~\frac{u^{2}}{8\pi E_{F}}\left[
\omega _{m}^{2}~\ln \frac{E_{F}}{\omega _{m}-i\epsilon
_{k}}+\frac{1}{4}~(\omega _{m}+i\epsilon _{k})^{2}\ln \frac{\omega
_{m}-i\epsilon _{k}}{\omega _{m}+i\epsilon _{k}}\right] .
\label{t1}
\end{equation}
We neglected regular terms of order $\omega _{m}^{2}$ in
(\ref{t1}). Although the second term in   Eq.~(\ref{t1}) is also
formally of order $\omega ^{2} $ for a generic ratio of $\omega
_{m}$ and $\epsilon _{k}$, we have to keep
it in order to compensate for a superficial divergence in the first term at $%
\omega _{m}=\pm i\epsilon _{k}$ in   Eq.~(\ref{t1}). Converting to real
frequencies and taking the imaginary part, we obtain
\begin{equation}
\text{\textrm{Im}}\Sigma_{g_{2}}^{R}(k,\omega )=\frac{u^{2}}{8\pi E_{F}}%
\left[ \omega ^{2}~\ln \frac{E_{F}}{|\omega |}+\omega ^{2}\ln \Bigg|\frac{%
\omega }{\omega +\epsilon _{k}}\Bigg|-\frac{1}{4}~(\omega
-\epsilon
_{k})^{2}\ln \Bigg|\frac{\omega -\epsilon _{k}}{\omega +\epsilon _{k}}\Bigg|%
\right] .  \label{t2}
\end{equation}
We note in passing that while the logarithmic factors in
Eq.~(\ref{t1}) comes from the integration over boson momenta $Q$
that exceeds $\omega _{m}/v_{F}$, the boson frequency ($\Omega _{m}$)
is smaller than the fermion one ($\omega _{m}$); \emph{e.g.}, for
positive $\omega _{m}$, the non-analytic contribution comes from
$-\omega _{m}<\Omega _{m}<0$. This implies that a non-analytic
part of the self-energy cannot be obtained within a
renormalization-group scheme, in which internal energies in the
self-energy diagram are assumed to be larger than the external ones.
Notice also that the self-energy in   Eq.~(\ref{t2}) is regular at
$\omega =\pm \epsilon _{k}$, since the superficial divergences in the
second and third terms cancel each other.
Both of these terms are of order $\omega ^{2}$ for a generic ratio of $%
\omega $ and $\epsilon _{k}$, and hence, to logarithmic accuracy,
one can neglect them compared to the first term.   Eq.~(\ref{t2})
then simplifies to
\begin{equation}
\text{Im}\Sigma^R _{g_{2}}(\omega ,k)=\frac{u^{2}}{8\pi E_{F}}\omega
^{2}~\ln \frac{E_{F}}{|\omega |}.  \label{t3}
\end{equation}
Next, we calculate the contribution of the $2k_{F}$-scattering to
the self-energy. Substituting   Eq.~(\ref{apr20_3}) for the
polarization bubble near $Q=2k_{F}$ into the second-order
expression for self-energy
\begin{equation}
\Sigma (k,\omega _{m})=-U^{2}\int \int \frac{d^{2}Qd\Omega }{(2\pi )^{3}}%
G(\omega _{m}+\Omega _{m},\mathbf{k}+\mathbf{Q})\Pi _{m}(\Omega
_{m},Q), \label{t4}
\end{equation}
and expanding the quasi-particle spectrum $\epsilon _{\mathbf{k+Q}}$ near $%
Q=2k_{F}$ as
\begin{equation}
\epsilon _{\mathbf{k+Q}}=-\epsilon _{k}+v_{F}k_{F}\theta
^{2}+\left( 2k_{F}^{2}/m\right) q,  \label{t5}
\end{equation}
where $q=(Q-2k_{F})/2k_{F}$ and $\pi -\theta $ is the angle between $\mathbf{%
k}$ and $\mathbf{Q}$ ($|\theta |\ll \pi $), we obtain, after
re-scaling the variables,
\begin{equation}
\Sigma _{2k_{F}}(\omega _{m},k)=-\frac{mU^{2}k_{F}}{2\pi
^{4}}\int_{-\infty
}^{\infty }dq\int_{-\infty }^{\infty }d\bar{\Omega}_{m}\left[ q+\sqrt{q^{2}+{%
\bar{\Omega}}_{m}^{2}}\right] ^{1/2}~\int_{0}^{\infty }\frac{d\theta }{%
\theta
^{2}/2+q-\bar{\epsilon}_{k}-i({\bar{\omega}}_{m}+{\bar{\Omega}}_{m})},
\label{t6}
\end{equation}
where ${\bar{\omega}}_{m}=\omega _{m}/4E_{F}$ ,
${\bar{\Omega}}_{m}=\omega _{m}/4E_{F},$ and
$\bar{\epsilon}_{k}=\epsilon _{k}/4E_{F}$. Using
\begin{equation}
~\int_{0}^{\infty }\frac{dz}{z^{2}+a}=\frac{\pi }{2}~\frac{\text{{sgn}%
\textrm{Re}}(a)}{\sqrt{a}},  \label{t7}
\end{equation}
we can re-write   Eq.~(\ref{t6}) as
\begin{equation}
\Sigma _{2k_{F}}(\omega _{m},k)=-\frac{\sqrt{2}mU^{2}k_{F}}{4\pi ^{3}}%
\int_{-\infty }^{\infty }dq\int_{-\infty }^{\infty
}d{\bar{\Omega}_{m}}\left[
q+\sqrt{q^{2}+{\bar{\Omega}}_{m}^{2}}\right] ^{1/2}\frac{\text{{sgn}Re}(\sqrt{q-%
\bar{\epsilon}_{k}-i({\bar{\omega}}_{m}+{\bar{\Omega}}_{m})})}{\sqrt{q-\bar{%
\epsilon}_{k}-i({\bar{\omega}}_{m}+{\bar{\Omega}}_{m})}}.
\label{t8}
\end{equation}
As we are interested in a $\omega ^{2}\ln \omega $ contribution,
we need to check the behavior of the integrand for large $|q|$
(logarithms come from the integration over the range where
internal variables are larger that the external ones). For
$q\rightarrow +\infty $, the integrand approaches a finite limit
($\sqrt{2}$) as $\sqrt{2}(1+{\bar{\Omega}}^{2}_m/8q^{2})$; neither
the leading nor sub-leading terms in this asymptotic form produce
a logarithm upon integrating over $q.$  However, for negative and large $q$, $%
(q+\sqrt{q^{2}+{\bar{\Omega}}_{m}^{2}})^{1/2}\approx |{\bar{\Omega}}_{m}|/(%
\sqrt{2|q|})$, and $\sqrt{q-{\bar{\epsilon}}_{k}-i({\bar{\omega}}_{m}+{\bar{%
\Omega}}_{m})}\approx -i\sqrt{|q|}\mathrm{~sgn}({\bar{\Omega}_{m}}+{\bar{%
\omega}_{m}})$,  so that the integrand behaves as $\left| q\right|
^{-1};$ hence the logarithmic singularity does come from this
region of $q.$   To logarithmic accuracy, we obtain
\begin{equation}
\Sigma _{2k_{F}}(\omega _{m},k)=-i\frac{mU^{2}k_{F}}{4\pi
^{3}}\int_{-\infty
}^{\infty }d{\bar{\Omega}_{m}}|{\bar{\Omega}_{m}}|~{\text{sgn}}({\bar{\Omega}%
_{m}}+{\bar{\omega}_{m}})~\int^{-|{\bar{\omega}}_{m}|/v_{F}}_{-E_{F}/v_{F}}%
\frac{dq}{q}.  \label{t9}
\end{equation}
Evaluating the integrals, we find
\begin{equation}
\Sigma _{2k_{F}}(\omega _{m},k)=-i\frac{u^{2}}{8\pi }~\frac{\omega _{m}^{2}}{%
E_{F}}~\ln \frac{E_{F}}{|\omega _{m}|}.  \label{t10}
\end{equation}
Continuing to real frequencies and taking the imaginary part
yields
\begin{equation}
\text{\textrm{Im}}\Sigma _{2k_{F}}^{R}(\omega,k)=\frac{u^{2}}{8\pi E_{F}%
}\omega ^{2}~\ln \frac{E_{F}}{|\omega |}.  \label{t11}
\end{equation}
Comparing this result with   Eq.~(\ref{t3}), we see that the
non-analytic parts of \textrm{Im}$\Sigma ^{R}$ resulting from the
$g_{2}-$ and $2k_{F}-$processes are identical within logarithmic
accuracy. The result for the sum of the two terms is quoted in
  Eq.~(\ref{c1c}). Observe that, according to   Eq.~(\ref{t11}), a
non-analytic, $\omega ^{2}\ln |\omega |$ part of $\Sigma
_{2k_{F}}(\omega _{m},k)$ comes from negative $q$, \emph{i.e.},
from $Q<2k_{F}$, where the static polarization bubble is just a
constant; thus a singular correction is entirely dynamical. A
singular, $(Q-2k_{F})^{1/2},$ behavior of the static bubble $\Pi
(\Omega =0,Q)$ for $Q>2k_{F}$ does not give rise to an imaginary
part of the self-energy-- this result follows due to the fact that  a static density fluctuation
cannot decay the quasi-particles. However, the fact that
even for a dynamic bubble, only the region of $Q<2k_{F}$ is
responsible for a non-analytic part of $\Sigma (k,\omega )$ is
rather peculiar and has not been emphasized explicitly in earlier
work~\cite{chm}. Note in this regard that both regions of
$Q<2k_{F}$ and $Q>2k_{F}$ contribute to a non-analytic behavior of
the spin susceptibility~ \cite{millis,chm}.

Now, we evaluate the integrals in   Eq.~(\ref{t8}) beyond logarithmic
accuracy. The result is
\begin{equation}
\text{\textrm{Im}}\Sigma _{2k_{F}}^{R}(k,\omega )=\frac{u^{2}}{8\pi E_{F}}~%
\left[ \omega ^{2}\ln \frac{E_{F}}{|\omega |}+\omega ^{2}\ln \left| \frac{%
\epsilon _{k}}{\omega }\right| -\left( \omega +\epsilon
_{k}\right) ^{2}\ln \left| \frac{\epsilon _{k}}{\omega +\epsilon
_{k}}\right| \right] . \label{t12}
\end{equation}
Again, the last two terms are of order $\omega ^{2}$ for a generic ratio of $%
\omega $ and $\epsilon _{k}$ but we keep them in order to
demonstrate that \textrm{Im}$\Sigma _{2k_{F}}^{R}(\omega ,k)$
remains finite at $\omega =\pm \epsilon _{k}$.
\section{Non-symmetrized vertex}
\label{sec:appendixA} In this Appendix, we derive an  expression
for forward-scattering part of the non-symmetrized vertex,
${\bar{\Gamma%
}}$, summing up diagrams with the maximum number of polarization
bubbles to all orders in contact interaction, $U$. We then
anti-symmetrize the vertex, and substitute the result into the
Dyson equation [  Eq.~(\ref{dyson})] to obtain the corresponding part
of the self-energy. The diagrams for a non-symmetrized vertex up
to the third order are presented in Fig.~\ref{fig:vertex2}. In the
Matsubara technique, we associate a factor of $-U$ with each of
the interaction lines, and a factor $-2$ with each of the
polarization bubble. There is also an extra factor of
$%
-1$ for exchange processes in which the two outgoing legs are permuted
(the
last diagrams of second and third order in Fig.~\ref{fig:vertex2}).
We present here a general recipe for calculating the $\nu ^{th}$ ($\nu
>1)$
order vertex diagram. As is seen from the Fig.~\ref{fig:vertex2},
the vertex consists of two parts. The first part comes from the
direct interaction and contains a spin factor $\delta _{\alpha
\gamma }\delta _{\beta \varepsilon }$. The second part is due to
the exchange interaction, and comes with a spin factor
$%
\delta _{\alpha \varepsilon }\delta _{\beta \gamma }$. At each
order, there is only one exchange diagram whose contribution is
$-(-U)^{\nu }\Pi ^{\nu -1}\delta _{\alpha \varepsilon }\delta
_{\beta \gamma }$. (At second and third orders, these are the
first and third diagrams in the third column of
Fig.~\ref{fig:vertex2}, respectively.) The rest of diagrams are
due to the direct interaction and contain various number of
bubbles. At order $\nu $, the number of these bubbles ($R$) varies
from $0$ to $\nu -1$. For a diagram with $R$ bubbles, $R+1$
interaction lines are used up in making a chain of bubbles and
connecting it to two external solid lines. The remaining $N=\nu
-R-1$ interaction lines can be arranged anywhere either at the two
ends of the chain of bubbles or inside the $R$ bubbles. There are
$%
S=R+2$ sites where $N$ interaction lines can be placed. The number of
diagrams with $R$ bubbles is equal to the number of ways to arrange $N$
lines among $S$ sites:
\begin{equation}
\frac{(S+N-1)!}{(S-1)!N!}=\frac{\nu !}{(R+1)!(\nu -R-1)!}.
\end{equation}
Consequently, the contribution to ${\bar{\Gamma}}$ from diagrams with $R$
bubbles is
\begin{equation}
\frac{\nu !}{(R+1)!(\nu -R-1)!}(-U)^{\nu }(-2)^{R}\Pi ^{\nu -1}\delta
_{\alpha \gamma }\delta _{\beta \varepsilon }.
\end{equation}
The total contribution from all bubble diagrams at the order $\nu $ is
then
\begin{equation}
\sum_{R=0}^{\nu -1}\frac{\nu !(-2)^{R}}{(R+1)!(\nu -R-1)!}(-U)^{\nu }\Pi
^{\nu -1}\delta _{\alpha \gamma }\delta _{\beta \varepsilon }={}-\frac{%
(1-(-1)^{\nu })}{2}U^{\nu }\Pi ^{\nu -1}\delta _{\alpha \gamma }\delta
_{\beta \varepsilon }
\end{equation}
where we have used an identity
\begin{equation}
\sum_{R=0}^{\nu }\frac{\nu !(-2)^{R}}{R!(\nu -R)!}=(1-2)^{\nu }=\left(
-1\right) ^{\nu }.
\end{equation}
Adding up direct and exchange terms, we obtain the following form
for non-symmetrized vertex at order $\nu >1$:
\begin{equation}
{\bar{\Gamma}}_{\alpha \beta ,\gamma \varepsilon }^{\nu }=-\frac{1}{2}%
(1-(-)^{\nu })U^{\nu }\Pi ^{\nu -1}\delta _{\alpha \gamma }\delta _{\beta
\varepsilon }-(-U)^{\nu }\Pi ^{\nu -1}\delta _{\alpha \varepsilon }\delta
_{\beta \gamma }  . \label{f7}
\end{equation}
The vertex function can now be readily summed up to all orders,
with the result
\begin{equation}
{\bar{\Gamma}}_{\alpha \beta ,\gamma \varepsilon
}(p_{1},p_{2};p_{1}-q,p_{2}+q) ={\bar{\Gamma}}_{\alpha \beta ,\gamma \varepsilon
} (q) = -U\delta _{\alpha \gamma }\delta _{\beta
\varepsilon }+\sum_{\nu =1}^{\infty }\bar{\Gamma}_{\alpha \beta ,\gamma
\varepsilon }^{\nu }=-\delta _{\alpha \gamma }\delta _{\beta \varepsilon
}%
\frac{U}{1-(U\Pi )^{2}}-\delta _{\alpha \varepsilon }\delta _{\beta \gamma
}%
\frac{U^{2}\Pi (q)}{1+U\Pi (q)}.
\end{equation}
Using an SU$\left( 2\right) $ identity
\begin{equation}
\delta _{\alpha \varepsilon }\delta _{\beta \gamma }=\left(
1/2\right) \left( \sigma _{\alpha \gamma }^{a}\sigma _{\beta
\varepsilon }^{a}+\delta _{\alpha \gamma }\delta _{\beta
\varepsilon }\right),
\end{equation}
and introducing dimensionless spin and charge vertices $\mathcal{G}_{\rho
}$
and spin $\mathcal{G}_{\sigma }$, defined in   Eq.~(\ref{f8}), we obtain
  Eq.~(%
\ref{b4}). Anti-symmetrizing the vertex as prescribed by
  Eq.~(\ref{f9}) and substituting the result into the Dyson equation
(\ref{dyson}), we obtain for the self-energy
\begin{eqnarray}
\Sigma _{\alpha \beta }(p) &=&\delta _{\alpha \beta }\int_{q}U
G(p-q)-\int_{p'}U\overline{\Gamma }_{\gamma \alpha ;\gamma \beta
}G(p')\Pi (p-p')+\int_{p''}U\overline{\Gamma }_{\gamma \alpha
;\beta \gamma }G(p'')\Pi (p-p''){}  \notag \\
&=&{}\delta _{\alpha \beta }\int_{q}U G(p-q)+2\delta _{\alpha
\beta
}\int_{p'}\left\{ \frac{U^{2}}{1-(U\Pi )^{2}}+\frac{U^{3}}{1+U\Pi }%
\right\} G(p')\Pi (p-p')  \notag \\
&&-\delta _{\alpha \beta }\left\{ \int_{p''}\frac{U^{2}}{1-(U\Pi )^{2}}+%
\frac{U^{3}}{1+U\Pi }\right\} G(p'')\Pi (p-p''){}  \notag \\
&=&\delta _{\alpha \beta }\int_{q}U G(p-q)+\delta _{\alpha \beta }\int_{q}%
\frac{U^{2}\Pi (q)}{1-(U\Pi )^{2}}G(p-q)+\delta _{\alpha \beta }\int_{q}%
\frac{U^{3}\Pi (q)^{2}}{1+U\Pi }G(p-q).  \label{f10}
\end{eqnarray}
\textrm{Re}-arranging the result, we obtain the self-energy in the form of
  Eq.~(\ref{b5}).

\section{Relation between Eqs.(\ref{feb5_3T0}) and
(\ref{feb5_3_11}).} \label{app:extra_1}

In this Appendix, we discuss the relation between the two results
for the specific heat: the one derived in Ref.~\cite{agd}
[Eq.(\ref{feb5_3T0})] and the one following from the
Luttinger-Ward functional for the thermodynamic potential
\cite{luttinger} [Eq.(\ref{feb5_3_11})]. We show that the two
expressions yield identical results for the $T^2$-term in the
specific heat, provided  that one uses in temperature-dependent
self-energy in Eq.(\ref
{feb5_3T0}) (the recipe in Ref.~\cite{agd} is to use the self-energy at $T=0$%
).

For the sake of simplicity, we focus on the contact interaction
and consider only a perturbative part of the thermodynamic
potential, neglecting the interaction with the zero-sound mode.

We begin with the Luttinger-Ward approach. To second order in $u$,
the Luttinger-Ward formula for the entropy is
\begin{equation}
S = S_0 + \frac{\partial}{\partial T} \left[\frac{N_0}{2} T \sum_m
\int d \epsilon_k \Sigma (\omega_m,k,T) G_0 (\omega_m,k)\right].
\label{e_1}
\end{equation}
Here $S_0$ is the entropy of an ideal gas, $N_0 = m/2\pi$ is the
density of states per spin projection, $G_0 = (i\omega_m -
\epsilon_k)^{-1}$ is the bare Green's function, and $\Sigma
(\omega_m,k,T)$ is the self-energy, which depends on $T$ both
implicitly, via $\omega_m = \pi T (2m+1)$, and also explicitly.
The second-order self-energy in Matsubara frequencies is, to
logarithmic accuracy, ~\cite{chm}
\begin{equation}
\Sigma(\omega_m,k,T) = - \frac{i}{2E_F} u^2 T \sum_{\Omega_m} |\Omega_m| {%
\text{sign}}(\omega_m + \Omega_m) \ln\frac{E_F}{\epsilon_k -
i(2\Omega_m +
\omega_m)} -\frac{i}{4E_F} u^2 T \sum_{\Omega_m} |\Omega_m| {\text{sign}}%
(\omega_m + \Omega_m) \ln\frac{E_F} {\epsilon_k -i \omega_m}.
\label{e_3}
\end{equation}
The first term in Eq.(\ref{e_3}) comes from backscattering and
includes the sum of contributions from the $g_2-$ and
$2k_F$-processes. The second term in Eq.(\ref{e_3}) comes from
forward scattering.

Had there been no dependence on $\epsilon _{k}$ under the
logarithm in (\ref {e_3}), the integration over $\epsilon _{k}$ in
(\ref{e_1}) would have been straightforward, as it would have
involved only $G_{0}(\omega _{m},k)$. Using the familiar relations
\begin{equation}
\mathcal{P}\int d\epsilon _{k}{\text{Re}}G_{0}^{R}(\omega
,\epsilon _{k})=0,\int d\epsilon _{k}{\text{Im}}G_{0}^{R}(\omega
,\epsilon _{k})=-\pi ; \label{e_11}
\end{equation}
and converting the Matsubara sums into the contour integrals, one
could
readily verify that the expression for the entropy would have involved only $%
\mathrm{Re}\Sigma ^{R}(\omega ,k,T)$ on the mass shell and that
both terms in (\ref{e_3}) would have contributed to the entropy.

Because of the $k-$dependence under the logarithms in (\ref{e_3}),
the actual situation is different. Indeed, substituting the second
term in (\ref {e_3}) into (\ref{e_1}) we find that the integral
over $\epsilon _{k}$ reduces to
\begin{equation}
\int_{-E_{F}}^{E_{F}}\frac{d\epsilon _{k}}{\epsilon _{k}-i\omega
_{m}}\ln
\frac{E_{F}}{\epsilon _{k}-i\omega _{m}}=\frac{1}{2}\ln ^{2}\frac{%
E_{F}-i\omega }{-E_{F}-i\omega }  \label{ee_1}
\end{equation}
The result in (\ref{ee_1}) vanishes in the limit of $|\omega _{m}|\ll E_{F}$%
. Thus, to logarithmic accuracy, forward scattering does not
contribute to the entropy. It is instructive, however, to see how
the zero for the integral in (\ref{ee_1}) comes about. In fact,
there are two contributions to this integral, each of order $\ln
{E_{F}/|\omega _{m}|}$. The first comes from the region of small
$\epsilon _{k}$: $\epsilon _{k}\simeq \omega _{m}$. For this
contribution, the dependence on $\epsilon _{k}$ under the
logarithm can be neglected, \emph{i.e.}, the logarithmic factor
can be approximated by $\ln E_{F}/|\omega _{m}|$. The integration
over $\epsilon _{k}$ then yields
\begin{equation}
~\ln \frac{E_{F}}{|\omega _{m}|}\int \frac{d\epsilon
_{k}}{\epsilon _{k}-i\omega _{m}}=i\pi {\mathrm{sgn}}\omega
_{m}\ln \frac{E_{F}}{|\omega _{m}|}.  \label{e_5}
\end{equation}
The second contribution comes from the region of large $\epsilon _{k}:$ $%
\epsilon _{k}\gg \omega _{m}$. Here, we obtain, combining the
contributions from positive and negative $\epsilon _{k}$
\begin{eqnarray}
\int_{-E_{F}}^{E_{F}}\frac{d\epsilon _{k}}{\epsilon _{k}-i\omega
_{m}}~\ln \frac{E_{F}}{\epsilon _{k}-i\omega _{m}} &\approx
&\int_{|\omega _{m}|}^{E_{F}}\frac{d\epsilon _{k}}{\epsilon
_{k}}~\ln {\frac{-\epsilon
_{k}-i\omega _{m}}{\epsilon _{k}-i\omega _{m}}}  \notag \\
&=&-i\pi {\mathrm{sgn}}\omega _{m}\ln \frac{E_{F}}{|\omega _{m}|}.
\label{e_6}
\end{eqnarray}
Combining (\ref{e_5}) and (\ref{e_6}), we find that for the
forward scattering contribution to $\Sigma $, the net integral
over $\epsilon _{k}$ vanishes.

For the backscattering contribution, the integral over $\epsilon
_{k}$ is
\begin{equation}
\int \frac{d\epsilon _{k}}{\epsilon _{k}-i\omega _{m}}\ln \frac{E_{F}}{%
\epsilon _{k}-i\omega _{m}-2i\Omega _{m}}.  \label{ee_2}
\end{equation}
There are again two contributions to this integral to order $\ln
E_{F}/|\omega _{m}|$. The contribution from $\epsilon _{k}\simeq
\omega _{m}$ is the same as (\ref{e_5}), while the one  from
$\epsilon _{k}\gg \omega _{m} $ is
\begin{equation}
-i\pi {\mathrm{sgn}}\Omega _{m}\ln \frac{E_{F}}{|\omega _{m}|}.
\label{ee_3}
\end{equation}
It follows from (\ref{e_3}) that, at low temperatures, the
dominant contribution to the self-energy comes from such $\Omega
_{m}$ that are of different sign compared to $\omega _{m}$. As a
result, for the
backscattering part of the self-energy, the small-$\epsilon _{k}-$ and large-%
$\epsilon _{k}$-contributions to the integral over $\epsilon _{k}$
add up instead of canceling out.

The outcome of this analysis is that, to logarithmic accuracy, one
can still formally neglect the dependence on $\epsilon _{k}$ in
the logarithm in (\ref{e_3}); however, one should simultaneously
neglect the forward-scattering contribution to the self-energy,
and multiply backscattering contribution by a factor of two.
Converting the Matsubara sum in Eq.~(\ref{e_1}) into a contour
integral, one can express the entropy via the real part of the
self-energy on the mass shell, as
\begin{equation}
S=S_{0}-N_{0}\int_{-\infty }^{\infty }d\omega \left[ \frac{\omega }{T}\frac{%
\partial n_{0}}{\partial \omega }\mathrm{Re}\Sigma ^{R}(\omega ,T)-\left(
n_{0}-\frac{1}{2}\right) ~\frac{\partial }{\partial
T}\mathrm{Re}\Sigma ^{R}(\omega ,T)\right]   \label{e_7}
\end{equation}
We emphasize again that this expression is two times larger than
the one that one would obtain by just neglecting the dependence on
$\epsilon _{k}$ in the forward scattering self-energy.

Next, we find the relation between the two terms in (\ref{e_7}).
If the self-energy were independent of $T$, the second term would
not contribute, and the entropy would be given just by  the first
term. We show, however, that the two terms in (\ref{e_7}) are in
fact equal, \emph{i.e.}, the entropy can be formally re-expressed
via only the first term in (\ref{e_7}), but with another extra
factor of two.

To demonstrate this, we need an expression for $\mathrm{Re}\Sigma
^{R}(\omega ,T)$. The easiest way to obtain it is to convert
Matsubara
self-energy into $\mathrm{Im}\Sigma ^{R}(\omega ,T)$ and then obtain $%
\mathrm{Re}\Sigma ^{R}(\omega ,T)$ by a Kramers-Kronig
transformation. Approximating the logarithm in the first term in
(\ref{e_3}) as $\ln E_{F}/|\Omega _{m}|$, we obtain
\begin{equation}
\mathrm{Im}\Sigma ^{R}(\omega ,T)=\frac{u^{2}}{4\pi
E_{F}}\int_{-\infty
}^{\infty }d\Omega \Omega \ln \frac{E_{F}}{|\Omega |}\left[ \coth \frac{%
\Omega }{2T}-\tanh \frac{\omega +\Omega }{2T}\right] .
\label{e_8}
\end{equation}
Substituting the Kramers-Kronig transform of (\ref{e_8}) into
(\ref{e_7}), and integrating in the second term in (\ref{e_7}) by
parts, we find after some algebra that the two terms in
(\ref{e_7}) contribute equally to the entropy.

As a result, the net expression for the entropy in the
Luttinger-Ward formalism is given by
\begin{equation}
S=S_{0}-2N_{0}\int_{-\infty }^{\infty }d\omega \frac{\omega }{T}\frac{%
\partial n_{0}}{\partial \omega }{e}\Sigma ^{R}(\omega ,T).  \label{e_9}
\end{equation}
It differs by a factor of four from what one would have obtained
neglecting
both $\epsilon _{k}$ under the logarithm in (\ref{e_3}) \emph{and} the $T$%
-dependence of $\mathrm{Re}\Sigma ^{R}(\omega ,T).$

Eq.~(\ref{e_9}) is very similar to the one from Ref.~\cite{agd}
which, to the lowest order in the interaction, reduces to
\begin{equation}
S=S_{0}+\frac{2}{\pi }N_{0}\int_{-\infty }^{\infty }d\omega \int
d\epsilon _{k}\frac{\omega }{T}\frac{\partial n_{0}}{\partial
\omega }\mathrm{Im}\left[ \Sigma (\omega ,k)G_{0}(\omega ,\epsilon
_{k})\right] .  \label{e_10}
\end{equation}
We first assume, and then verify,  that once  $\Sigma (\omega
,\epsilon _{k})
$ is evaluated \textit{in real frequencies} and substituted into (\ref{e_10}%
), the momentum dependence of the self-energy can be neglected.
Using (\ref {e_11}),  we then obtain that (\ref{e_9}) and
(\ref{e_10}) coincide if one substitutes the $T$\textit{-dependent
self-energy} in (\ref{e_10}).

Finally, we show that the momentum-dependent part of $\Sigma
(\omega ,k)$ does not contribute to the entropy. The proof is
somewhat tricky. The
backscattering term in the second-order self-energy [first term in (\ref{t2}%
] does indeed contain the dependence on $\epsilon _{k}$ under the
logarithm, just like in the Matsubara self-energy. Because of this
dependence, there is
a non-zero contribution to the entropy from \textrm{Im}$\Sigma ^{R}\mathrm{Re%
}G_{0}^{R}$ term in (\ref{e_10}). Should one keep it, this would
modify the answer for $S$ by a factor of two, compared to
(\ref{e_9}).

However, in contrast to the Matsubara self-energy, which is
strictly perturbative, the self-energy in real frequencies does
develop mass-shell singularities in the forward-scattering part.
We remind that the reason for these singularities is the presence
of the zero-sound mode that, within the perturbation theory,
coincides with the upper boundary of the particle-hole continuum.
These mass-shell singularities give rise a non-perturbative part
of the self-energy, $\Sigma _{\mathrm{ZS}}$ and also modify the
part of the
self-energy coming from  the interaction with the particle-hole continuum, $%
\Sigma _{\mathrm{ex}}+\Sigma _{\mathrm{PH}}$ [see Eq.(\ref{sum}].

We show that the two contributions to the entropy -- one from $\mathrm{Im}%
\Sigma _{B}^{R}\mathrm{Re}G_{0}^{R}$ and the other from
$\mathrm{Im}\Sigma _{F}^{R}\mathrm{Re}G_{0}^{R}$ -- cancel each
other. To this end, we demonstrate that
\begin{equation}
B(\omega )\equiv \frac{m}{2\pi }{}\mathcal{P}\int d\epsilon _{k}~\frac{\text{%
\textrm{Im}}\Sigma ^{R}(\omega ,\epsilon _{k})}{\omega -\epsilon
_{k}}. \label{y2}
\end{equation}
where $\Sigma ^{R}$ is a sum of backscattering and
forward-scattering contributions, does not contain terms
non-analytic in $\omega $ .

We begin with the backscattering part. As it was discussed in
Sec.~\ref
{sec:process}, there are two types of backscattering processes: $g_{2}$ and $%
2k_{F}$. The self-energy from the $g_{2}$-process is given by Eq.~(\ref{t2}%
). Terms that are independent of $\epsilon _{k}$ obviously do not
contribute
to $~B\left( \omega \right) .$ Substituting the rest of Eq.~(\ref{t2}) into (%
\ref{y2}), we obtain
\begin{equation}
B_{g_{2}}=-\frac{m}{32}\frac{u^{2}}{E_{F}}\omega \left| \omega
\right| . \label{b2g2}
\end{equation}
In (\ref{b2g2}), we neglected terms of order $\omega $, which
contribute just to the renormalization of the effective mass at
$T=0$. It can be also
verified that the subleading term to (\ref{b2g2}) is analytic and scales as $%
\omega ^{3}/E_{F}$.

The $2k_F$-contribution to the self-energy is given by
(\ref{t12}). Substituting this contribution into (\ref{y2}) and
evaluating the integral over $\epsilon _{k}$, we find that
$B_{2k_{F}}(\omega )$ contains regular terms of order $\omega $,
$\omega ^{3}$, etc., but no $\omega |\omega |$ term. Therefore,
for our purposes,
\begin{equation}
B_{2k_{F}}(\omega )=0.  \label{y7}
\end{equation}

Consider now the forward-scattering ($g_{4}$) part. The imaginary part of $%
\Sigma _{\text{ZS}}$ is analytic near the mass shell and does not
contribute
to $B$. The imaginary parts of $\Sigma _{\mathrm{PH}}$ and $\Sigma _{\mathrm{%
ex}},$ however, depend on $\Delta =\omega -\epsilon _{k}$
logarithmically away from the immediate vicinity of the mass
shell, where the logarithmic
dependence is regularized [see (\ref{h2}), (\ref{may_5_4_1}) and ~(\ref{h2_1}%
)]. Substituting the regularized expression, Eq.~(\ref{h2_1}) into (\ref{y2}%
) we find that the $B_{g_{4}}$ reduces to
\begin{equation}
B_{g_{4}}=\frac{m}{16\pi ^{2}}\frac{u^{2}}{E_{F}}~\left| \omega
\right| \omega \int_{-\infty }^{\infty }{}\frac{dx}{x-\delta }~\ln
|x|=\frac{m}{8\pi ^{2}}\frac{u^{2}}{E_{F}}~\delta {}\left| \omega
\right| \omega \int_{0}^{\infty }\frac{dx\ln x}{x^{2}-\delta
^{2}},  \label{y4}
\end{equation}
where $\delta \propto u$ is \textit{positive}. Performing the
integration, we obtain
\begin{equation}
B_{g_{4}}=\frac{m}{32}\frac{u^{2}}{E_{F}}\omega \left| \omega
\right| , \label{y6}
\end{equation}
We see that it is opposite in sign to the contribution from the
$g_{2}-$ process, Eq.~(\ref{b2g2}). Adding up the three
contributions, Eqs.~(\ref {b2g2}), (\ref{y7}) and (\ref{y6}), we
obtain that the total $B(\omega )=0$. This proves that the
momentum dependence of $\Sigma $ in (\ref{e_10}) can indeed be
neglected.

\section{Specific heat in Eliashberg-type theories}

In this Appendix, we consider the specific heat for the case when
the fermion self-energy depends only on the frequency but not on
momentum. Such a situation occurs, \emph{e.g.}, in Eliashberg-type
theories, which describes the interaction of fermions with slow boson
modes~\cite{eli_1}. We show that in order to obtain a correct form of the
linear-in-$T$, Fermi-liquid contribution to the specific heat, one
can use the approximate relation   Eq.~(\ref{ya_2}) (from Ref.~\cite{agd}), which expresses the thermodynamic potential solely in
terms of an exact fermion Green's function. However, a correct
form of the sub-leading, non-analytic part of the specific heat
can only be obtained using the full Luttinger-Ward expression,
while   Eq.~(\ref{ya_2}) gives an erroneous result, even if one uses
the self-energy at finite $T,$ $\Sigma (\omega _{m},T)$.

A general form of the thermodynamic potential, $\Xi ,$ for a
system of interacting fermions is given by Luttinger-Ward formula,
  Eq.~(\ref{o1}). It expresses $\Xi $ in terms of the full Green's
function and infinite series of skeleton self-energies. In many
cases, though, the interaction between fermions is strongly
enhanced in a particular interaction channel. In such a case,
multiple interactions between fermions in the same channel can be
adequately described as an exchange of corresponding low-energy
bosonic collective modes. If, in addition, these modes are slow
compared to fermions, the self-energy of fermions will be independent
of the momentum, $\Sigma (\omega _{m},k)=\Sigma (\omega _{m})$,
and corrections to fermion-boson vertex can be neglected due to
the  Migdal theorem. For these cases, the Luttinger-Ward formula
reduces to a closed-form expression in terms of full propagators
of fermions and low-energy bosons
\begin{eqnarray}
\Xi  &=&-2T\sum_{\omega _{m}}\int \frac{d^{D}k}{(2\pi )^{D}}\left[ \frac{1}{2%
}\ln \left[ \epsilon _{\mathbf{k}}^{2}+\{\omega _{m}+\Sigma
(\omega
_{m})\}^{2}\right] -i\Sigma (\omega _{m})G(k,\omega _{m})\right]   \notag \\
&+&\frac{1}{2}T\sum_{\Omega _{m}}\int \frac{d^{D}Q}{(2\pi
)^{D}}\left[ \ln [D^{-1}(Q,\Omega _{m})]+2\Pi (\Omega
_{m},Q)D(\Omega _{m},Q)\right]   \notag
\\
&+&T^{2}g^{2}\sum_{\omega _{m},\omega _{m}^{\prime }}\int \frac{%
d^{D}kd^{D}k^{\prime }}{(2\pi )^{2D}}G(k,\omega _{m})D(\mathbf{k}-\mathbf{%
k^{\prime }},\omega _{m}-\omega _{m}^{\prime })G(k^{\prime
},\omega _{m}^{\prime }).  \label{phononom}
\end{eqnarray}
Here $G(k,\omega _{m})=\left[ i\left\{ \omega _{m}+\Sigma (\omega
_{m})\right\} -\epsilon _{k}\right] ^{-1}$ is the full fermion
Green's function, $D(Q,\Omega _{m})=\left( D_{0}^{-1}(Q,\Omega
)-2\Pi (\Omega _{m},Q)\right) ^{-1}$ is the full boson propagator
(a factor of $2$ is due to the spin summation), $D_{0}(Q,\Omega )$ is
the bare boson propagator,  $\Sigma (\Omega _{m})$ and $\Pi
(\Omega _{m},Q)$ are the fermion and boson  self-energies,
respectively, and $g$ is the effective fermion-boson coupling. [In
this Appendix, the Matsubara self-energy is defined with an $i$ up
front.]   Eq.~(\ref{phononom}) was first obtained in the context of
the electron-phonon interaction~\cite{eliashberg,bardeen_stephen}
(in which case $D_{0}$ is the phonon propagator), and was applied
later to the electron-electron interaction mediated by
Landau-damped collective modes~\cite
{ioffe,grilli,acs,pepin,ferro,rob_1}. In the latter case, the frequency
dependence of $D(Q,\Omega _{m})$  comes predominantly from the
boson self-energy, and the bare boson propagator $D_{0}(Q,\Omega
_{m})$ can be approximated by its static form $D_{0}\left(
Q\right)\equiv D_{0}(Q,0)
 $~\cite{ioffe,acs,pepin,ferro}. Notice that
   Eq.~(\ref{phononom}) is applicable to the interaction with both
charge- and spin modes, except for the spin case it has to be
modified slightly due to the spin structure of the
interaction~\cite{rob_1}.

The fermion and boson self-energies, $\Sigma (\omega _{m})$ and
$\Pi (\Omega _{m})$, are obtained from the condition that $\Xi $
is stationary with respect to variations in $\Sigma (\omega _{m})$
and $\Pi (\Omega _{m})$. Conditions
\begin{equation*}
\delta \Xi /\delta \Sigma (k)=\delta \Xi /\delta \Pi (k)=0
\end{equation*}
yield~\cite{bardeen_stephen}
\begin{eqnarray}
&&\Sigma (\omega _{m})=iTg^{2}\sum_{\omega _{m}^{\prime }}\int \frac{%
d^{D}k^{\prime }}{(2\pi )^{D}}~G(k^{\prime },\omega _{m}^{\prime })D(\mathbf{%
k}-\mathbf{k}^{\prime },\omega _{m}-\omega _{m}^{\prime });  \notag \\
&&\Pi (\Omega _{m},Q)=-Tg^{2}\sum_{\omega _{m}}\int \frac{d^{D}k}{(2\pi )^{D}%
}G(k,\omega _{m})G(\mathbf{k}-\mathbf{Q},\omega _{m}-\Omega _{m}).
\label{2ndord}
\end{eqnarray}
Using    Eq.~(\ref{2ndord}), the last term in    Eq.~(\ref{phononom}) can be
re-written as
\begin{equation}
-T\sum_{\Omega _{m}}\int \frac{d^{D}Q}{(2\pi )^{D}}\Pi (\Omega
_{m})D(Q,\Omega _{m})  \label{oct_15_1}
\end{equation}
or, equivalently,
\begin{equation}
2iT\sum_{\omega _{m}}\int \frac{d^{D}k}{(2\pi )^{D}}\Sigma (\omega
_{m})G(k,\omega _{m}).  \label{oct_15_2}
\end{equation}
Accordingly,   Eq.~(\ref{phononom}) reduces to
\begin{eqnarray}
\Xi  &=&-2T\sum_{\omega _{m}}\int \frac{d^{D}k}{(2\pi )^{D}}\left[ \frac{1}{2%
}\ln \left[ \epsilon _{\mathbf{k}}^{2}+\left\{ \omega _{m}+\Sigma
(\omega _{m})\right\} ^{2}\right] -i\Sigma (\omega _{m})G(k,\omega
_{m})\right]
\notag \\
&+&\frac{1}{2}T\sum_{\Omega _{m}}\int \frac{d^{D}Q}{(2\pi
)^{D}}~\ln \left[ D^{-1}(Q,\Omega _{m})\right],
\label{phononom_1}
\end{eqnarray}
or, equivalently, to
\begin{eqnarray}
\Xi  &=&-T\sum_{\omega _{m}}\int \frac{d^{D}k}{(2\pi )^{D}}\ln
\left[ \epsilon _{\mathbf{k}}^{2}+\left\{ (\omega _{m}+\Sigma
(\omega _{m})\right\}
^{2}\right]   \notag \\
&+&\frac{1}{2}T\sum_{\Omega _{m}}\int \frac{d^{D}Q}{(2\pi
)^{D}}\left[ \ln [D^{-1}(Q,\Omega _{m})]+2\Pi (\Omega
_{m})D(Q,\Omega _{m})\right] . \label{phononom_2}
\end{eqnarray}
Each of the last two expressions can be simplified further. In
  Eq.~(\ref
{phononom_1}), we switch from the integration over momentum to that over $%
\epsilon _{k}$ using $\int d^{D}k/(2\pi )^{D}=N_{0}\int d\epsilon
_{k}$, where $N_{0}$ is the density of states. Integrating over
$\epsilon _{k}$, we find that the first line in    Eq.~(\ref{phononom_1})
reduces to
\begin{eqnarray}
&&-2T\sum_{\omega _{m}}\int \frac{d^{D}k}{(2\pi )^{D}}\left[ \frac{1}{2}\ln %
\left[ \epsilon _{\mathbf{k}}^{2}+\left\{ (\omega _{m}+\Sigma
(\omega _{m})\right\} ^{2}\right] -i\Sigma (\omega _{m})G(k,\omega
_{m})\right]
\notag \\
&=&-TN_{0}~\sum_{\omega _{m}}\left( |\omega _{m}+\Sigma (\omega
_{m})|-|\Sigma (\omega _{m})|\right).   \label{oct_15_3}
\end{eqnarray}
As the sign of $\Sigma (\omega _{m})$ coincides with that of
$\omega _{m} $, the self-energy drops out, and    Eq.~(\ref{oct_15_3})
reduces to the result for
non-interacting fermions, $\Xi _{\mathrm{FG}}$. Substituting    Eq.~(\ref{oct_15_3}%
) back into    Eq.~(\ref{phononom_1}), we obtain
\begin{equation}
\Xi =\Xi _{\mathrm{FG}}+\frac{1}{2}T\sum_{\Omega _{m}}\int \frac{d^{D}Q}{%
(2\pi )^{D}}~\ln [D^{-1}(Q,\Omega _{m})],  \label{oct_15_4}
\end{equation}
where
\begin{equation}
\Xi _{\mathrm{FG}}=-TN_{0}\sum_{\omega _{m}}|\omega _{m}|
\label{oct_15_4_1}
\end{equation}
is the thermodynamic potential of a free Fermi gas.   Eq.~(\ref{oct_15_4}) is often  called a Luttinger-Ward expression for
the thermodynamic potential in Eliashberg-type theories. For a
spin interaction, the factor $1/2$ in the second term in
   Eq.~(\ref{oct_15_4}) is replaced by $3/2$~\cite{rob_1}. We note in
passing that the frequency sum in $\Xi _{\mathrm{FG}}$ formally
diverges, but its temperature-dependent part, which is what we
 need, can be extracted by using the following spectral representation
\begin{equation}
|\omega _{m}|=-\frac{1}{\pi }\int \frac{dx~x}{x-i\omega _{m}}.
\label{oct_15_5}
\end{equation}
The above relation in conjuction with the following identity
\begin{equation}
T\sum_{\omega _{m}}~\frac{1}{x-i\omega
_{m}}=\frac{1}{2}-n_{F}(\omega _{m}),
\end{equation}
can be used to extract the $T$-dependent part of $\Xi _{\mathrm{FG}}$, we
thus obtain the familiar Fermi-gas result, $C_{{\rm
FG}}=(2\pi^2N_0/3)T$.

For the thermodynamic potential in the form of   Eq.~(\ref{phononom_2}), we can use the fact that for Landau-damped
collective modes $\Pi (\Omega _{m})\propto \left| \Omega
_{m}\right| $ and expand in $\Pi $, as higher powers of $\Omega $
in the summand in    Eq.~(\ref{phononom_2}) generally lead to higher
powers of $T$ in $\Xi $. Expanding the logarithm, we see that the
term linear in $\Pi $ drops out, and the thermodynamic potential
is given by
\begin{equation}
\Xi =-T\sum_{\omega _{m}}\int \frac{d^{D}k}{(2\pi )^{D}}\ln \left[
{\epsilon
_{\mathbf{k}}^{2}+}\left\{ {\omega _{m}+\Sigma (\omega _{m})}\right\} {^{2}}%
\right] =-2T\sum_{\omega _{m}}\int \frac{d^{D}k}{(2\pi )^{D}}\ln {%
G^{-1}(k,\omega ).}  \label{oct_20_1}
\end{equation}
This is the same expression as    Eq.~(\ref{o1}), obtained by a different
approach as compared to that in Ref.~\cite{agd}. In a Fermi liquid, $\Sigma (\omega
_{m})=\lambda \omega
_{m}$ at the lowest frequencies. Substituting this form into    Eq.~(\ref{oct_20_1}%
), we immediately obtain
\begin{equation}
C_{{\rm FL}}=C_{\mathrm{FG}}(1+\lambda ),  \label{oct_17_1}
\end{equation}
Thus, both approaches--the one based on the Luttinger-Ward
functional and
the one used in Ref.~\cite{agd}--give the same result for the  FL-part of $%
C(T)$.

The issue now is whether the approximate form of $\Xi $ [Eq.~(\ref{oct_20_1})], when modified to include a finite-temperature self-energy can  correctly describe the non-analytic corrections
to the Fermi liquid. We argue that it does not. Indeed, expanding
the logarithm to second order in $\Pi _{m}$ in   Eq.~(\ref
{phononom_2}), we obtain
\begin{equation}
\Xi =-T\sum_{\omega _{m}}\int \frac{d^{D}k}{(2\pi )^{D}}\ln \left[
{\epsilon
_{\mathbf{k}}^{2}+}\left\{ {\omega _{m}+\Sigma (\omega _{m})}\right\} {^{2}}%
\right] +T\sum_{\omega _{m}}\int \frac{d^{D}Q}{(2\pi )^{D}}\Pi
^{2}(Q,\Omega _{m})D_{0}^{2}(Q).  \label{oct_15_12}
\end{equation}
To avoid further complications with a long-range interaction, we
assume that $D_{0}(0)$ is finite. Then, as we showed in
Sec.~\ref{sec:sh_omega_m},  the integral over $Q$ gives $\Omega
_{m}^{2}\ln |\Omega _{m}|$ which, upon
summation over $\Omega _{m}$, results in a non-analytic term in $\Xi $ ($%
T^{3}$ in 2D and $T^{4}\ln T$ in 3D--see Appendix G). For these situations, the boson contribution to    Eq.~(\ref{phononom_2}) cannot be
neglected when evaluating the non-analytic term in the specific
heat.

We note in passing that expanding the logarithm in
   Eq.~(\ref{oct_15_4}) to order $\Pi ^{2}D_{0}^{2}$ results in
\begin{equation}
\Xi =\Xi _{FG}-\frac{1}{4}T\sum_{\omega }\int \frac{d^{D}q}{(2\pi
)^{D}}\Pi ^{2}(q,\Omega )D_{0}^{2}(q).  \label{oct_15_12_1}
\end{equation}
Comparing   Eqs.~(\ref{oct_15_12}) and   (\ref{oct_15_12_1}), we see
that the first term in    Eq.~(\ref{oct_15_12}) must be twice the
integral of $\Pi ^{2}D_{0}^{2}$. This is consistent with the
observation we made in Sec.~\ref{sec:cviasigma}, where we found a
factor of four difference between $-2T\ln G^{-1}$ [the first term
in    Eq.~(\ref{oct_15_12})] and   Eq.~(\ref{oct_15_12_1}). However, a
factor of two difference was attributed to neglecting the
temperature dependence of $\Sigma (\omega _{m})$ while converting
from Matsubara to real frequencies.

For completeness, we also demonstrate how one can obtain the
Fermi-liquid result,   Eq.~(\ref{oct_17_1}) from   Eq.~(\ref{oct_15_4}), which expresses $\Xi $ in terms of the boson
propagator. As we have already mentioned, in order to get a
Fermi-liquid, $T^{2}$-term in $\Xi ,$ one has to expand to first order in $%
\Pi (\Omega _{m},Q)\propto |\Omega _{m}|$. To reproduce   Eq.~(\ref{oct_17_1}), one therefore needs to relate the Landau damping
term to $\lambda $. This relation can be found for arbitrary
$D_{0}(Q)$. To shorten the presentation, we just consider a model
form of $D_{0}\left( Q\right) $--a Lorentzian peaked at $Q=0$
\begin{equation}
D_{0}(Q)=\frac{D_{0}}{Q^{2}+\xi ^{-2}}.  \label{ex_1}
\end{equation}
For this form of $D_{0}(Q)$, the fermion self-energy at the lowest
frequencies is readily obtained from    Eq.~(\ref{2ndord}):
\begin{equation}
\Sigma (\omega _{m})=\lambda \omega _{m},~~~\lambda
=\frac{g^{2}}{4\pi v_{F}\xi ^{-1}}.  \label{oct_17_6}
\end{equation}
The polarization bubble at low frequencies and small momenta is
also obtained from    Eq.~(\ref{2ndord}):
\begin{equation}
\Pi (\Omega _{m},Q)=\frac{\gamma }{D_{0}}\frac{|\Omega _{m}|}{Q},
\label{oct_17_2}
\end{equation}
where $\gamma =mg^{2}/(\pi v_{F})$ (we set interatomic distance
$a=1$). Substituting    Eq.~(\ref{oct_17_2}) into    Eq.~(\ref{oct_15_4}),
integrating over boson momentum and collecting terms of order
$T^{2}$ in $\Xi $, we obtain
\begin{equation}
\Xi =-mT\sum_{\omega _{m}}|\omega _{m}|+\frac{1}{8}\gamma \xi
T\sum_{\Omega _{m}}|\Omega _{m}|.  \label{oct_17_3}
\end{equation}
Summation over boson frequencies is performed by using spectral
representation, in the same way as the sum over fermion
frequencies in    Eq.~(\ref {oct_15_4_1}). For temperature-dependent
parts of the two sums in    Eq.~(\ref {oct_17_3}) we find
\begin{equation}
T\sum |\omega _{m}|\rightarrow \frac{\pi }{6}T^{2};~~~~~T\sum
|\Omega _{m}|\rightarrow -\frac{\pi }{3}T^{2}.  \label{oct_17_4}
\end{equation}
Substituting this into    Eq.~(\ref{oct_17_3}), we obtain
\begin{equation}
\Xi =-\frac{m\pi }{6}T^{2}\left[ 1+\frac{g^{2}}{4\pi v_{F}\xi
^{-1}}\right]. \label{oct_17_5}
\end{equation}
Comparing    Eq.~(\ref{oct_17_5}) and    Eq.~(\ref{oct_17_6}), we see that
\begin{equation}
\Xi =-\frac{m\pi }{6}T^{2}\left[ 1+\lambda \right],
\label{oct_17_7}
\end{equation}
\emph{i.e.},   Eq.~(\ref{oct_17_1}) is reproduced.

Finally, we discuss the specific heat near a Quantum Critical
Point (QCP) which, formally,  corresponds to the limit of $\xi
=\infty $ in    Eq.~(\ref{ex_1}). Here, we find
\begin{equation}
\Xi =\Xi _{\mathrm{FG}}+\frac{1}{4\sqrt{3}}\gamma ^{2/3}T\sum
|\Omega _{m}|^{2/3}.  \label{oct_17_8}
\end{equation}
Evaluating the sum using
\begin{equation}
T\sum |\Omega _{m}|^{2/3}\rightarrow -\frac{T}{u}\sum\int \frac{dxx^{4}\mathrm{%
sgn}x}{x^{3}+i\Omega _{m}}  \label{oct_17_9}
\end{equation}
where $u=\int_{0}^{\infty }dz/(z^{3}+1)=2\pi /(3\sqrt{3})$, we
obtain
\begin{equation}
\Xi =\Xi _{\mathrm{FG}}-\frac{0.4803}{\pi }\gamma ^{2/3}T^{5/3}.
\label{oct_17_10}
\end{equation}
Using    Eq.~(\ref{comega}), we obtain
\begin{equation}
C(T)=C_{\mathrm{FG}}+\frac{0.5337}{\pi }\gamma ^{2/3}T^{2/3}.
\label{oct_17_11}
\end{equation}

  Eq.~(\ref{oct_17_10}) also allows one to verify the conjecture in
Ref.~\cite{agd} that   Eq.~(\ref{oct_20_1}) can be used to evaluate
the specific heat beyond the Fermi-liquid term. Evaluating the
self-energy at QCP, we find~\cite {pepin,ferro}
\begin{equation}
\Sigma (\omega _{m})=\omega _{m}^{2/3}\omega _{0}^{1/3},~~~~\omega
_{0}^{1/3}=\frac{1}{2\pi \sqrt{3}}~\frac{g^{2}}{v_{F}\gamma
^{1/3}}. \label{oct_15_15}
\end{equation}
Substituting this result into    Eq.~(\ref{oct_20_1}), and evaluating the
momentum integral and the frequency sum, we obtain
\begin{equation}
\Xi =-T\sum_{\omega _{m}}\int \frac{d^{D}k}{(2\pi )^{D}}\log \left[ {%
\epsilon _{\mathbf{k}}^{2}+}\left\{ {\omega }_{m}{+\Sigma (\omega }_{m}{)}%
\right\} {^{2}}\right] =\Xi _{\mathrm{FG}}-\frac{0.3546}{\pi
}\gamma ^{2/3}T^{5/3},  \label{oct_15_16}
\end{equation}
which differs from    Eq.~(\ref{oct_17_10}) by a numerical prefactor. We
see that using the zero-temperature self-energy in
   Eq.~(\ref{oct_20_1}), we do not reproduce    Eq.~(\ref{oct_17_10}). This is
another indication that   Eq.~(\ref {oct_20_1}) is not valid for the
calculations of $C(T)$ beyond the Fermi-liquid term.

To avoid confusion, we emphasize that   Eq.~(\ref{phononom_2}) is
valid only for Eliashberg-type theories. For the problem that we
considered in the main text, the self-energy is $k-$dependent, and
this eventually makes   Eq.~(\ref {o1})  valid to second order in
$U$, provided that one uses a finite-$T$ self-energy instead of a
zero-$T$ one. In view of the above consideration, however, we do
not expect   Eq.~(\ref{o1}) to remain valid at higher orders in $U$.
In any event, it is always safe to use   Eq.~(\ref{phononom_1}) in
the calculations of the specific heat.

\section{Evaluation of the thermodynamic potential in real
frequencies}
\label{sec:omega_re_freq}

In this Appendix, we find the specific heat by calculating the
thermodynamic potential which is expressed in  real frequencies. The evaluation of the
second-order contribution to $\Xi$ is straightforward: we just
replace the Matsubara sum in    Eq.~(\ref{e2}) by a contour integral.
Using   Eq.~(\ref{comega}), we obtain
\begin{equation}
\delta C\left( T\right) /T=-\frac{\partial ^{2}\Xi _{2}}{\partial T^{2}}%
=U^{2}\frac{\partial }{\partial T}\int_{0}^{\infty }\frac{d\Omega }{4\pi
}%
\frac{\Omega }{T^{2}\sinh ^{2}\Omega /2T}\int \frac{d^{2}Q}{\left( 2\pi
\right) ^{2}}\Pi _{2}^{R}(\Omega ,Q),
\end{equation}
where $\Pi _{2}^{R}(\Omega ,Q)=$\textrm{Im}$\Pi ^{2}\left( \Omega
+i0^{+},Q\right) $ [see   Eq.~(\ref{a5})]. When differentiating $\Xi
$ with respect to $T$, we assumed that the non-analytic part of
the particle-hole bubble does not depend on temperature--keeping
this source of the temperature dependence would result only in
analytic, $T^{2}$-terms in $C\left( T\right) /T.$ Using
  Eq.~(%
\ref{a6}) for the singular part of  $\Pi _{2}^{R}(\Omega ,Q)$ near
$Q=0$, performing elementary integrations, and multiplying the
result by $2$ to account for the contribution from $2k_{F}$, we
indeed reproduce   Eq.~(\ref{aa1_1}).

Next, we demonstrate how non-perturbative contributions to
$C\left( T\right)$ cancel out in the real-frequency formalism. To
this end, we use series    Eq.~(\ref{e1}) for the thermodynamic potential
\begin{equation}
\Xi =\Xi_{0}+\int_{q}\left[ -2U\Pi _{m}+\frac{1}{2}\left( U\Pi
_{m}\right) ^{2}-\frac{1}{2}\ln \mathcal{G}_{\rho }-\frac{3}{2}\ln \left(
-%
\mathcal{G}_{\sigma }\right) \right] ,  \label{o4}
\end{equation}
where the effective vertices of charge and spin channels,
$\mathcal{G}_{\rho
}$ and $\mathcal{G}_{\sigma },$ are defined in   Eq.~(\ref{f8}). We recall
that the charge term --$\left( 1/2\right) \ln \mathcal{G}_{\rho }$
--contains the contribution from the zero-sound mode. It is convenient to
single out first- and second-order contributions in $U$, \emph{i.e}., to
re-arrange    Eq.~(\ref{o4}) as
\begin{equation}
\Xi =\Xi _{0}+\Xi _{1}+\Xi _{2}+\Xi ^{\prime },
\end{equation}
where
\begin{equation}
\Xi _{1}=-\int_{q}U\Pi ,~~\Xi _{2}=-\frac{1}{2}\int_{q}\left( U\Pi
_{m}\right) ^{2},  \label{o55}
\end{equation}
and
\begin{equation}
\Xi ^{\prime }=\Xi _{1}-2\Xi _{2}+\int_{q}\left[ -\frac{1}{2}\ln
\mathcal{G}%
_{\rho }-\frac{3}{2}\ln \left( -\mathcal{G}_{\sigma }\right) \right]
\label{o5}
\end{equation}
contains a combined contribution of all higher orders. Our goal is
to show that there is no non-analytic contribution to the specific
heat from $\Xi ^{\prime }$. Converting Matsubara sums into contour
integrals, we obtain for the non-perturbative part of the specific
heat corresponding to $\Xi ^{\prime }$
\begin{eqnarray}
C^{\prime }\left( T\right) /T &=&-\frac{\partial ^{2}\Xi ^{\prime }}{%
\partial T^{2}}=\frac{\partial }{\partial T}\int_{0}^{\infty
}\frac{d\Omega
}{2\pi }\frac{\Omega }{T^{2}\sinh ^{2}\Omega /2T}\int \frac{d^{2}Q}{\left(
2\pi \right) ^{2}}  \notag \\
&&\times \left[ \frac{1}{2}\arg \mathcal{G}_{\rho }+\frac{3}{2}\arg \left(
-%
\mathcal{G}_{\sigma }\right) -U^{2}\Pi _{2}^{R}\right]   \label{apr20_12}
\end{eqnarray}
\begin{figure}[tbp]
\begin{center}
\epsfxsize=0.4\columnwidth
\epsffile{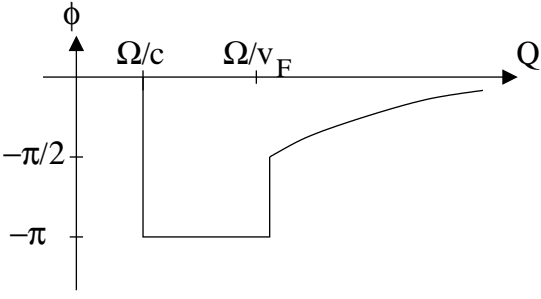}
\end{center}
\caption{ Argument of charge vertex, $\protect\phi =\arg \mathcal{G}_{%
\protect\rho }$, where $\mathcal{G}_{\protect\rho }$ is defined in
  Eq.~(\ref {f8}), as a function of the boson momentum $Q$ at fixed
frequency
$\Omega $%
. $Q=\Omega /c$ corresponds to the position of zero-sound pole, whereas $%
Q=\Omega /v_{F}$ corresponds to the position of particle-hole continuum
boundary.}
\label{fig:argument}
\end{figure}
When differentiating  Eq.~(\ref{apr20_12}), we assumed again that
the only $T-$ dependence comes from the Bose distribution
function. The $T-$ dependence of vertices becomes important either
near a finite-temperature critical point \cite{brinkman} or in 1D;
neither of the cases are considered in this paper.

The momentum dependence of $\arg \mathcal{G}_{\rho }$ at fixed
$\Omega
>0$ is shown in Fig.~\ref{fig:argument} (solid line). For
$Q<\Omega /c,$ where $c$ is the zero-sound velocity, \textrm{Re}$%
\mathcal{G}_{\rho }>0$ and \textrm{Im}$\mathcal{G}_{\rho }=0,$
hence $\arg \mathcal{G}_{\rho }=0.$ At the zero-sound pole,
$Q=\Omega /c,$ $\arg \mathcal{G}_{\rho }$ jumps from zero to $-\pi
.$ It remains equal to $-\pi $ until
$%
Q$ reaches the boundary of the particle-hole continuum, $Q=\Omega
/v_{F}$. At this point, $\arg \mathcal{G}_{\rho }$ jumps from
$-\pi $ to $-\pi /2$ and decreases monotonically upon further
increase in $Q$. On the other
hand, $%
\arg \mathcal{G}_{\sigma }$ is nonzero only inside the
particle-hole continuum ($Q>\Omega /v_{F})$. Finally, as $\Pi
_{2}^{R}\propto \delta (\Omega ^{2}-v^2_{F}Q^{2})$, this term is
only relevant at the boundary of the particle-hole region. This
behavior of $\arg \mathcal{G}_{\rho }$, $\arg \mathcal{G}_{\sigma
}$, and $\Pi _{2}^{R}$ suggests that it is convenient to split the
momentum integral into the one over $Q<\Omega /v_{F}$ and another
one over $Q>\Omega /v_{F}$, and consider three terms in
   Eq.~(\ref{apr20_12}) separately. In the region $Q<\Omega /v_{F}$, only
the collective mode contributes to the momentum integral in
   Eq.~(\ref{apr20_12}), and this contribution is given by
\begin{equation}
\int_{Q<\Omega /v_{F}}\frac{d^{2}Q}{\left( 2\pi
\right) ^{2}}\frac{1}{2}\arg
\mathcal{G}_{\rho }=\frac{1}{2\pi }\int_{\Omega /c}^{\Omega
/v_{F}}dQQ\left(
-\frac{\pi }{2}\right) =-\left( \frac{1}{v_{F}^{2}}-\frac{1}{c^{2}}\right)
\frac{\Omega ^{2}}{8}\approx -\frac{u^{2}}{8}\frac{\Omega
^{2}}{v_{F}^{2}}.
\label{o7}
\end{equation}
At the last step, we have expanded the full result to order
$u^{2}.$ The contribution from $\arg \mathcal{G}_{\rho }$ in the
particle-hole region $Q>\Omega /v_{F}$ is
\begin{eqnarray}
&&\int_{Q>\Omega /v_{F}}\frac{d^{2}Q}{\left( 2\pi
\right) ^{2}}\frac{1}{2}%
\left[ -\tan ^{-1}\left( \frac{u}{1+u}\frac{\Omega }{\sqrt{%
v_{F}^{2}Q^{2}-\Omega ^{2}}}\right) +\frac{u}{1+u}\frac{\Omega }{\sqrt{%
v_{F}^{2}Q^{2}-\Omega ^{2}}}\right]  \notag \\
&=&\left( \frac{u}{1+u}\right) ^{2}\frac{\Omega ^{2}}{16v_{F}^{2}}\approx
u^{2}\frac{\Omega ^{2}}{16v_{F}^{2}}.  \label{o9}
\end{eqnarray}
To ensure the convergence of the $Q$-integral at large momenta, we
subtracted off a term proportional to \textrm{Im}$\Pi ^{R}$, which
gives no
$%
T^{2}-$ contribution to $C\left( T\right) $, and used that $%
\int_{0}^{\infty }dx\left( \tan ^{-1}x^{-1/2}-x^{-1/2}\right) =-\pi /2.$
The contribution from $\arg \mathcal{G}_{\sigma }$ comes only from the
particle-hole region and is equal to
\begin{eqnarray}
&&\int_{Q>\Omega /v_{F}}\frac{d^{2}Q}{\left( 2\pi
\right) ^{2}}\frac{3}{2}%
\left[ \tan ^{-1}\left( \frac{u}{1-u}\frac{\Omega }{\sqrt{%
v_{F}^{2}Q^{2}-\Omega ^{2}}}\right) -\frac{u}{1-u}\frac{\Omega }{\sqrt{%
v_{F}^{2}Q^{2}-\Omega ^{2}}}\right]  \notag \\
&=&-\left( \frac{u}{1-u}\right) ^{2}\frac{3\Omega
^{2}}{16v_{F}^{2}}\approx
-u^{2}\frac{3\Omega ^{2}}{16v_{F}^{2}}.  \label{o10}
\end{eqnarray}
Again, a term proportional to \textrm{Im}$\Pi ^{R}$ has been
subtracted off to ensure convergence. Finally, the $U^{2}\Pi
_{2}^{R}$-term in   Eq.~(\ref{apr20_12}), coming from the boundary
of the particle-hole region, yields
\begin{equation}
\frac{1}{4}u^{2}\frac{\Omega ^{2}}{v_{F}^{2}}.  \label{o11}
\end{equation}
Adding up Eqs.~(\ref{o7}-\ref{o11}), we find that
\begin{equation}
C^{\prime }\left( T\right) =0,
\end{equation}
as it was anticipated. Once again, this means that
non-perturbative corrections do not change the result at order
$u^{2}$--it is still given by   Eq.~(\ref{aa1}).

\section{Evaluation of the diagrams for the thermodynamic potential}
\label{app:extra}

\subsection{second-order in the interaction}
We have shown in the main text that, to second order in $U(Q)$,
the
 thermodynamic potential, $\Xi$,  contains a non-analytic, $T^3$-term
 whose magnitude depends only on  $U(0)$ and $U(2k_F)$.
 To this order in the interaction, $\Xi$ consists of two particle-hole bubbles, and the argument
 for the non-analyticity in $\Xi$ was that it originated from both
 $Q=0$ and $Q= 2k_F$ non-analyticities of the bubbles.
 This argument,  however, does not specify the
 relation between fermion momenta in the two bubbles, and therefore
 it does not distinguish between the cases when the total incoming momentum in the two vertices
 of diagrams 2(b) and 3(a) in  Fig.~\ref{fig:omega} is near zero or
 near $2k_F$.

Now, we look into
 diagrams 2(b) and 3(a) in  Fig.~\ref{fig:omega} in more detail, and show that
the non-analytic term in $\Xi$  involves
 only vertices with ``1D'' momentum structure $({\bf k},-{\bf k};{\bf k},-{\bf k})$ ($Q=0$ contribution)
  and $({\bf k},-{\bf k};-{\bf k},{\bf k})$ ($Q=2k_F$ contribution). For definiteness, we consider
 diagram 2(b) and focus on the $Q=0$ contribution.
The $2k_F$-contribution to diagram 2(b) and  diagram 2(a)
 can be treated in a similar manner.

The argument why the momenta in the two bubbles  are related to each other
 is based on the
 observation that  in order to have
$\Xi = T \sum_{\Omega_m} \Xi_\Omega \propto T^3$,
 the summand $\Xi_\Omega\propto \int d^2Q \Pi^2 (\Omega_m,Q)$ must be non-analytic in $\Omega$.
  The momentum integral does, indeed, diverge
 logarithmically at $v_F Q \gg \Omega_m$, as
 $\Pi^2 (\Omega_m,Q)$ contains a term $\Omega_m^2/Q^2$, which is just the product of
 the $\Omega_m/Q$ terms in  each  of the bubbles. Then
$\Xi_\Omega \propto \Omega_m^2 \ln |\Omega_m|$, and the summation over
 Matsubara frequencies yields $\Xi \propto T^3$.

 We now demonstrate that the $|\Omega_m|/Q$ term
in $\Pi (\Omega_m,Q)$ comes from
 integration over internal momenta ${\bf k}$ in
a narrow range around ${\bf k}\cdot{\bf Q} =0$.  To see this, we
 recall that the bubble has the following form -
 \begin{equation}
 \Pi (\Omega_m,Q)=(1/2\pi^2) T \sum_{\omega_m} \int d^2k G_0(\omega_m,{\bf k})
 G_0(\omega + \Omega,{\bf k} + {\bf Q}).
 \end{equation}
  Expanding
 $\epsilon_{{\bf k} + {\bf Q}}$
near the Fermi surface  as $\epsilon_k + v_F Q\cos \theta$ and
replacing the integration over $d^2k$
 by that over $d\epsilon_k d \theta $, we obtain
\begin{equation}
\Pi (\Omega_m,Q) \propto T \sum_{\omega_m}  \int d \epsilon_k\int
d\theta \frac{1}{\left(i\omega_m - \epsilon_k\right)
\left(i\omega_m + i\Omega_m - \epsilon_k - v_F Q \cos
\theta\right)}. \label{k_1}
\end{equation}
For the dynamic part of $\Pi (\Omega_m,Q)$, the order in which the
integration  over $\epsilon_k$ and frequency summation  are
performed does not matter. Integrating over
 $\epsilon_k$ first, and then summing over frequency,  we obtain
 \begin{equation}
\Pi (\Omega,Q) \propto \int d \theta \frac{\Omega_m}{i\Omega_m  -
v_F Q \cos \theta}. \label{k_2}
\end{equation}
Integration over $\theta$ gives
\begin{equation}
\Pi (\Omega_m,Q) \propto \frac{|\Omega_m|}{\sqrt{\Omega_m^2  + v^2
_F Q^2}}. \label{k_33}
\end{equation}
It is important that typical $|\cos \theta|$ in the angular
integral  are of order
 $|\Omega_m|/v_F Q$. For large $Q$,    Eq.~(\ref{k_33}) reduces to $\Pi (\Omega_m,Q) \propto |\Omega_m|/Q$,
 whereas typical angles are near $\pm\pi/2$: $|\theta\pm\pi/2|\simeq |\Omega_m|/v_FQ$,
 {\it i.e.}, ${\bf k}$ and ${\bf Q}$ are nearly orthogonal.

Since typical momenta in both bubbles in diagram 2(b) of Fig.~
\ref{fig:omega} are nearly orthogonal to the same vector, ${\bf
Q}$, they are either almost parallel or antiparallel. In the first
case, the momentum structure of both vertices  in diagram 2(b) is
$({\bf k},{\bf k};{\bf k},{\bf k})$, in the second, it is $({\bf
k},-{\bf k};{\bf k},-{\bf k})$.
\begin{figure}[tbp]
\begin{center}
\epsfxsize=0.6\columnwidth \epsffile{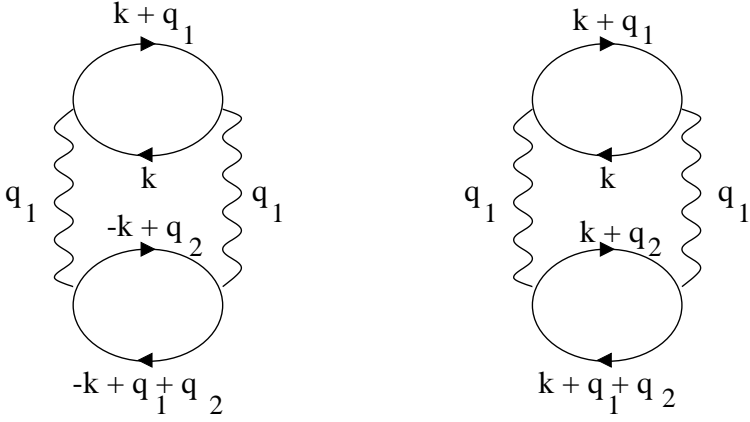}
\end{center}
\caption{A second-order ring diagram for the thermodynamic
potential. Momenta ${\bf q}_1$ and ${\bf q}_2$ are small: ${\bf
q}_1,{\bf q}_2\ll |{\bf k}|\approx k_F$. Labelling on the left and
right diagrams corresponds to backscattering and forward
scattering, correspondingly.} \label{fig:app2}
\end{figure}
We now demonstrate that only the backscattering vertex, $({\bf
k},-{\bf k};{\bf k},-{\bf k})$, contributes to
 the non-analyticity in $\Xi$. To see this, we evaluate diagram
2(b) in a different way, namely, via integrating the product of
four Green's functions over a common fermion momentum (momentum
${\bf k}$ in Fig.~\ref{fig:app2}) rather than pairing them into
bubbles. As shown in Fig.~\ref{fig:app2}, the momenta of the four
Green's functions involved are labelled as either $G_{\bf k}
G_{{\bf k}+{\bf q}_1} G_{-{\bf k}+{\bf q}_2} G_{-{\bf k}+ {\bf
q}_1 + {\bf q}_2}$ (backscattering) or $G_{\bf k} G_{{\bf k}+{\bf
q}_1} G_{{\bf k}+{\bf q}_2} G_{{\bf k}+ {\bf q}_1 + {\bf q}_2}$
(forward scattering). The arguments presented above for a single
bubble suggest that the non-analyticity in $\Xi$ comes from the
region of small
 $q_1$ and $q_2$. We show that the integral over small  $q_1$ and $q_2$ indeed
 gives the correct result, but only in the first case, when two fermion
momenta are near ${\bf k}$ and another two are near $-{\bf k}$,
while in the second case, when all four momenta are near ${\bf
k}$, the integral over small
  $q_1$ and $q_2$ vanishes.

We begin with the first case. We first assume, and then verify,
that one can expand the quasi-particle spectra to second order in
$q_1$ and $q_2$. For a circular Fermi surface, we have
\begin{equation}
\epsilon_{{\bf k}+{\bf q}_1} = \epsilon_k + v_F q_1 \cos \theta_1
+ \frac{q^2_1}{2m}; \epsilon_{-{\bf k}+{\bf q}_2} = \epsilon_k -
v_F q_2 \cos \theta_2 + \frac{q^2_2}{2m}; \epsilon_{-{\bf k}+{\bf
q}_1 +{\bf q}_2} =
 \epsilon_k - v_F (q_1 \cos \theta_1 + q_2 \cos \theta_2) +
\frac{({\bf q}_1 + {\bf q}_2)^2}{2m}. \label{k_3}
\end{equation}
We also assume, following the reasoning for a single bubble, that
both ${\bf q}_1$ and ${\bf q}_2$ are nearly orthogonal to ${\bf
k}$. This implies that ${\bf q}_1$ and ${\bf q}_2$ are either
nearly parallel or nearly antiparallel. Collecting the overall
factor for diagram 2(b), expanding in $\theta_1$ and $\theta_2$
near
 $\pm \pi/2$ and introducing new variables $x = v_F q_1 \cos \theta_1, ~y = v_F q_2 \cos \theta_2$,
 we obtain the contribution to diagram 2(b) from small $q_1$ and $q_2$
\begin{eqnarray}
\Xi^{\rm B}_{Q=0} &=& -\frac{2 m U^2(0)}{(2\pi)^5 v^2_F} T^3
\sum_{\omega_m,\Omega_m,\omega_m^\prime}~ \int_{-\infty}^\infty
d\epsilon_k  \int_{-\infty}^\infty dx  \int_{-\infty}^\infty dy
\int_0^\infty dq_1
 \int_0^\infty dq_2\nonumber\\
 &&\times\frac{1}{(i\omega_m - \epsilon_k) (i\omega + i\Omega_m - \epsilon_k -x - q^2_1/2m)}
~\frac{1}{i\omega_m^\prime - \epsilon_k + y - q^2_2/2m}\nonumber\\
&&\times\left[\frac{1}{i\omega_m^\prime + i\Omega_m - \epsilon_k
+x +y - (q_1 +q_2)^2/(2m)} + \frac{1}{i\omega_m^\prime + i\Omega_m
- \epsilon_k +x +y - (q_1 -q_2)^2/2m}\right]. \label{k_4}
\end{eqnarray}
Integrating over $y$ first, and then over $x$, we obtain
\begin{eqnarray}
\Xi^{\rm B}_{Q=0} &=&  \frac{m U^2(0)}{(16\pi^3 v^2_F)} T^3
\sum_{\omega_m,\Omega_m,\omega_m^\prime}~ \int_{-\infty}^\infty
d\epsilon_k \int_0^\infty dq_1 \int_{-\infty}^\infty dq_2
 \left[\text{sgn} (\omega_m + \Omega_m) + \text{sgn} \Omega_m\right] \left[\text{sgn} \omega_m^\prime -
\text{sgn} (\omega_m^\prime + \Omega_m)\right]\nonumber\\
 &&\times\frac{1}{i\omega_m -
\epsilon_k} ~\frac{1}{i\omega_m + 2 i\Omega_m) - \epsilon_k - q_1
(q_1 +q_2)/m}. \label{k_5}
\end{eqnarray}
Integrating next over $\epsilon_k$, then over $q_2$, and finally
over $q_1$ using
\begin{equation}
 \int_{-\infty}^\infty dq_2 \frac{1}{q_1 (q_1 +q_2)/(m) - 2 i \Omega_m} = i \frac{\pi m}{q_1} \text{sgn}\Omega~
 \left[\text{sgn} \omega_m - \text{sgn} (\omega_m + 2
 \Omega_m)\right],
\end{equation}
we obtain, to logarithmic accuracy
\begin{equation}
\Xi^{\rm B}_{Q=0} = -\frac{m^2 U^2(0)}{(16\pi v^2_F)} T
\sum_{\Omega_m} \ln{\frac{E_F}{|\Omega_m|}}~ S_\Omega, \label{k_7}
\end{equation}
where
\begin{equation}
S_\Omega = T \sum_{\omega_m} \left[\text{sgn} \omega_m -
\text{sgn} (\omega_m + 2 \Omega_m)\right]~ \left[\text{sgn}
(\omega_m + \Omega_m) + \text{sgn} \Omega_m\right] T
\sum_{\omega_m^\prime} \left[\text{sgn} \omega^\prime_m -
\text{sgn} (\omega_m^\prime + \Omega_m)\right]. \label{k_8}
\end{equation}
Performing the frequency summation, we finally obtain
\begin{equation}
\Xi^{\rm B}_{Q=0} = -\frac{u^2_0}{(2\pi v^2_F)} T \sum_{\Omega_m}
\Omega_m^2
 \ln{\frac{E_F}{|\Omega_m|}},
\label{k_9}
\end{equation}
where $u_0 = mU(0)/(2\pi)$. In Sec.~\ref{sec:sh_omega_m}, we have shown
that
\begin{equation}
T \sum_\Omega  \Omega^2 \ln{\frac{E_F}{|\Omega|}} = -2 T^3 \zeta
(3) + ...,
\end{equation}
where the dots stand for regular $T^2, T^4$, etc terms [see
   Eq.~(\ref{e3}) and    Eq.~(\ref{e33})]. Thus,
\begin{equation}
\Xi^{\rm B}_{Q=0} = \zeta (3)u^2_0\frac{T^3}{\pi v^2_F}.
\label{k_10}
\end{equation}
The evaluation of the $2k_F$-contribution from  diagram 2(b) and
of the entire contribution of diagram 2(a) (the latter involves
only a process in which one of the vertices carries momentum near
zero and the other one near $2k_F$) proceeds in the same way. The
net result from the two diagrams is
\begin{equation}
\Xi^{\rm B}= \zeta (3)(u^2_0 + u^2_{2k_F} - u_0 u_{2k_F})
\frac{T^3} {\pi v^2_F}. \label{k_11}
\end{equation}
In all three contributions, the total momentum in each vertex is
near zero, {\it i.e.}, both the vertices with momentum transfers
near zero and near $2k_F$, have the structure of $({\bf k},-{\bf
k};{\bf k},-{\bf k})$ and $({\bf k},-{\bf k};-{\bf k},{\bf k})$,
respectively. Calculating the specific heat corresponding to $\Xi$
in    Eq.~(\ref{k_11}), we find that the result for $C(T)$ coincides with
that in    Eq.~(\ref{genu}).

The remaining task is to show that the contribution from forward
scattering, {\it i.e.}, from processes of the type $({\bf k},{\bf
k};{\bf k},{\bf k})$, vanishes. To see this, we repeat the same
calculation, assuming now that the momenta
 in all four Green's functions in Fig.~\ref{fig:omega} 2(b) are nearly parallel to each other.
 Instead of     Eq.~(\ref{k_4}), we then obtain
\begin{eqnarray}
&&\Xi^{{\rm F}}_{Q=0} = -\frac{2 m U^2(0)}{(2\pi)^5 v^2_F} T^3
\sum_{\omega_m,\Omega_m,\omega_m^\prime}~
\int_{-\infty}^\infty d\epsilon_k  \int_{-\infty}^\infty dx  \int_{-\infty}^\infty dy \int_0^\infty dq_1 \int_0^\infty dq_2 ~ \frac{1}{(i\omega - \epsilon_k) (i(\omega + \Omega) - \epsilon_k -x - q^2_1/(2m))} \nonumber \\
&&~\times\!\!\frac{1}{i\omega_m^\prime - \epsilon_k -y -
q^2_2/2m}~\!\!\left[\frac{1}{i\omega_m^\prime + i\Omega_m
 - \epsilon_k -x -y - (q_1 +q_2)^2/2m} + \frac{1}{i\omega_m^\prime + i\Omega_m - \epsilon_k -x -y - (q_1 -q_2)^2/2m}
 \right].
\nonumber
\end{eqnarray}
Integrating first over $y$, and then over $x$, we obtain
\begin{equation}
\Xi^{{\rm F}}_{Q=0} \propto \int_{-\infty}^\infty d\epsilon_k
\frac{1}{i\omega_m - \epsilon_k} \left(\frac{1}{i\omega_m -
\epsilon_k + q_1 q_2/m} + \frac{1}{i\omega_m - \epsilon_k - q_1
q_2/m}\right). \label{k_13}
\end{equation}
This integral vanishes as the poles of the integrand are located
in the same half-plane of $\epsilon_k$. This completes our proof
of the statement that, to second order in the interaction, the
singular part of $\Xi$ comes only from backscattering,{\it i.e.},
from the diagrams containing vertices with zero total momentum.

\subsection{third order in the interaction}

Next, we consider the diagrams for $\Xi$ to third order in $U$. As
 we have shown in Sec.~\ref{sec:sh_omega_m}, the non-analyticity in $\Xi$
 results from the logarithmic singularity in the momentum
 integral, followed by the Matsubara sum
 \[T\sum_{\Omega_m}\int_{|\Omega_m|} dQQ\frac{\Omega_m^2}{Q^2}
 \propto T\sum_{\Omega_m} \Omega^2_m\ln|\Omega_m| \propto T^3.\]
  At the second order, the singular term in the momentum integral,
$\Omega_m^2/Q^2$ is obtained by multiplying the dynamic parts of
two polarization bubbles. Higher-order diagrams contain higher
powers of the polarization bubbles; however,  we still need to
select terms that behave only as
 $\Omega_m^2/Q^2$: both less or more divergent terms do not result in a logarithmic non-analyticity of the
 momentum integral, and hence in a $T^3$ non-analyticity in $\Xi$.
 At the third order, such terms are
obtained by selecting the dynamic parts of two out of  three
polarization bubbles, while putting $\Omega_m=0$ in the third
bubble.  A subtle point here is that the static part of the bubble
involves the integration over {\it all} internal fermion momenta,
${\bf l}$, {\it i.e.}, in  contrast to the dynamic part, there is
no correlation between the directions of boson momentum, ${\bf
Q}$, and ${\bf l}$. Hence, vertices that appear in the third order
diagrams are generally not ``one-dimensional'' in a sense that the
four fermion momenta are not directed along the same line.
Furthermore, despite the fact that the static polarization for
free
 fermions in 2D $\Pi (\Omega =0, Q)$ is a constant ($= -m/2\pi$) for
 all $Q\leq 2k_F$, only $\Pi (\Omega =0, Q \rightarrow 0)$ comes
 from the states in the immediate vicinity of the Fermi surface. For generic $Q$,
  the static polarization bubble involves fermion states
 away from the Fermi surface~\cite{chubukov}. As a result, the convolution
 of  $U(Q)$ and two fermion propagators, which form
 a polarization bubble with non-zero external momentum,
 cannot be expressed as the
 angular average of bare interaction between the particles on the Fermi surface.
As an example, consider
 the ``$Q=0$'' contribution from diagram 3(b) in Fig.~\ref{fig:omega}. This contribution  involves only bubbles
  with small
 external momenta, \emph{i.e.}, all internal  fermion
 momenta--${\bf p}$, ${\bf k}$, and ${\bf l}$--are located near the Fermi surface.
Two of these momenta, \emph{e.g.}, ${\bf p}$ and ${\bf k}$,
 must be nearly antiparallel; otherwise there is no non-analyticity in $\Xi$.
However, the direction of the third momentum--${\bf l}$--with
respect to ${\bf k}$ is arbitrary. As a result, depending on the
choice of two antiparallel momenta, the ``$Q=0$'' contribution
from diagram 3(b)
 contains a term proportional to $u^2_0 u_\pi$, where
 \begin{equation}
 u_\theta =
 (m/2\pi) U(2k_F \sin{\theta/2}),
 \end{equation}
  as well as another one proportional to $u^2_0 \langle
u_\theta\rangle$, where
\begin{equation}
\langle u_\theta\rangle = \frac{1}{\pi} \int_0^\pi u(\theta) d
\theta . \label{10_9_4}
\end{equation}
In this last term,  the integration goes over the entire Fermi
surface. [It is understood that $u_{\pi}=u_{2k_F}$.]

The same reasoning applies to  the $2k_F$-contribution from this
diagram.
 In addition,  the $2k_F$-contribution
  involves the convolutions of the interaction potential with the
  Green's functions forming  bubble. As we just said, the corresponding momentum
integral is not confined to the Fermi surface. Evaluating the
integrals, we find that the $2k_F$-part of diagram 3(b) contains a
term proportional to $u^2_\pi u_0$ and another one proportional to
$u^2_\pi \langle \langle u_\theta\rangle\rangle$, where
\begin{equation}
\langle \langle u_\theta\rangle\rangle = \frac{1}{\pi} \int_0^\pi
d\theta ~u_\theta~ \cos{\frac{\theta}{2}}
 ~\ln{\frac{\sqrt{1 + \sin{\theta/2}} +\sqrt{1 - \sin{\theta/2}}}
 {\sqrt{1 + \sin{\theta/2}}- \sqrt{1 - \sin{\theta/2}}}}.
\label{10_9_5}
\end{equation}
(These two terms occur for different choices of two anti-parallel
momenta.) If $u$ does not depend on $\theta$, then $\langle
\langle u_\theta\rangle\rangle = u$. The first term --proportional
to $u^2_\pi u_0$-- involves the static bubble, $\Pi (0,2k_F)$,
which, once again, is determined by the states far away from the
Fermi surface. This term
 is similar to the ``$Q=0$''  contribution from the states near the Fermi
 surface,
simply because for free fermions in $D=2$,
 $\Pi (2k_F) = \Pi (0)$. To distinguish between the Fermi-surface and non-Fermi
 surface contributions, we multiply $\Pi (2k_F)$ by $\langle \langle 1\rangle\rangle$
  (according to    Eq.~(\ref{10_9_5}, $\langle \langle 1\rangle\rangle =1$) to emphasize that the
   integration is not confined to the
 Fermi surface.

Applying this reasoning to all third order diagram, and combining
all choices of choosing two dynamic and one static bubble, we find
\begin{subequations}
\begin{eqnarray}
&&\Xi_{3a} = -\left (u_0 \langle u_\theta u_{\pi -\theta}\rangle +
2 u_0 u_{\pi} \langle u_{\theta}\rangle\right) K + \left(u_{\pi}
\langle \langle u^2_{\theta}\rangle\rangle + 2 u_0 u_{\pi} \langle
\langle u_{\theta}\rangle\rangle\right) K \label{k_14} \label{3a}\\
&& \Xi_{3b} = \left (4 u^2_0 \langle u_\theta\rangle + 2 u^2_0
u_{\pi} \right) K + \left[ 4 u^2_\pi \langle \langle
u_\theta\rangle\rangle +  2 u^2_{\pi} u_0 \langle \langle
1\rangle\rangle \right] K
 \label{3b}\\
&&\Xi_{3c} = - 4 \left[u^3_0  + u^3_{\pi} \langle \langle 1\rangle\rangle\right] K \label{3c}\\
&&\Xi_{3d} = 2 u_{\pi} \langle u_\theta u_{\pi -\theta}\rangle K +
2 u_0 \langle \langle u^2_\theta\rangle\rangle K, \label{3d}
\end{eqnarray}
\end{subequations}
where $K\equiv \zeta(3)T^3/\pi v_F^2$. Adding up
Eqs.~(\ref{3a}-\ref{3d}), we obtain a total third-order
contribution to $\Xi$
\begin{eqnarray}
\Xi_{3} &=& \left[- 4u^3_0 +  2 u^2_0 u_{\pi} + (4 u^2_0 - 2 u_0 u_{\pi})
  \langle u_\theta\rangle - (u_0 -2u_{\pi})  \langle u_\theta u_{\pi -\theta}\rangle \right] K \nonumber \\
&& + \left[(-4 u^3_\pi + 2 u_0 u^2_\pi) \langle \langle
1\rangle\rangle + (4 u^2_\pi - 2 u_0 u_\pi) \langle \langle
u_\theta\rangle\rangle - (u_\pi - 2 u_0) \langle \langle
u^2_\theta\rangle\rangle\right] K.
 \label{k_16}
\end{eqnarray}

Combining the last expression with the second-order result,   Eq.~(\ref{k_11}),
 and using relation    Eq.~(\ref{comega}) between $\Xi$ and $C(T)$, we obtain for $C(T)/T$ to third order in $u$
\begin{equation}
\delta C\left( T\right) /T=- ~\frac{3 m\zeta (3)}{4
\pi}~\frac{T}{E_{F}} \left[\left(2{\tilde u}_{0}- {\tilde u}_{\pi}
+ 2 \langle \langle u^2_\theta\rangle\rangle - \langle u_\theta
u_{\pi-\theta}\rangle\right)^2 + 3 \left({\tilde u}_{\pi} +
\langle u_\theta u_{\pi-\theta}\rangle \right)^2\right],
\label{genu_1_01}
\end{equation}
where
\begin{equation}
{\tilde u}_0 = u_0 ~\left(1 - 2 u_0 + 2 \langle
u_\theta\rangle\right),~~~
 {\tilde u}_{\pi} = u_{\pi}~\left(1 - 2 u_{\pi} \langle \langle 1\rangle\rangle + 2 \langle
 \langle u_\theta\rangle\rangle\right).
\label{genu_11}
\end{equation}

Next, we verify whether   Eq.~(\ref{genu_11}) can be obtained by
substituting the renormalized static vertices into the
second-order result for the specific heat,   Eq.~(\ref{k_11}). To
first order in $u$, we simply have: $\Gamma^k({\bf k},-{\bf
k};{\bf k},-{\bf k})=u_0$ and $\Gamma^k({\bf k},-{\bf k};-{\bf
k},{\bf k})=u_{2k_F}$. To evaluate the third-order contribution to
$C(T)$, we need to renormalize the vertices up to second
order--the third order terms will then result as cross-products of
first and second-order terms. Evaluating
 the vertex corrections, presented diagrammatically in Fig.~\ref{fig:vertex2},
 in the same way as we evaluated the diagrams for $\Xi$, we obtain
\begin{subequations}
\begin{eqnarray}
 \Gamma^k({\bf k},-{\bf k};{\bf
k},-{\bf k})&=&\frac{2\pi}{m} \left[{\tilde u}_{0} + \langle \langle u^2_\theta\rangle\rangle\right]; \label{10_9_6a} \\
\Gamma^k({\bf k},-{\bf k};-{\bf k},{\bf k})&=&\frac{2\pi}{m}\left[
{\tilde u}_{\pi} + \langle u_\theta u_{\pi -\theta}\rangle\right].
\label{10_9_6}
\end{eqnarray}
\end{subequations}
 Replacing $u_0$ and $u_{2k_F}$ in   Eq.~(\ref{k_11}) by renormalized vertices,    Eq.~(\ref{10_9_6a}) and    Eq.~(\ref{10_9_6}),
 correspondingly,  we find after simple manipulations that it does indeed
 reproduce   Eq.~(\ref{genu_1_01}). This proves, to order $u^3$, that the non-analytic
 term in the specific heat is expressed in terms of renormalized
 static vertices $\Gamma^k({\bf k},-{\bf k};{\bf
k},-{\bf k})$ and $\Gamma^k({\bf k},-{\bf k};{\bf k},-{\bf k})$,
\emph{i.e.}, in terms of $\Gamma^k (\pi)$.

Using now the relations between $\Gamma({\bf k},-{\bf k};{\bf
k},-{\bf k})$ and $\Gamma({\bf k},-{\bf k};{\bf k},-{\bf k})$ and
the spin and charge components of $\Gamma^k (\pi)$
\begin{equation}
\Gamma(k,-k;k,-k) = \Gamma^k_c (\pi) - \Gamma^k_s (\pi), \Gamma (k,-k;-k,k)
 = -2\Gamma^k_s (\pi),
\label{10_9_7}
\end{equation}
and restoring quasi-particle $Z$ factors and $m^*/m$, which come
from self-energy insertions not considered above, we obtain   Eq.~(\ref{jul3_4}).

It is also instructive to re-express $\Gamma^k (\pi)$ not in terms
of  the bare interaction potential, but in terms of another
vertex--$\Gamma^\omega(\theta)$,
  which, we remind, is the static vertex renormalized by the states away from the Fermi surface.
   To this end, we separate the Fermi-surface and
 non-Fermi-surface contributions to   Eq.~(\ref{10_9_6a},~\ref{10_9_6}), \emph{i.e.}, re-write
   Eq.~(\ref{10_9_6a},~\ref{10_9_6}) to order $u^3$ as
\begin{subequations}
\begin{eqnarray}
\Gamma^k({\bf k},-{\bf k};{\bf k},-{\bf k})&=& \frac{2\pi}{m}~\left[u_{0} +
\langle \langle u^2_\theta\rangle\rangle\right] \left[1 -2 u_0 + 2
\langle u_\theta\rangle\right];
 \label{10_9_8a}\\
\Gamma^k({\bf k},-{\bf k};-{\bf k},{\bf k}) &=& \frac{2\pi}{m}~u_{\pi}
\left [1 -2
u_\pi \langle \langle 1\rangle\rangle + 2 \langle \langle
u_\theta\rangle\rangle\right) ( 1 + \frac{\langle u_\theta
u_{\pi -\theta}\rangle}{u_\pi} )]. \label{10_9_8}
\end{eqnarray}
\end{subequations}
(For simplicity, we neglected the $Z$-factor and effective mass
renormalizations here and in what follows.) The first brackets in
both formulas come from the states away from the Fermi surface,
\emph{i.e}, they give $\Gamma^{\omega}(\pi)$. The second brackets
come from states near the Fermi surface and account for the
difference between $\Gamma^k$ and $\Gamma^{\omega}$. Introducing
spin and charge components of $\Gamma^\omega$ in the same way as
 in   Eq.~(\ref{10_9_7}), \emph{i.e.}, as
\begin{eqnarray}
&& \Gamma^k (k,-k, k,-k) =\frac{\pi}{m}\left[ f_c (\pi )-f_s (\pi
)\right] ,~~ \Gamma^k (k,-k;-k,k) =
-2\frac{\pi}{m}~f_s (\pi ) \nonumber \\
&& \Gamma^\omega (k,-k, k,-k) =\frac{\pi}{m}\left[ \gamma_c (\pi
)-\gamma_s (\pi )\right] ,~~ \Gamma^\omega (k,-k;-k,k) =
-2\frac{\pi}{m}~\gamma_s (\pi ),
  \label{new_new}
\end{eqnarray}
 we obtain
\begin{eqnarray}
\gamma_c^\omega (\pi) &=& u_0 - \frac{1}{2} u_\pi \left (1 - 2 u_\pi \langle \langle 1\rangle\rangle + 2 \langle \langle u_\theta\rangle\rangle \right) + \langle \langle u^2_\theta\rangle\rangle \nonumber \\
\gamma_s^\omega (\pi) &=& - \frac{1}{2} u_\pi  \left (1 - 2 u_\pi
\langle \langle 1\rangle\rangle + 2 \langle \langle
u_\theta\rangle\rangle \right). \label{10_9_9}
\end{eqnarray}
Substituting   Eq.~(\ref{10_9_9}) into   Eq.~(\ref{10_9_8}) and
(\ref{10_9_7}), we
 obtain the relation between $f_a$ and $\gamma^a$ presented in the main the text [  Eq.~(\ref{10_9_1})].

Finally, when the interaction is strongly peaked at $Q=0$, so that
$u_0$ is much larger than $u_\theta$ for a generic $\theta$,
including $\theta = \pi$, only corrections to $u_0$ matter. These
corrections come from  the ring diagrams and can be summed up
exactly.
 The full combinatoric factor for the $u^n_0$ term
from the ring diagram of order $n$ is $(-1)^n 2^{n-2} (n-1)$.
Evaluating the sum over $n$, we reproduce   Eq.~(\ref{genu_111}).


\begin{thebibliography}{99}
\bibitem[*]{perm} Permanent address.
\bibitem{agd}  A.\ A.\ Abrikosov, L.\ P.\ Gorkov, and I.\ E.\
Dzyaloshinski,
\emph{Methods of quantum field theory in statistical physics}, (Dover
Publications, New York, 1963).
\bibitem{statphys}  E. M. Lifshitz and L. P. Pitaevski, \emph{Statistical
Physics}, (Pergamon Press, 1980).
\bibitem{tom} see e.g., T. Timusk and B. Statt, Rep. Prog. Phys. {\bf 62}, 61 (1999).
\bibitem{piers} see, e.g.,  P. Coleman , C. Pepin, Q. Si , R. Ramazashvili, Journal of
Physics: Condensed Matter {\textbf 13}, 723-738, (2001);
Q. Si et al, Nature, \textbf{413}, 804 (2001).
\bibitem{abanov}
A. Abanov, A. Chubukov and J. Schmalian, Advances in Physics {\bf 52}, 119 (2003).
\bibitem{eliashberg}  G. M. Eliashberg, Sov. Phys. JETP \textbf{16, }780
(1963).
\bibitem{doniach}  S. Doniach and S. Engelsberg,
Phys. Rev. Lett. \textbf{17}
, 750 (1966).
\bibitem{brinkman}  W. F. Brinkman and S. Engelsberg,
Phys. Rev. \textbf{169}%
, 417 (1968). \bibitem{amit} D. J. Amit, J. W. Kane, and H.
Wagner, Phys. Rev. \textbf{175}, 313 (1968); \emph{ibid.}
\textbf{175}, 326 (1968).
\bibitem{comment_1}  This terms are often associated in the literature
with the interaction of fermions with phonons \cite{eliashberg} or
spin fluctuations (``paramagnons'')
\cite{doniach},\cite{brinkman}). However, they arise already at
the second order in the fermion-fermion interaction, with no boson
modes involved--cf. Ref.~\cite{amit}.
\bibitem{belitz}  D. Belitz, T. R. Kirkpatrick, and T. Vojta, Phys. Rev. B
\textbf{55}, 9452 (1997)
\bibitem{bedell}  D. Coffey and K. S. Bedell,
Phys. Rev. Lett. \textbf{71},
1043 (1993).
\bibitem{chm}  a) A. V. Chubukov and D. L. Maslov, Phys. Rev. B
\textbf{68, }%
155113 (2003); b) \emph{ibid. }\textbf{69}, 121102 (2004)
\bibitem{dassarma}  V. M. Galitski and S. Das Sarma, cond-mat/0311559
\bibitem{marenko}  M. A. Baranov, M. Yu. Kagan, and M. S. Mar'enko, JETP
Lett. \textbf{58}, 709 (1993).
\bibitem{millis}  G. Y. Chitov and A. J. Millis, \prl {\bf 86}, 5337
(2001); %
\prb {\bf 64}, 0544414 (2001).
\bibitem{pepin}  A. V. Chubukov, C. P\'{e}pin, and J. Rech,
Phys. Rev. Lett. \textbf{92}, 147003 (2004).
\bibitem{experiment}  D. S. Greywall, Phys. Rev. B {\bf 27}, 2747
(1983); G. R. Stewart, Rev. Mod. Phys. {\bf 86}, 755 (1984); A.
Casey, H. Patel, J. Nyeki, B. P. Cowan, and J. Saunders, Phys.
Rev. Lett. {\bf 90}, 115301 (2003).
\bibitem{belitz_qc}  D. Belitz, T. R. Kirkpatrick, and T. Vojta,
\prl {\bf 82}, 4707 (1999); D. Belitz and T. Vojta, \textit{ibid.}
\textbf{89}, 247202 (2002).
\bibitem{qc}  J. A. Hertz, Phys. Rev. B \textbf{ 14}, 1165
(1976); A. Millis, Phys. Rev. B \textbf{ 48}, 7183 (1993); T.\
Moriya, {\it Spin Fluctuations in Itinerant Electron Magnets}
(Springer-Verlag, Berlin, 1985).

\bibitem{rudin}  A.\ M.\ Rudin, I.\ L.\ Aleiner, and L.\ I.\ Glazman,
\prb
{\bf 55}, 9322 (1997).
\bibitem{zna}  G.\ Zala, B.\ N.\ Narozhny, I.\ L.\ Aleiner, Phys. Rev. B
\textbf{64}, 214204 (2001).
\bibitem{castellani} C. Castellani, C. Di Castro, and W. Metzner,
\prl {72}, 316 (1994).

\bibitem{fukuyama} H. Fukuyama and M. Ogata, J. Phys. Soc. Jpn.
{\bf 63}, 3923 (1995).

\bibitem{metzner} C. Halboth and W. Metzner, \prb {\bf 57}, 8873
(1998).
\bibitem{1D}  Yu. A. Bychkov, L. P. Gor'kov, and I. E. Dzyaloshisnkii,
Zh.\ Eksp. Teor. Fiz. \textbf{50}, 738 (1966) [Sov.\ Phys.\ JETP
\textbf{23}, 489 (1966)]; I.\ E.\ Dzyaloshinskii and A.\ I.\
Larkin, Zh.\ Eksp. Teor. Fiz. \textbf{65}, 411 (1973) [Sov.\
Phys.\ JETP \textbf{38}, 202 (1974)].
\bibitem{anderson}  P. W. Anderson, Phys. Rev. Lett. \textbf{65, }2306
(1990).
\bibitem{2D}  A. V. Chaplik, Zh. Eksp. Teor. Fiz. \textbf{60, }1845 (1971)
[Sov. Phys. JETP \textbf{33}, 997 (1971)]; C. Hodges, H. Smith and J. W.
Wilkins, Phys. Rev. B \textbf{4}, 302 (1971); P. Bloom, Phys. Rev. B
\textbf{%
\ 12}, 125 (1975).
\bibitem{photo} J. C. Campuzano,  M. Randeria,   M. R. Norman and H. Ding,  in
''The Gap Symmetry and Fluctuations in High Tc Superconductors '', edited by J.Bok \emph{et al}., (Plenum, 1998).
\bibitem{eisenstein}  J. P. Eisenstein, T. J. Gramila, L. N. Pfeiffer, and K. W. West
Phys. Rev. B \textbf{44}, 6511-6514 (1991); J. P. Eisenstein, L. N. Pfeiffer, and K. W. West
Phys. Rev. Lett. \textbf{69}, 3804-3807 (1992); S. Q. Murphy, J. P. Eisenstein, L. N. Pfeiffer, and K. W. West Phys. Rev. B \textbf{52}, 14825-14828 (1995).
\bibitem{aleiner} G. Catelani and I. L. Aleiner, cond-mat/04053333.
\bibitem{solyom} J. S{\'o}lyom, Adv. Phys. {\bf 28}, 201 (1979).
\bibitem{DL72}I. E. Dzyaloshinskii and A. I. Larkin, Zh.\ Eksp. Teor. Fiz.
\textbf{61}, 791 (1972) [Sov.\ Phys.\ JETP \textbf{34}, 202
(1972)].
\bibitem{nersesyan} G. I. Japaridze and A. A. Nersesyan, Phys.
Lett. {\bf 94} A, 224 (1983).
\bibitem{maslov} R. Saha and D. L. Maslov (unpublished).
\bibitem{stern}  T. Ando, A. B. Fowler, and F. Stern, Rev. Mod. Phys.
\textbf{54}, 437 (1982).
\bibitem{quinn} G. Guiliani and J. J. Quinn, \prb {\bf 26}, 4421
(1982). \bibitem{kink} We have chosen to present the result for
the Coulomb potential as a function of $\epsilon_k $ rather than
of $\Delta$, in contrast to what we did for the short-range case.
The reason is that, in terms of $\Delta$, the kink in
Eq.(\ref{kink_c}) is at $\Delta=\omega$, \emph{i.e.}, not close to
the mass shell, so that $\Delta$ is not a convenient variable any
more.
\bibitem{anton}  E. G. Mishchenko and A. V. Andreev, Phys. Rev. B
\textbf{65}%
, 235310 (2002).
\bibitem{reizer}  D. V. Khveshchenko and M. Yu. Reizer, Phys. Rev. B \textbf{57},
4245 (1998).

\bibitem{eli_1}  G. M. Eliashberg, Sov. Phys. JETP \textbf{11, }696
(1960).

\bibitem{luttinger}  J. M. Luttinger and J. C. Ward,
Phys. Rev. \textbf{118}%
, 1417 (1960).
\bibitem{wasserman}  A. Wasserman and M. Springford,
Adv. Phys. \textbf{45}, 471 (1996).
\bibitem{bardeen_stephen} J. Bardeen and M. Stephen, Phys. Rev. B {\bf 136},
 A1485 (1964); Y. Wada, Phys. Rev. {\bf 135}, A1481 (1964);
 D. J. Scalapino in ``Superconductivity''
 R. D. Parks, ed., Marcel Dekker, New York, 1969, vol. 1, p.449.
For a recent review, see F. Marsiglio and J.P. Carbotte in ``The physics of Superconductors'', K.H. Bennemann and J.B. Ketterson eds, Springer, 2003, vol. 1, p. 233 and R. Haslinger and A. Chubukov, Phys. Rev. {\bf 68}, 214508 (2003).
\bibitem{ioffe} B. Altshuler, L. B. Ioffe and A. J. Millis, Phys. Rev. B {\bf 52}, 5563 (1995).

\bibitem{grilli} C. Castellani, C. DiCastro, and
M. Grilli, Z. Phys. B {\bf 103}, 137 (1997).
S. Caprara, M. Sulpizi, A. Bianconi, C. Di Castro and M. Grilli, Phys. Rev. B {\bf 59}, 14980 (1999); A. Perali, C. Castellani, C. Di Castro, M. Grilli, E. Piegari and A. A. Varlamov, Phys. Rev. B {\bf 62}, R9295 (2000); S. Andergassen, S. Caprara, C. Di Castro and M. Grilli, Phys. Rev. Lett. {\bf 87}, 056401-1 (2001).

\bibitem{acs} A. Abanov, A. V. Chubukov and J. Schmalian,
 Adv. Phys. \textbf{52}, 119 (2003); A. V. Chubukov, D. Pines and
 J. Schmalian, in
``The physics of Superconductors'', K. H. Bennemann and
J. B. Ketterson eds, Springer, 2003, vol. 1, p. 495.

\bibitem{ferro}  R. Roussev and A. J. Millis, Phys. Rev. B {\bf 63},
 140504 (2001),  Z. Wang, W. Mao and K. Bedell,  Phys. Rev. Lett.
{\bf 87}, 257001 (2001),  A. Chubukov, A. Finkelstein,  R. Haslinger and D. Morr, \prl {\bf 90}, 077002 (2003).



\bibitem{rob_1} R. Haslinger and A. Chubukov, Phys. Rev. {\bf 67}, 140504
 (2003).

\bibitem{chubukov} A. V. Chubukov, Phys. Rev. {\bf 48}, 1097 (1993).


\end{thebibliography}
\end{document}